\documentclass[12pt,a4paper,twoside]{book}
\PassOptionsToPackage{british}{babel}

\usepackage{times}

\usepackage{color}				
\usepackage{graphicx}			

\usepackage{fancyhdr}			
\usepackage{enumerate}			
\usepackage{cite}				

\usepackage{algorithm}
\usepackage{algorithmic}
\usepackage{subfigure}
\usepackage{amsmath}
\usepackage{amssymb}

\usepackage{tabularx}
\usepackage{array}
\usepackage{CV}

\definecolor{RED}{RGB}{255,0,0}
\definecolor{GREEN}{RGB}{0,255,0}
\definecolor{BLUE}{RGB}{0,0,255}
\definecolor{GRAY}{RGB}{128,128,128}
\definecolor{BLACK}{RGB}{0,0,0}

\input{letterspacing.tex} 
\usepackage[nottoc]{tocbibind} 

\usepackage{xspace}

\makeatletter
\def\cleardoublepage{\clearpage\if@twoside \ifodd\c@page\else
\hbox{}
\thispagestyle{empty}
\newpage
\if@twocolumn\hbox{}\newpage\fi\fi\fi}
\makeatother

\usepackage{sectsty}
\allsectionsfont{\large \usefont{T1}{fvs}{b}{n}\selectfont}
\subsectionfont{\normalsize \usefont{T1}{fvs}{b}{n}\selectfont}
\subsubsectionfont{\small \usefont{T1}{fvs}{b}{n}\selectfont}
\paragraphfont{\small \usefont{T1}{fvs}{b}{n}\selectfont}

\definecolor{NUMCOLOR}{rgb}{0.22,0.37,0.56}

\usepackage{multirow} 

\makeatletter
\def\@makechapterhead#1{%

  \line(1,0){0}
  \newline
  {
	 \begin{tabular}{@{}lr@{}@{}}
	 \linethickness{ 4px }\color{NUMCOLOR}\line(1,0){245}
	& \multirow{2}{*}{\fontsize{100}{62}\usefont{OT1}{ptm}{m}{n}\selectfont \color{NUMCOLOR} \thechapter}\\ 
	 & \\
	\scshape \LARGE \usefont{T1}{fvs}{sc}{n}\selectfont \letterspace to 2.5\naturalwidth{CHAPTER} \hspace{3cm}
	& \\
	\end{tabular}

\vskip 100\p@
\raggedleft
    \interlinepenalty\@M
    \scshape \fontsize{24}{30} \usefont{T1}{fvs}{n}{n}\selectfont \scshape \MakeUppercase{#1}\par\nobreak
      \vskip 80\p@
  }
  }

\makeatletter
\def\@makeschapterhead#1{%

  \line(1,0){0}
  \newline
  {
	 \begin{tabular}{@{}lr@{}@{}}
	 \linethickness{ 4px }\color{NUMCOLOR}\line(1,0){245}
	& 
	\\ 
	 & \\
	\scshape \LARGE \usefont{T1}{fvs}{sc}{n}\selectfont \letterspace to 2.5\naturalwidth{} \hspace{3cm} \vspace{0.17cm}
	& \\
	\end{tabular}

\vskip 100\p@
\raggedleft
    \interlinepenalty\@M
    \scshape \fontsize{24}{30} \usefont{T1}{fvs}{n}{n}\selectfont \scshape \MakeUppercase{#1}\par\nobreak
      \vskip 80\p@
  }
  }

\usepackage[titles]{styles/tocloft}
\renewcommand{\cftchapfont}{\fontsize{11}{13}\usefont{T1}{fvs}{b}{n}\selectfont}

\usepackage[font=small,format=plain,labelfont=bf,up,textfont=it,up]{caption}

\usepackage[intoc]{nomencl}
\usepackage{ifthen}

\renewcommand{\nomgroup}[1]
{
	\ifthenelse{\equal{#1}{A}}
	{\item[\fontsize{11}{13}\usefont{T1}{fvs}{b}{n}\selectfont{Acronyms}]}
 	{
	\vspace{1cm} 

	\ifthenelse{\equal{#1}{B}}
	{\item[\fontsize{11}{13}\usefont{T1}{fvs}{b}{n}\selectfont{Operators and Functions}]}
	{

	\ifthenelse{\equal{#1}{C}}
	{\item[\fontsize{11}{13}\usefont{T1}{fvs}{b}{n}\selectfont{Physics}]}
	{

	\ifthenelse{\equal{#1}{D}}
	{\item[\fontsize{11}{13}\usefont{T1}{fvs}{b}{n}\selectfont{Fluid Particles}]}
	{

	\ifthenelse{\equal{#1}{E}}
 	{\item[\fontsize{11}{13}\usefont{T1}{fvs}{b}{n}\selectfont{Solid Particles}]}
 	{

	\ifthenelse{\equal{#1}{F}}
	{\item[\fontsize{11}{13}\usefont{T1}{fvs}{b}{n}\selectfont{Surface Points}]}
	{
		{}
	}
	}
	}
	}
	}
	}

	\item[]\hspace*{-\leftmargin}%
	\rule[2pt]{1\textwidth}{1pt}%
}

\makenomenclature

\usepackage{pbox}
\usepackage{tabularx}
\usepackage{wrapfig}
\usepackage{newfloat}
\usepackage{csvsimple}
\usepackage{paralist}
\usepackage{multibib}
\usepackage{hyperref}
\usepackage{csquotes}
\usepackage{styles/myapalike}
\usepackage{caption}
\usepackage{latexsym}
\usepackage{titlesec}
\usepackage[british]{babel}

\definecolor{red}{rgb}{1,0,0}

\newcommand{\fig}[1]{Figure~\ref{#1}}
\newcommand{\bench}[1]{Benchmark~\ref{#1}}
\newcommand{\cref}[1]{Chapter~\ref{#1}}
\newcommand{\sref}[1]{Section~\ref{#1}}

\def\quad{\hspace*{1em}}
\def\qquad{\hspace*{2em}}
\DeclareFloatingEnvironment[
  fileext=lor,
  listname={List of Benchmarks},
  name=Benchmark,
  placement=tp,
]{benchmark}

\newcites{conf}{Conference Publications}
\newcites{journal}{Journal Articles}

\begin{document}
\thispagestyle{empty}
\setcounter{tocdepth}{2}


\begin{titlepage}

\setlength{\parindent}{0pt} 

\includegraphics[width=6cm]{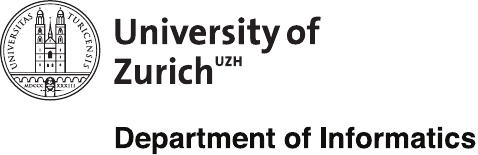}

\vspace{-2.25cm}

\begingroup
\leftskip7cm

\vspace{-0.75cm}
\Large
\textbf{Parallel Rendering and\\Large Data Visualization}

\vspace{0.70cm}

\normalsize
Dissertation submitted to the Faculty of Business, Economics and Informatics of the University of Zurich\\

\vspace{0.70cm}

to obtain the degree of\\
Doktor / Doktorin der Wissenschaften, Dr. sc.\\
(corresponds to Doctor of Science, PhD)

\vspace{0.70cm}

presented by\\
Stefan Eilemann\\
from Neuch\^atel, NE, Switzerland

\vspace{3.7cm}

Approved in February 2019 \\
\\
at the request of\\
Prof. Dr. Renato Pajarola\\
Prof. Dr. Markus Hadwiger

\vspace{2.2cm}

\endgroup

\end{titlepage}


\begin{titlepage}

\setlength{\parindent}{0pt} 

\normalsize
The Faculty of Business, Economics and Informatics of the University of Zurich hereby authorizes the printing of this dissertation, without indicating an opinion of the views expressed in the work.

\vspace{3ex}
Zurich, February 13, 2019

\vspace{3ex}
Chairman of the Doctoral Board: Prof. Dr. Thomas Fritz
\newpage
\thispagestyle{empty}
\quad
\newpage
\setcounter{page}{1}

\end{titlepage}

\pagenumbering{roman}

\chapter*{Abstract} \addcontentsline{toc}{chapter}{Abstract}

We are living in the big data age: An ever increasing amount of data is being
produced through data acquisition and computer simulations. While large scale
analysis and simulations have received significant attention for cloud and
high-performance computing, software to efficiently visualise large data sets
is struggling to keep up.

Visualization has proven to be an efficient tool for understanding data, in
particular visual analysis is a powerful tool to gain intuitive insight into
the spatial structure and relations of 3D data sets. Large-scale visualization
setups are becoming ever more affordable, and high-resolution tiled display
walls are in reach even for small institutions. Virtual reality has arrived in
the consumer space, making it accessible to a large audience.

This thesis addresses these developments by advancing the field of parallel
rendering. We formalise the design of system software for large data
visualization through parallel rendering, provide a reference implementation of
a parallel rendering framework, introduce novel algorithms to accelerate the
rendering of large amounts of data, and validate this research and development
with new applications for large data visualization. Applications built using
our framework enable domain scientists and large data engineers to better
extract meaning from their data, making it feasible to explore more data and
enabling the use of high-fidelity visualization installations to see more
detail of the data.


\chapter*{Kurzfassung}
\addcontentsline{toc}{chapter}{Kurzfassung}

Daten sind das Gold des 21. Jahrhunderts: Computersimulationen, bildgebende
Verfahren und andere Datenerfassungssysteme generieren immer gr\"ossere
Datenmengen. Visualisierungssoftware zur Darstellung grosser Datenmengen ist,
relativ zu Simulationssoftware und verteilten Systemen f\"ur Cloudumgebungen,
in der Forschung und Entwicklung vernachl\"assigt.

Visualisierung ist ein effizientes Mittel um grosse Datenmengen zu analysieren.
Insbesondere die Visualisierung von dreidimensionalen Datens\"atzen erlaubt ein
intuitives Verst\"andnis der r\"aumlichen Zusammenh\"ange und ihrer Struktur.
Visualisierungshardware steht immer mehr Benutzern zur Verf\"ugung,
insbesondere hochaufl\"osende Monitorw\"ande sind mittlerweile auch f\"ur
kleine Institutionen erschwinglich.

Diese Doktorarbeit besch\"aftigt sich mit paralleler Software und Algorithmen
zur Visualisierung dreidimensionaler Datens\"atze, um diesen Entwicklungen
Folge zu tragen. Als Grundlage f\"ur Forschung und Entwicklung formalisieren
wir die Softwarearchitektur f\"ur paralleles Rendering und stellen unsere
Referenzimplementierung vor. Auf dieser Basis pr\"asentieren wir neue
Forschungsergebnisse und Algorithmen zur schnelleren Visualisierung grosser
Datenmengen. Visualisierungssoftware, welche mit unserer Bibliothek entwickelt
wurde, validiert unseren Ansatz, und erlaubt Benutzern mehr Daten mit besserer
Detail zu analysieren.


\chapter*{Acknowledgments}
\addcontentsline{toc}{chapter}{Acknowledgments}
The research leading to this proposal was supported in part by the Blue Brain
Project, the Swiss National Science Foundation under Grant 200020-129525, the
European Union Seventh Framework Programme (FP7/2007-2013) under grant agreement
no. 604102 (Human Brain Project), the Hasler Stiftung grant (project number
$12097$), and the King Abdullah University of Science and Technology (KAUST)
through the KAUST-EPFL alliance for Neuro-Inspired High Performance Computing.

I would like to take the opportunity to thank the Blue Brain Project and its
visualization team, RTT AG (now part of Dassault Systems), KAUST, University of
Siegen, the Electronic Visualization Laboratory at the University of Illinois
Chicago, and all the other contributors for their support in the research and
development leading to this thesis.

I would like to thank Prof. Renato Pajarola and the VMML for his long-term
commitment to my research work and Patrick Bouchaud for putting me onto the
path taken by this thesis. A special gratitude goes to all collaborators who
joined me in this endeavour: Daniel Nachbaur, Cedric Stalder, Maxim Makhinya,
Christian Marten, Dardo D. Kleiner, Carsten Rohn, Daniel Pfeifer, Sarah
Amsellem, Juan Hernando, Marwan Abdellah, Raphael Dumusc, Lucas Peetz Dulley,
Jafet Villafranca, Philippe Robert, Ahmet Bilgili, Tobias Wolf, Dustin Wueest,
and Martin Lambers.

\tableofcontents
\newpage

\markboth{NOTATIONS}{}


\nomenclature[A]{CPU}{Central Processing Unit}
\nomenclature[A]{FPS}{Frames per Second}
\nomenclature[A]{GPU}{Graphics Processing Unit}
\nomenclature[A]{GUI}{Graphical User Interface}
\nomenclature[A]{LB}{Load Balancing}
\nomenclature[A]{LOD}{Level of Detail}

\printnomenclature

\listoffigures
\listofbenchmarks

\newpage

\newpage
\thispagestyle{empty}
\quad
\newpage
\setcounter{page}{1}

\pagenumbering{arabic}

\chapter{Background}

\section{Motivation}

\begin{figure}[h!t]%
\includegraphics[width=0.497\textwidth]{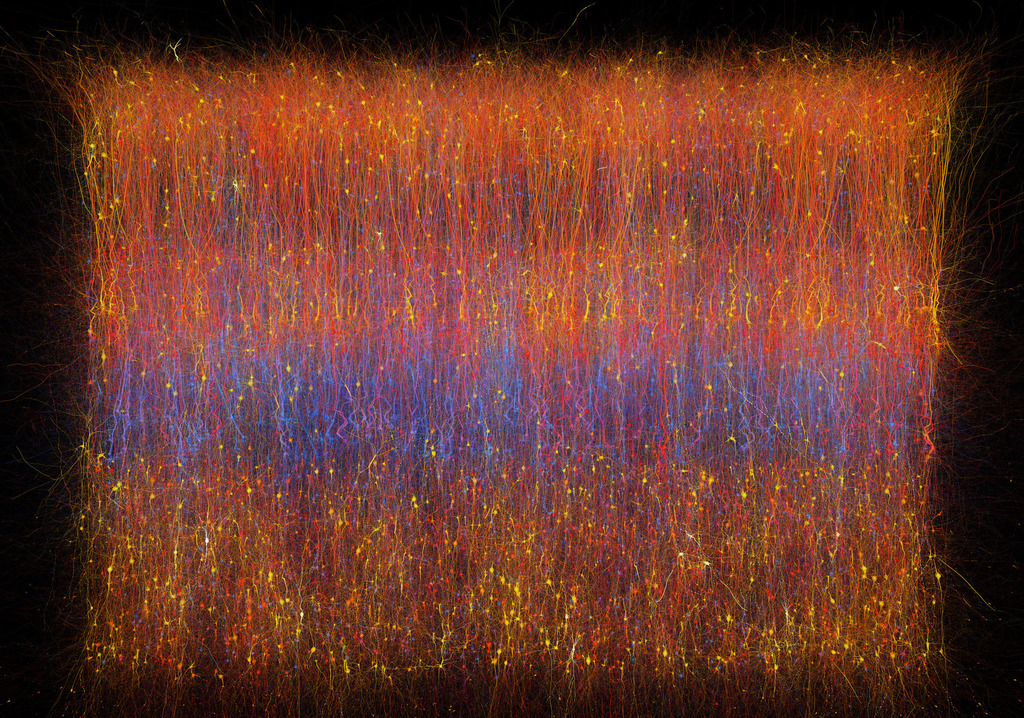}\hfil\includegraphics[width=0.497\textwidth]{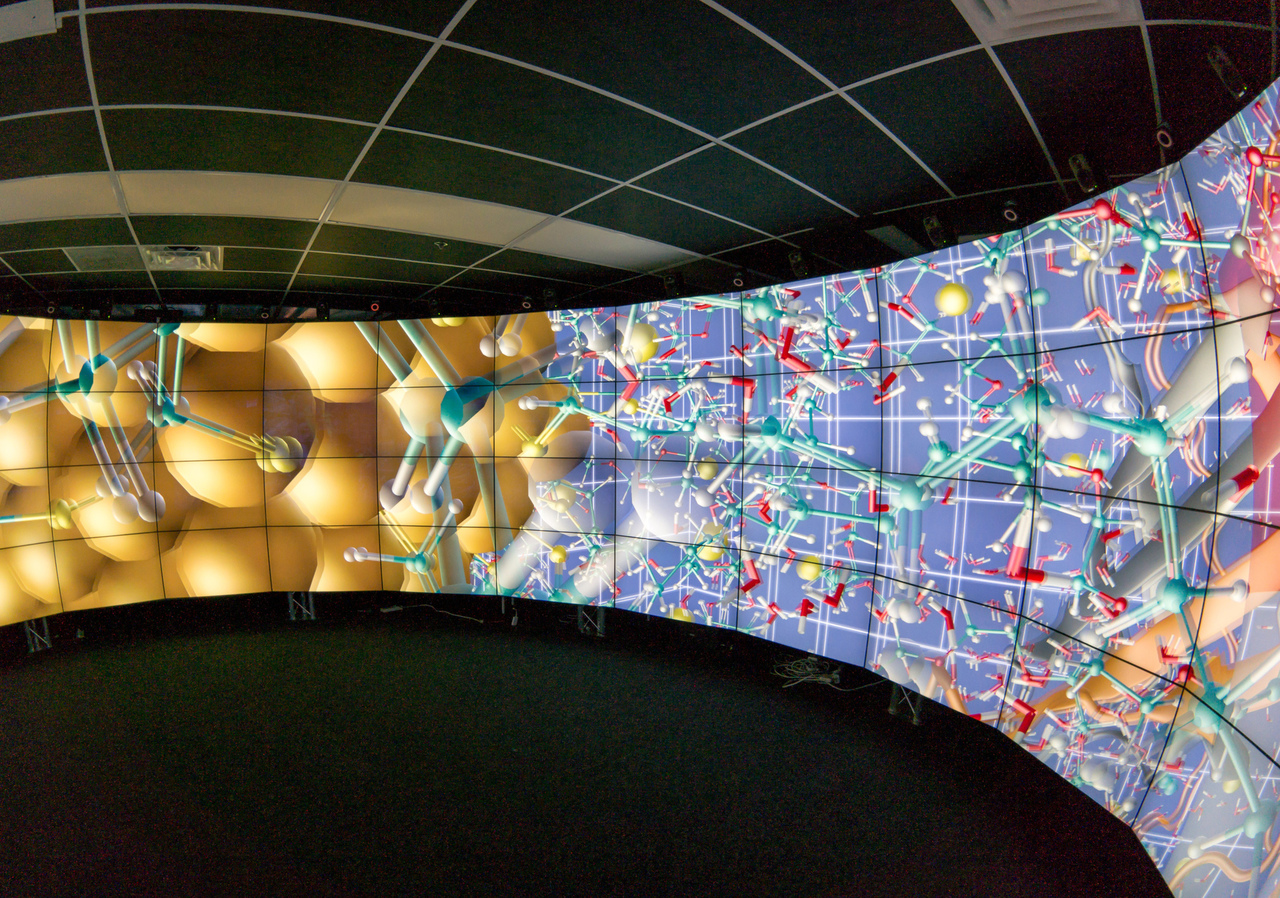}\\%
\includegraphics[width=0.497\textwidth]{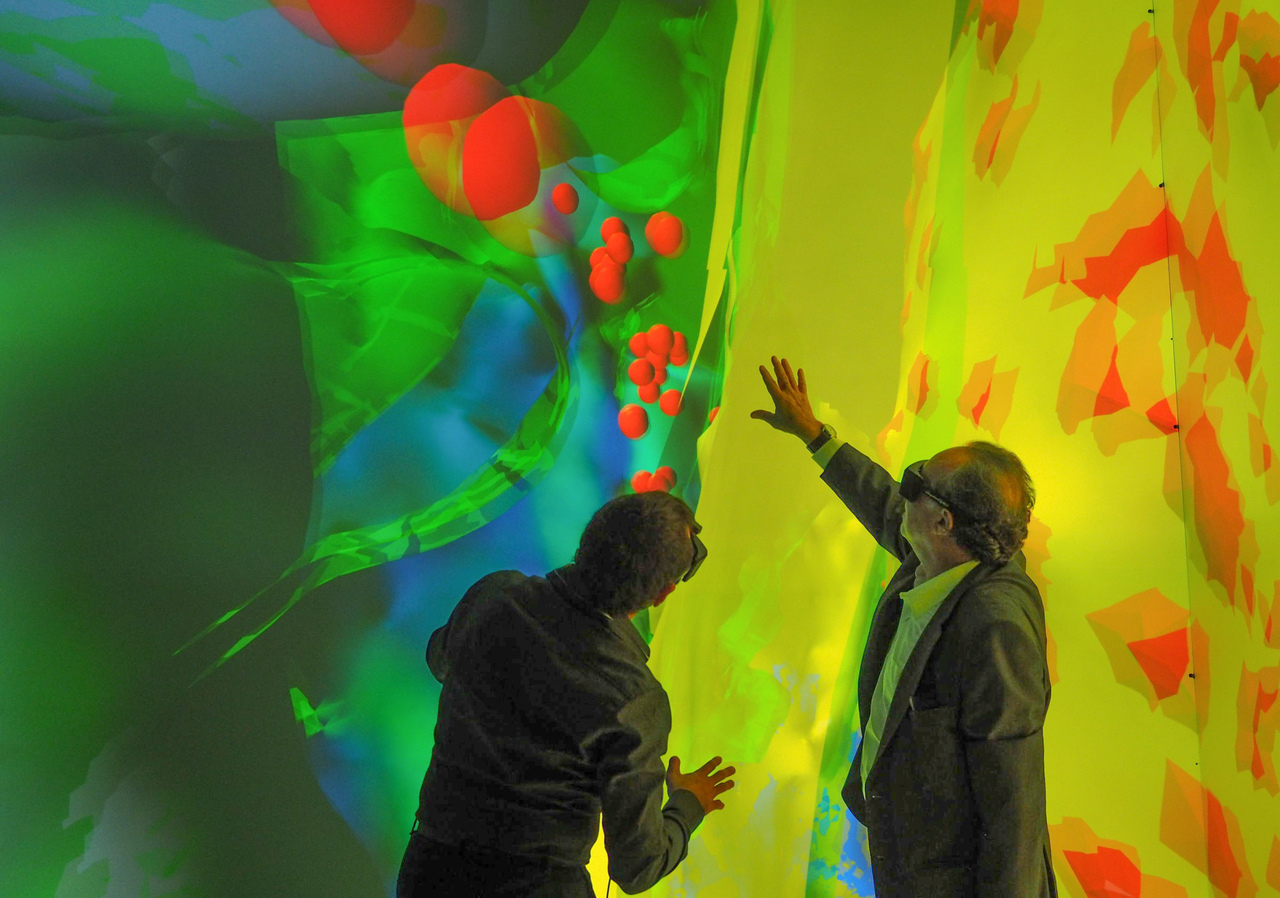}\hfil\includegraphics[width=0.497\textwidth]{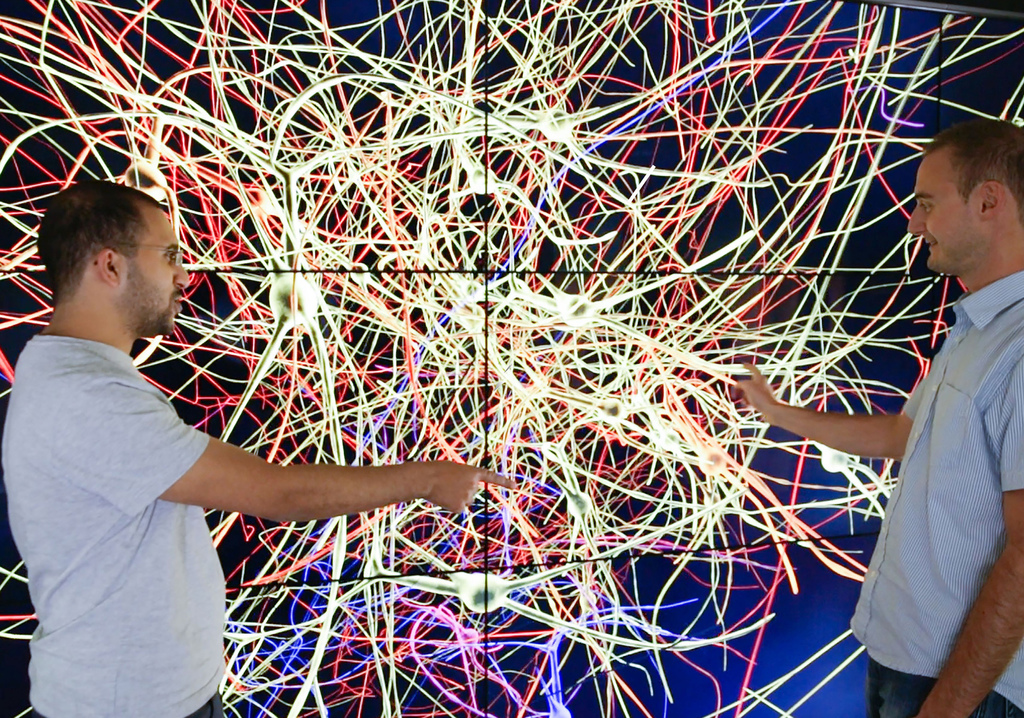}\\%
\caption{\label{FIG_teaser}Large Data Visualisation of a
  Brain Simulation, Molecular Visualisation in the Cave~2, Exploration of EM
  Stack Reconstructions in a Cave, Collaborative Data Analysis on a Tiled
  Display Wall}
\end{figure}

After decades of exponential growth in computational performance, storage and
data acquisition, computing is now well in the big data age, where future
advances are measured in our capability to extract meaningful information from
the available data. Visual analysis based on the interactive rendering of
three-dimensional data has been proven to be a particularly efficient approach
to gain intuitive insight into spatial structures and the relations of very
large 3D data sets. For example, the electrical slice simulation in
\fig{FIG_teaser}~(top left) contains millions of voltage samples per time step.
A visualisation makes this electrical activity immediately understandable, and
highlights eventual anomalies in the simulation. These developments create new,
unique challenges for applications and system software to enable users to fully
exploit the available resources to gain insight from their data.

The quantity of computed, measured or collected data is growing exponentially,
fuelled by the pervasive diffusion of digitalisation in modern life. Moreover,
the fields of science, engineering and technology are increasingly defined by a
data-driven approach to research and development. High-quality and
large-scale data is continuously generated at a growing rate from sensor and
scanning systems, as well as from data collections and numerical simulations in
a number of science and technology domains.

Display technology has made significant progress in the last decade:
High-resolution screens and tiled display walls are now affordable for most
organisations, and are getting deployed at an increasing rate. This increased
resolution and display size helps with understanding the data through higher
fidelity, but causes a quadratic increase in pixels to be rendered, which in
turn challenges rendering algorithms to deliver an interactive frame rate. Such
large-scale visualisation systems are often driven by multiple GPUs and
workstations, making it natural, and most times necessary, to drive them using
parallel and distributed applications.

However, not only applications are becoming more and more data-driven, but also
the technology used to tackle these kinds of problems has been witnessing a
paradigm shift towards massively parallel on-chip and distributed parallel
cluster solutions. On one hand, parallelism within a system has increased
massively, with tenths of CPU cores, thousands of GPU cores and multiple CPUs
and GPUs in a single system. On the other hand, massively parallel distributed
systems are easily accessible from various cloud infrastructure providers, and
are also affordable for on-site hosting for many organisations.

System software to exploit the available hardware parallelism capable of
performing efficient interactive data exploration has not kept up with the pace
in hardware developments and data gathering capabilities. Mostly, this is
due to an inherent delay between hardware and software capabilities, as
development typically only starts once the hardware is available. Secondly,
existing software is often engineered for different design parameters and
has a significant inertia to change, to the extreme cost of having to rewrite
it from scratch.

In the context of emerging data-intensive knowledge discovery and data analysis,
efficient interactive data exploration methodologies have become critical.
Visual analysis by means of interactive visualisation and inspection
of three-dimensional data is a particularly efficient approach to gain intuitive
insight into the spatial structure and relations of very large 3D data sets.
However, defining visual and interactive methods scaling with problem size and the
degree of parallelism, as well as generic applicability of high-performance
interactive visualisation methods and systems, are recognised among the major
current and future challenges.

\section{Challenges}

Increased display fidelity and faster rendering performance help to visualise
large data sets efficiently. Parallel rendering is one approach to achieve this
goal by using multiple GPUs, and often multiple computers, to improve the
rendering performance. It creates a new set of research challenges, which can
be broken down in more concrete challenges, starting with formalising and
implementing the architecture of a parallel rendering framework.

These sub-challenges to build better scalable parallel rendering applications
can be identified as finding better task decompositions, decreasing the cost
for the result composition, reducing the latency of the overall system, and
minimising synchronisation between the parallel execution threads.

Interactive visualisation poses its own unique set of challenges. The goal is to
present a believable alternate universe to the visual system of the user. This
process turns interactive visualisation into a powerful tool, by utilising
the brains' native capabilities to interpret and understand data. Virtual
Reality (VR) takes this goal to the extreme, and when done right, makes the user
forget that he interacts with a virtual world.

To achieve this goal, visualisation has the daunting task to transform large
amount of data into coloured pixels in a short amount of time. Believable
visualisation has to minimise the latency between user input and the resulting
output, and to maximise the number of frames rendered per second. With
increased immersion in the data, these parameters become more important --
for Virtual Reality, a 60~Hz refresh rate and a latency below 50~ms is required,
whereas for non-immersive desktop visualisation 10~Hz and 200~ms are acceptable.

When starting from a given rendering problem, the first task of a parallel
rendering system is to decompose (parallelise) this task into independent
sub-tasks, each rendered by a separate resource in parallel. While the basics
of this decomposition have been researched extensively, there are architectural
challenges to make these decompositions easily available in a generic and
structured manner. Load balancing these tasks for an optimal
parallelisation present many still unaddressed challenges for modern
visualisation cluster sizes, consisting of tens to hundreds of GPUs, and increasingly
affordable high-fidelity visualisation systems with tens of displays and
hundreds of millions of pixels.

By scaling up the amount of resources employed to accelerate the rendering
task, the task of combining the partial results from each resource becomes more
challenging. For some decomposition algorithms, the amount of data to composite
grows linearly with the amount of parallel resources used, and keeping the
compositing time within the available budget is a non-trivial problem.

For parallel rendering, these constraints make building a parallel and
distributed application harder compared to other distributed applications for
simulations and cloud computing. In particular, one has to be careful with
synchronisation and pipelining of operations to minimise latency. In addition,
an interactive application has different requirements when it comes to resource
allocation compared to other large-scale distributed computing domains.

Last, but not least, a significant challenge is how to make all this research
available to the large data scientists with the actual needs and use cases for
parallel rendering.

\section{Parallel Rendering}

Parallel rendering utilises multiple rendering units (GPUs), often on
different computers, to generate images for one or more output displays.
Scalable rendering is the subset of parallel rendering which uses multiple
resources to accelerate the rendering of one or more outputs. The goal of
parallel rendering is to increase the output resolution, rendering performance
or rendering quality. Traditionally the focus has been on the first two goals,
often in isolation of each other, e.g., algorithms and implementations for Cave
systems tend to be different from scalable rendering for large data
visualisation.

\begin{figure}[h!t]\center
 \includegraphics[width=\textwidth]{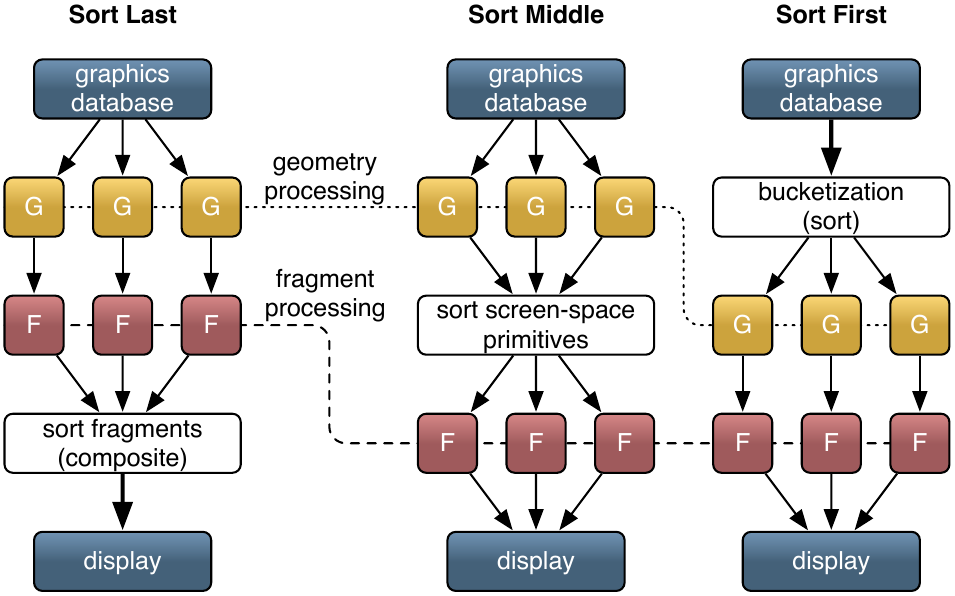}%
 \caption{Sort-Last, Sort-Middle and Sort-First Parallel Rendering\label{fSorts}}
\end{figure}

The main performance indicator for Large Data Interactive Rendering is the
performance of the rendering algorithm, that is, the framerate with which the
program produces new images. This framerate can be improved by either using
faster or more hardware, or by better algorithms exploiting existing
hardware and data. This thesis primarily focuses on the first approach, using
parallel rendering to exploit the CPU and GPU parallelism available on a single
system, or a distributed cluster. The early fundamental concepts have been laid
out in \cite{MCEF:94} and \cite{Crockett:97} (\fig{fSorts}). A number of
domain specific parallel rendering algorithms and special purpose hardware
solutions have been proposed in the past, however, only few generic parallel
rendering frameworks have been developed.

Sort-last rendering decomposes the rendering task in data space, that is, each
resource renders a part of the data. In the end partial fragments from each
resource are composited into a final result image. Sort-middle rendering also
decomposes the rendering at the data level, but collects and sorts the unshaded
primitives before or after rasterisation, and then performs the fragment
shading on the sorted data. Sort-first rendering decomposes the rendering task
in screen space, and the application needs either to be fill-rate bound or have
efficient view frustum culling to scale the rendering performance. We will
focus on sort-last and sort-first rendering, since sort-middle architectures
are only feasible in a hardware implementation due to the large amount of data
processed and transferred in the sorting stage.

\subsection{Domain Specific Solutions}

Cluster-based parallel rendering has been commercialised for off-line rendering
(i.e. distributed ray-tracing) for computer generated animated films or special
effects, since the typically used ray-tracing technique is inherently amenable
to parallelisation for off-line processing. Other special purpose solutions
exist for parallel rendering in specific application domains such as volume
rendering \cite{LWMT:97,Wittenbrink:98,HSCSM:00,SL:02,GS:02,NSJLYZ:05} or
geo-visualisation \cite{VR:91,AG:95,LDC:96,JLMV:06}. However, such specific
solutions are typically not applicable as a generic parallel rendering paradigm
and do not translate to arbitrary scientific visualisation and distributed
graphics problems.

In \cite{NC:07} parallel rendering of hierarchical level-of-detail (LOD) data
has been addressed and a solution specific to sort-first tile-based parallel
rendering has been presented. While the presented approach is not a generic
parallel rendering system, basic concepts presented in \cite{NC:07}, such as load
management and adaptive LOD data traversal, can be carried over to other
sort-first parallel rendering solutions.

\subsection{Special Purpose Architectures}

Historically, high-performance real-time rendering systems have relied on an
integrated proprietary system architecture, such as the early SGI graphics
supercomputers. Special purpose solutions have become a niche product as their
graphics performance did not keep up with off-the-shelf workstation graphics
hardware and scalability of clusters.

Due to its conceptual simplicity, a number of special purpose image compositing
hardware solutions for sort-first parallel rendering have been developed. The
proposed hardware architectures include Sepia \cite {MHS:99a,sepia}, Sepia~2
\cite{LMSBHa:01,LMSBH:01}, Lightning~2 \cite{Stoll01}, Metabuffer
\cite{Blanke00,Zhang01}, MPC Compositor \cite{Muraki01} and PixelFlow
\cite{Molnar92,Eyles97}, of which only a few have reached the commercial
product stage (i.e. Sepia~2 and MPC Compositor). However, the inherent
inflexibility and setup overhead have limited their distribution and
application support. Moreover, with the recent advances in the speed of CPU-GPU
and GPU-GPU interfaces, such as PCI Express, NVLink and other modern
interconnects, combinations of software and GPU-based solutions offer more
flexibility at a comparable performance.

\subsection{Generic Approaches}

A number of algorithms and systems for parallel rendering have been developed in
the past. Some general concepts applicable to cluster parallel
rendering have been presented in \cite{Mueller:95,Mueller:97} (sort-first
architecture), \cite{SZFLS:99,SFLS:00} (load balancing), \cite{SFL:01} (data
replication), or \cite{CMF:05,CM:06} (scalability). On the other hand, specific
algorithms have been developed for cluster based rendering and compositing such
as \cite{AP:98}, \cite{CKS:02} and \cite{YYC:01,SMLAP:03}. However, these
approaches do not constitute APIs and libraries that can be readily integrated
into existing visualisation applications, although the issue of the design of a
parallel graphics interface has been addressed in \cite{Igehy98}.

\begin{figure}[h!t]
 \includegraphics[width=\textwidth]{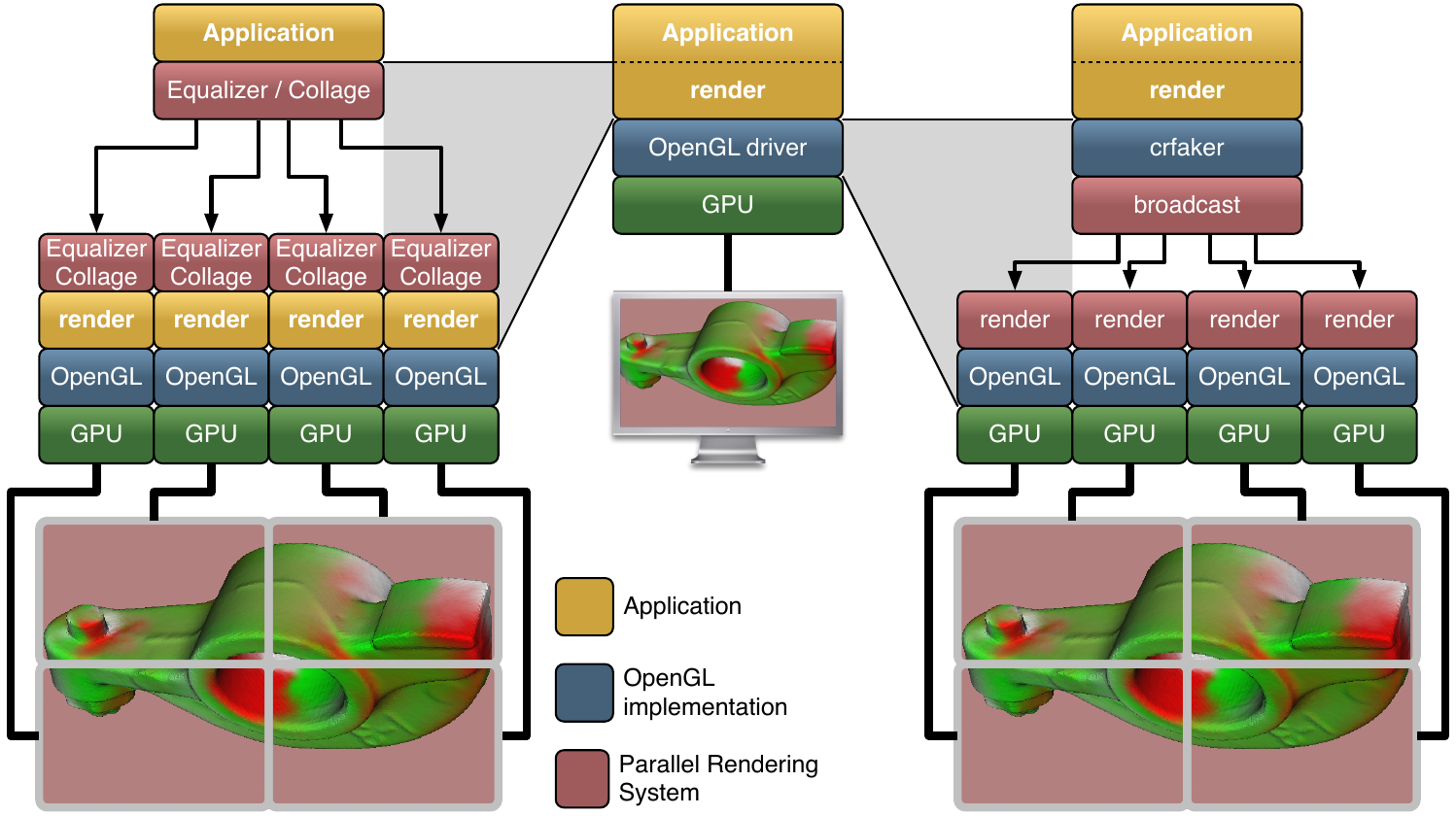}
 \caption{Parallel Execution  (left) versus Transparent OpenGL Interception (right)\label{fChromium}}
\end{figure}

Only few generic APIs and (cluster-)parallel rendering systems exist,
including VR Juggler \cite{BJHMBC:01} (and its derivatives), Chromium
\cite{HHNFAKK:02} (an evolution of \cite{Humphreys99,Humphreys00,HEBSEH:01}),
{ClusterGL}~\cite{NHM:11} and OpenGL Multipipe SDK
\cite{JDBJBCER:04,BRE:05,MPK}. These approaches can be categorised into
transparent interception and distribution of the OpenGL command stream and into
the parallelisation of the application rendering code (\fig{fChromium}).

\subsubsection{VRJuggler}

VR Juggler \cite{BJHMBC:01,JBBC:98} is a graphics framework for virtual reality
applications, shielding the application developer from the underlying hardware
architecture, devices and operating system. Its main aim is ease of use in
virtual reality configurations and use, without the need to know about the
devices and hardware configuration details, but not specifically to provide
scalable rendering. Extensions of VR Juggler, such as for example
ClusterJuggler \cite{BC:03} and NetJuggler \cite{AGLMR:02}, are typically based
on the replication of application and data on each cluster node and only take
care of synchronisation issues, but fail to provide a flexible and powerful
configuration mechanism that efficiently supports scalable rendering as also
noted in \cite{SWNH:03}. VR Juggler does not support scalable parallel
rendering such as sort-first and sort-last task decomposition and image
compositing, nor does it provide other important features for parallel
rendering, such as network swap barriers (synchronisation), distributed
objects, image compression and transmission, or multiple rendering threads per
process.

\subsubsection{Chromium}

Chromium \cite{HHNFAKK:02} provides a powerful and transparent abstraction of
the OpenGL API allowing a flexible configuration of display resources. It is
limited in scalability, due to its focus on streaming OpenGL commands through a
network of nodes, often initiated from a single source. This has also been
observed in \cite{SWNH:03}, and is caused by the size of the OpenGL stream.
This data stream not only contains OpenGL calls, but also geometry and image
data. Only if the geometry and textures are mostly static and can be kept in
GPU memory on the graphic card, no significant bottleneck can be expected, as
the OpenGL stream is then composed of a relatively small number of rendering
instructions. For typical real-world visualisation applications, display and
object settings are interactively manipulated, data and parameters may change
dynamically, and large data sets do not fit statically in GPU memory, but are
often dynamically loaded from out-of-core and/or multi-resolution data
structures. This can lead to frequent updates not only of commands and
parameters which have to be distributed, but also of the rendered data itself
(geometry and texture), thus causing the OpenGL stream to expand dramatically.
Furthermore, this stream of function calls and data must be packaged and
broadcast in real-time over the network to multiple nodes for each rendered
frame. This makes CPU performance and network bandwidth more likely the
limiting factor.

The performance experiments in \cite{HHNFAKK:02} indicate that Chromium is
working well when the rendering problem is fill-rate limited. This is due
to the fact that the OpenGL commands and a non-critical amount of rendering data
can be distributed to multiple nodes without significant problems. The
critical fill-rate work is then performed locally on the graphics hardware.

Chromium also provides some facilities for parallel application development: A
sort-last, binary-swap compositing stream processing unit and an OpenGL
extension providing synchronisation primitives, such as a barrier and
semaphore. It leaves problems like configuration, task decomposition, process
and thread management unaddressed. Parallel Chromium applications tend to be
written for one specific parallel rendering use case and configuration, e.g.
the sort-first distributed memory volume renderer in \cite{BHPB:03}, or the
sort-last parallel volume renderer raptor \cite{Raptor}. We are not aware of a
generic Chromium-based application using many-to-one sort-first or stereo
decompositions.

The concept of transparent OpenGL interception popularised by WireGL and
Chromium has received further contributions. While some commercial
implementations such as {TechViz} and {MechDyne Conduit} continue to exist, on
the research side only {ClusterGL}~\cite{NHM:11} has been presented recently.
{ClusterGL} employs the same approach as {Chromium}, but delivers a
significantly faster implementation of transparent OpenGL interception and
distribution for parallel rendering. Transparent OpenGL interception is an
appealing approach for some applications, as it requires no code changes.
It has inherent limitations due to the fact that eventually the bottleneck
becomes the single-threaded application rendering code, the amount of
application data the single application instance can load or process, or the the
size of the OpenGL command stream sent over the network.

\subsubsection{CGLX}

{CGLX}~\cite{DK:11} aims to bring parallel execution transparently to OpenGL
applications, by emulating the GLUT API and intercepting certain OpenGL calls.
Its target use case are multi-display installations, i.e., static sort-first
rendering with no compositing. In contrast to frameworks like {Chromium} and
{ClusterGL}, which distribute OpenGL calls, {CGLX} follows the distributed
application approach. This works transparently for trivial applications, but
quickly requires the application developer to address the complexities of a
distributed application, when mutable application state needs to be
synchronised across processes. For production applications, writing parallel
applications remains the only viable approach for scalable rendering, as shown
by the success of {Paraview}, {Visit} and {Equalizer}-based applications.

\subsubsection{OpenGL Multipipe SDK}

OpenGL Multipipe SDK (MPK) \cite{BRE:05} implemented an effective parallel
rendering API for a shared memory multi-CPU/GPU system. It is similar to IRIS
Performer \cite{RH:94} in that it handles multi-GPU rendering by a lean
abstraction layer via a callback mechanism, and that it runs
different application tasks in parallel. However, MPK is not designed nor meant
for rendering nodes separated by a network. MPK focuses on providing a parallel
rendering framework for a single application, parts of which are run in
parallel on multiple rendering channels, such as the culling, rendering and
final image compositing processes. The author used to be the technical lead
developer of OpenGL Multipipe SDK, therefore Equalizer is in many ways an
evolution of MPK for distributed execution, improved performance and better
configurability.

\subsubsection{Tiled Display Walls}

Software for driving and interacting with tiled display walls has received
significant attention, in particular {Sage}~\cite{Sage} and
 {Sage~2}~\cite{Sage2}. {Sage} was built entirely
around the concept of a shared framebuffer where all content windows are
separate applications using pixel streaming. It is no longer actively supported.
{Sage 2} is a complete, browser-centric reimplementation where each
application is a web application distributed across browser instances.
{DisplayCluster}~\cite{DisplayCluster}, and its continuation
 {Tide}~\cite{tide}, also implement the shared framebuffer concept of
 {Sage}, but provide a few native content applications integrated into the
display servers. These solutions implement a scalable display environment and
are a target display platform for scalable 3D graphics applications.

\section{Thesis Structure}

In the next chapter, we
give a summary of the contributions of this thesis, listing relevant
publications and the contributions of the author to these publications.
\cref{sArchitecture} introduces the architecture of a parallel rendering
framework, the foundation for this thesis. \cref{sScalable} presents new
algorithms for the task decomposition in parallel rendering.
\cref{sCompositing} focuses on optimisations to reduce the cost of recombining
the results of a parallel rendering decomposition. \cref{sLoadBalancing}
describes better approaches to balance the task assignment to rendering
resources. \cref{sCollage} describes the design and architecture of a network
library tailored to parallel rendering. Before a conclusion in
\cref{sConclusion}, \cref{sApplications} provides an overview of the major
Equalizer applications.

\chapter{Contributions}

This chapter summarises the main contributions of this thesis. In each section,
we list the relevant publications and specify the contributions of the author.

\section{Parallel Rendering Architecture}

A major contribution of this thesis is the formalisation of the architecture for
a parallel rendering framework and its reference implementation, which advances
the state of the art in many aspects:

\begin{compactdesc}

\item[Minimally invasive API:] The guiding principle for the API design was to
allow applications to retain all their rendering code and application logic.
The programming interface is based on a set of C++ classes, modelled closely to
the resource hierarchy of a graphics rendering system. The application
subclasses these objects and overrides C++ task methods, similar to C
callbacks. These task methods will be called in parallel by the framework,
depending on the current configuration. The contract for the implementation of
the task methods does not assume any specific rendering library, algorithm or
technology, thus facilitating the adaptation of existing applications for
parallel rendering. This parallel rendering interface is significantly
different from Chromium \cite{HHNFAKK:02} and more similar to VRJuggler
\cite{BJHMBC:01} and MPK \cite{BRE:05}.

\item[Runtime configuration:] The architecture of our parallel rendering
framework makes a clear separation between the rendering algorithm and the runtime
configuration. It provides a contract between the framework and the application
code based on a rendering context, and uses this context to drive the
application output depending on the runtime configuration. Application
developers are unaware of parallel rendering setups and make no assumptions on
how the rendering code will be executed. This clear separation yields parallel
rendering applications which can be deployed on a wide set of installations,
and are often configured in new ways unforeseen during their deployment.

\item[Display abstraction:] Large scale visualisation systems cover a wide set
of use cases from classical workstation setup to monoscopic tiled display
walls, stereoscopic, edge-blended multi-projector walls to fully immersive
installations CAVE systems. Consequently, applications running on these
systems serve many different use cases. Our novel canvas-layout
abstraction provides a simple configuration for all these installations and
empowers applications using these installations with 2D and 3D contextual
information, runtime stereo configuration, and head tracking.

\item[Compound trees:] The introduction of compounds, and their underlying
contract, provides a formalisation of a flexible task decomposition and result
recomposition for parallel rendering. Compound trees allow for easy
specification of complex parallel task decomposition strategies, which are
implemented and executed by the Equalizer system. They generalise parallel
rendering principles without hardcoding a specific parallel rendering
algorithm, thus proposing an orthogonal parameter set for decomposing rendering
tasks, assembling results, and adapting these parameters at runtime.
Furthermore, they facilitate new parallel rendering research due to their
flexibility and extensibility.

\item[Equalizers:] The namesake of our framework, they are active components
hooked into a compound tree, and modify compound tree parameters at runtime.
For example, a sort-first load balancer adapts the sub-viewports assigned to
each resource at runtime. Compounds are the passive configuration, and
equalizers are the active component to optimize this configuration dynamically.
This makes their implementation independent of the rest of the framework,
providing a powerful abstraction for research and development of better
resource usage for parallel rendering.

\item[Modular architecture:] Our architecture uses layered abstractions that
gradually provide higher level abstractions. On the lower level, a network
library for distributed abstractions provides the substrate for Equalizer and its
applications. Within each library, a clear separation of responsibilities
allows an easy combination of existing algorithms. For example, an advanced feature
like a cross-segment equalizer relies on per-segment load equalizers, and
both equalizers reconfigure the underlying compound tree each frame.

\end{compactdesc}

\cite{EMP:09} and \cite{ESP:18} publish the architectural foundations of
parallel rendering frameworks. Any algorithmic implementation and
architectural contributions in these publications are contributed by the
author, while experimental results have significant contributions from the
secondary authors.

\cite{BRE:05} provides in many ways the foundation for Equalizer, to which the
author was a contributor for the implementation of a parallel rendering
framework for shared memory systems.

\section{Scalable Rendering and Compositing}

Based on the flexible system architecture we implemented new scalable rendering
algorithms, introduced in \cite{EMP:09} and \cite{ESP:18}. A particular focus
was given on reducing cost for the expensive compositing step of sort-last
rendering.

\cite{EP:07} provides an analysis of parallel compositing algorithms in an
early version of our parallel rendering framework, and we show that direct send
compositing has advantages on commodity visualisation clusters. The
implementation, algorithm and experimental analysis in this paper are
contributed by the author.

\cite{MEP:10} introduced more sort-last compositing optimisations, most notably
automatic region-of-interest detection and new compression algorithms. The
author contributed the foundations for using region of interest, and the fast
RLE compression with optimised data preconditioning.

\cite{EBAHMP:12} introduces many algorithmic optimisations for modern
visualisation clusters, ranging from asynchronous compositing, thread and
memory placement on NUMA architectures, region of interest, and an analysis of
real-world application performance. We show that careful system design and
detailed optimisations are necessary to achieve scalability on larger
visualisation clusters. For this publication, the author contributed the
algorithms, large parts of the implementation in Equalizer, and some
experimental analysis.

\section{Load Balancing}

Optimal resource usage in larger visualisation clusters relies on an even
distribution of work over the available resources. This load balancing problem
requires real-time algorithms based on imperfect knowledge of the system and
application behaviour. In our architecture, load balancing is achieved by modifying
the compound tree parameters at runtime. For example, a sort-first load balancer
adapts the sub-viewports assigned to each resource at runtime. These so-called
{\em equalizers} are a component hooked into the compound tree, which makes
their use and implementation independent of the rest of the framework.

\cite{EMP:09} and \cite{ESP:18} provide experimental results on the
effectiveness of our sort-first and sort-last load balancing implementations.
In the latter publication, we compare two different reactive load balancing
algorithms and show that the theoretically superior algorithms do not necessarily
provide better performance in realistic scenarios.

\cite{EEP:11} introduces a novel algorithm for load balancing an arbitrary set
of rendering resources to drive visualisation installations with many output
displays, like tiled display walls or multi-projector systems. The author
contributed the algorithm and implementation for this publication.

\cite{SPEP:16} provides an implementation and detailed analysis of central task
queueing with work packages and different task affinity modes for sort-first and
sort-last rendering. The author provided the base queueing infrastructure for
this publication.

\chapter{Parallel Rendering Architecture}\label{sArchitecture}

\section{Overview}

A generic parallel rendering framework has to cover a wide range of use cases,
target systems, and configurations. This requires a strong separation between
the implementation of the application and its configuration, linked with a
careful design to allow the resulting program to scale up to hundreds of nodes,
while providing a minimally invasive API for the developer. In this section we
present the system architecture of the Equalizer parallel rendering framework,
and motivate its design in contrast to related work.

The motivation to use parallel rendering is either driven by the need to drive
multiple displays or projectors from multiple GPUs and potentially multiple
nodes, or by the need to increase rendering performance to visualise
more data, or use a more demanding rendering algorithm for higher visual quality.
Occasionally both needs coincide, e.g., for the analysis for large data
sets on high fidelity visualisation systems.

Parallel rendering has similarities to other distributed computing domains like
cloud computing and high-performance computing (HPC). It aims to accelerate
the completion of a task by parallelising a time-consuming algorithm, or to
allow the computation of a larger problem by employing multiple resources. Certain
aspects are shared across these distributed computing domains, such as the need
to load balance the parallel task execution, minimise synchronisation and
communication overhead, as well as to find a task decomposition which allows
to produce correct results.

Parallel rendering has one significant additional constraint: serving
an interactive use case. Depending on the application domain and visualisation
system, typically a framerate between 10~Hz and 120~Hz is required for
useful user interaction. In turn, this translates to a budget of 8~ms to
100~ms to decompose the task to render a frame, perform parallel rendering,
and to composite and display the result. In comparison, cloud computing and
HPC typically have turnaround times of seconds to hours. Therefore, many
algorithms for parallel rendering compute a suboptimal solution,
but do so in at most a few milliseconds.

Fundamentally two approaches enable applications to use multiple
GPUs: transparent interception at the graphics API (typically OpenGL), or
extending the application to support parallel rendering natively
(\fig{fChromium}). The first approach has been extensively explored by Chromium
and others, while the second is the foundation for this thesis. The
architecture of Equalizer is founded on an in-depth requirements analysis of
typical visualisation applications, existing frameworks, and previous work on
OpenGL Multipipe SDK.

The task of parallelising a visualisation application boils down to configuring
the applications' rendering code differently for each resource, enabling this
rendering code to access the correct data, and synchronising execution. For
scalable rendering, when multiple GPUs are used to accelerate a single output,
partial results need to be collected from all contributing resources, combined,
and send to the output.

The architecture of our parallel rendering framework addresses the following research questions:
\begin{compactitem}
  \item How can we reduce end-to-end system latency for better user experience?
  \item In a generic parallel rendering framework, how can we schedule the different rendering stages to minimise the latency for the user?
  \item How can we architect the parallel rendering framework to minimise synchronisation between threads?\\[\smallskipamount]

\sref{sAsyncExec} introduces our asynchronous execution model, which has been
carefully designed to minimise synchronisation points, maximise pipelining and
enables early display of rendered images.\\[\smallskipamount]

\item How can we maximise the impact of this research on large data
scientists?\\[\smallskipamount]

Ultimately, accessible applications determine the impact for large data
research. With Equalizer we provide the base building blocks: a minimally
invasive API and distributed execution layer to lower the entry barrier for
application developers, a flexible configuration with a clear separation from
the implementation, comprehensive VR features and tiled display wall
integration addressing a wide set of visualisation installations. Various
applications, introduced in \cref{sApplications}, have been developed using
Equalizer.

\end{compactitem}

In this chapter, we will first describe the execution model and resource
configuration, followed by how the generic configuration is used to model
the desired visualisation setup, and finally introduce specifics of scalable
and distributed rendering.

\section{Asynchronous Execution Model}\label{sAsyncExec}

The core execution model for parallel rendering was pioneered by CAVELib
\cite{DACNCCGHPSNS:97}, refined by OpenGL Multipipe SDK for shared memory
systems and scalable rendering, and substantially extended by Equalizer for
asynchronous and distributed execution. By analysing the typical architecture
of a visualisation application we observe an initialization phase, a main
rendering loop, and an exit phase. Equalizer decomposes these steps for
parallel execution.

\begin{figure}[ht]\center
 \includegraphics[width=.9\columnwidth]{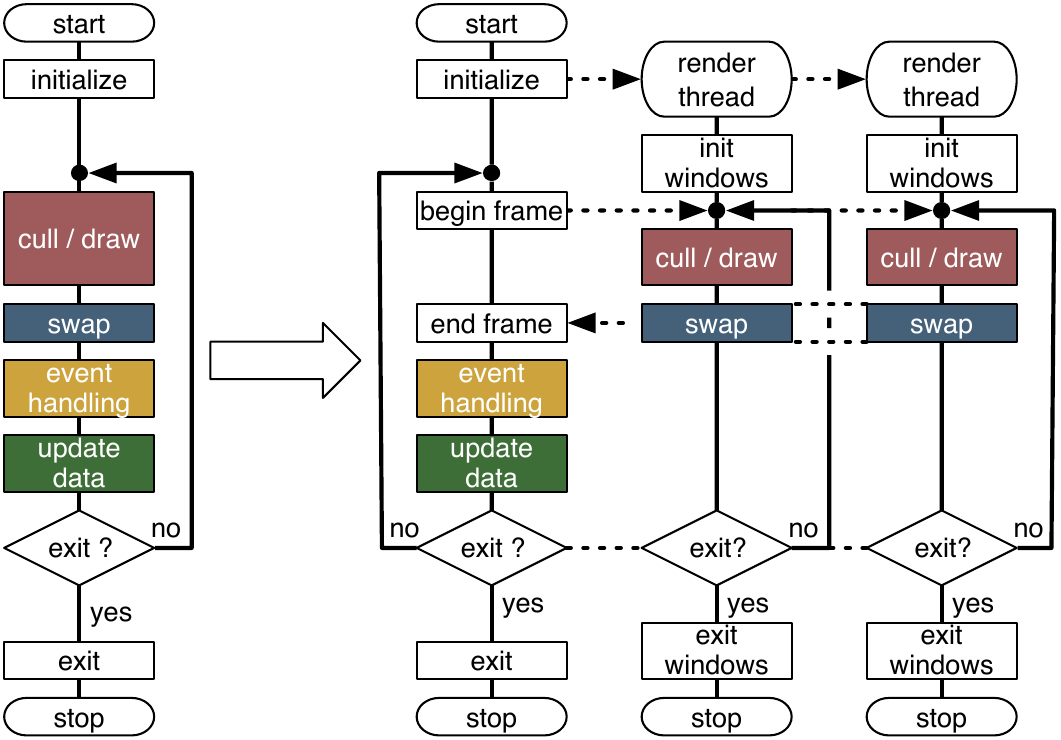}
 \caption{Simplified Execution Flow of a Classical Visualisation Application
  and an Equalizer Application}
 \label{FIG_execution}
\end{figure}

The main rendering loop typically consists of four phases: submitting the
rendering commands to the graphics subsystem, displaying the rendered image,
retrieving events from the operating system, and updating the application state
before a new image is rendered. Usually, the configuration of the rendering is
largely hard-coded, with a few configurable parameters such as field of view or
stereo separation. For parallel execution, we need to separate the rendering
code from this main loop, and execute it in parallel with different rendering
parameters, as shown in \fig{FIG_execution}. Similarly, the initialisation and
exit phase also need to be decomposed to allow managing of multiple distributed
resources.

\begin{figure}[h!t]\center
 \includegraphics[width=\textwidth]{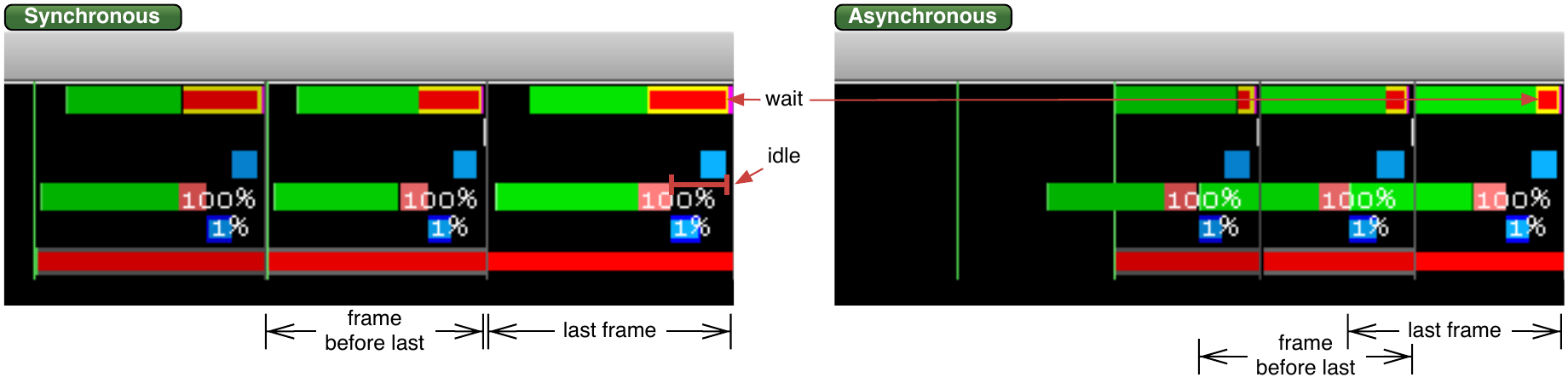}
 {\caption{\label{fSyncAsync}Synchronous and Asynchronous Execution}}
\end{figure}

\fig{fSyncAsync} shows the execution of the rendering tasks of a two-node
sort-first compound without latency and with a latency of one frame. The
asynchronous execution pipelines rendering operations and hides imbalances in
the load distribution, resulting in an improved framerate. We have observed a
speedup of 15\% on a five-node rendering cluster when using a latency of one
frame instead of no latency in a sort-first configuration.

Another critical design parameter are synchronisation points. Most
implementations, including OpenGL Multipipe SDK, use a per-frame barrier or
similar synchronisation to manage parallel execution. In larger installations,
this is detrimental to scalability, as even slight load imbalances limit
parallel speedup. The Equalizer execution model is fully asynchronous, and only
introduces synchronisation points when strictly required. The main
synchronisation points are: configured swap barriers between a set of output
channels which have to display simultaneously, the availability of input frames
for scalable rendering, and a task synchronisation to prevent runaway of the
main loop execution. By default, Equalizer keeps up to one frame of latency in
execution, that is, some resources might render the next frame while others are
still finishing the current frame. Nonetheless, finished resources will
immediately display their result. This asynchronous execution architecture,
coupled with a frame of latency, allows pipelining of many operations, such as
the application event processing, task computation and load balancing,
rendering, image readback, compression, network transmission, and compositing.
It also hides small imbalances in the task distribution, as they usually average
out over multiple frames.

In practical scenarios, application initialisation and exit is also a factor
for usability. Consequently, these phases are also parallelised in Equalizer. A
first pass identifies the resources to be launched or terminated, kicks off the
tasks, and then uses a second pass to synchronize their execution and results.

\begin{wrapfloat}{benchmark}{O}{.618\textwidth}
 \includegraphics[width=.618\textwidth]{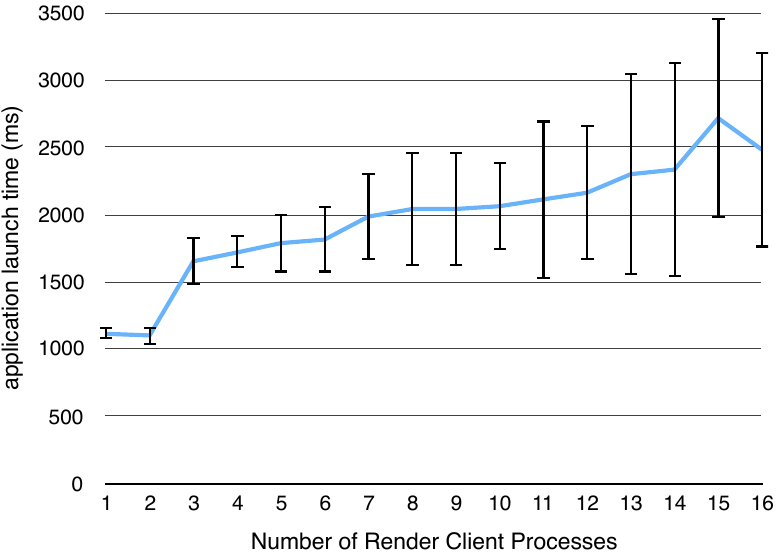}
 {\caption{\label{rLaunch}Parallel Application Startup}}
\end{wrapfloat}

\bench{rLaunch} shows the startup time of eqPly, our parallel polygon renderer.
This benchmarks simply measures the time taken by {\em Config::init}, which
includes the render client process creation using ssh from the application
node, library loading from a shared filesystem, network setup, OpenGL and
window initialisation, and object data mapping for the Equalizer resource
instances and a few internal objects used by eqPly. The benchmark confirms that
the application launch is scaling nicely to a medium cluster size. A slight
increase in startup time with larger configurations is expected, since more
processes increase the load on the shared filesystem and worsen distribution
and synchronisation overheads. Due to the shared filesystem used for the
executable, the startup times observe a large uncertainty, shown by the
standard deviation bars.

\begin{wrapfloat}{benchmark}{O}{.618\textwidth}
 \includegraphics[width=.618\textwidth]{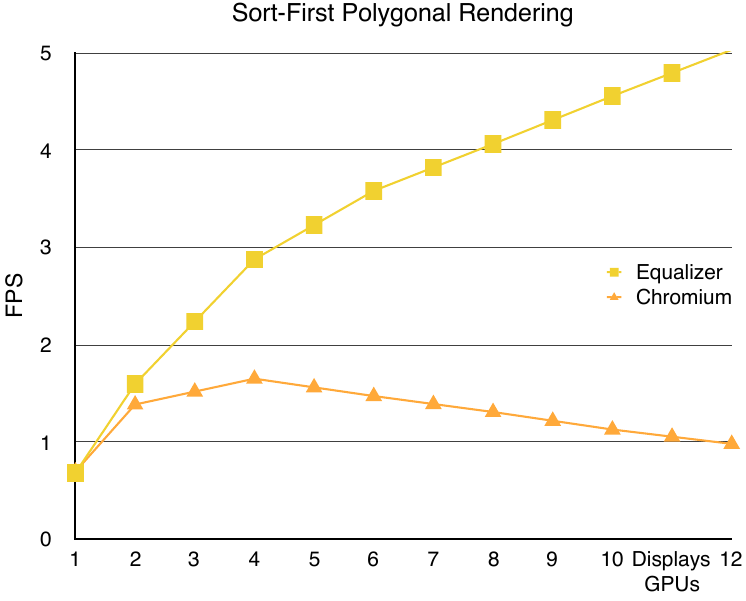}
 {\caption{\label{rCR}Driving a Tiled Display Wall}}
\end{wrapfloat}

In comparison to interception approaches as used by Chromium, our asynchronous
programming model inherently provides better performance. \bench{rCR} tests
the rendering performance for driving a simple tiled display wall configuration
with a static model, rotating about its vertical axis, placed such that it
nicely covers the different screens.

A standard tile-sort Chromium configuration is comparable to a simple Equalizer
display wall setup, where in each case a single GPU and node is responsible for
driving the attached display. The polygonal model is rendered using eqPly and
uses display lists for the static geometry. Using display lists allows Chromium
to send geometry and texture data only once to the rendering nodes (retained mode
rendering) and display them repeatedly using {\em glCallLists()}, which is
inexpensive in terms of network overhead. This setup is favourable for Chromium,
because the display lists are transmitted only once over the network,
and only simple display calls will be processed and distributed by Chromium for
each rendered frame.

Chromium initially increases performance when adding nodes, but it quickly
stagnates, and even decreases, when more nodes are added. In contrast, Equalizer
continually improves performance with more added nodes and exhibits a smooth
drop-off in speed-up, due to the expected synchronisation and network overhead
as the rendered data gets negligible in size per node. This performance
difference is also due to the fact that Equalizer can benefit from distributed
parallel view frustum culling on each render thread.

\subsection{Programming Interface}

\begin{wrapfloat}{figure}{O}{.618\textwidth}
 \includegraphics[width=.618\textwidth]{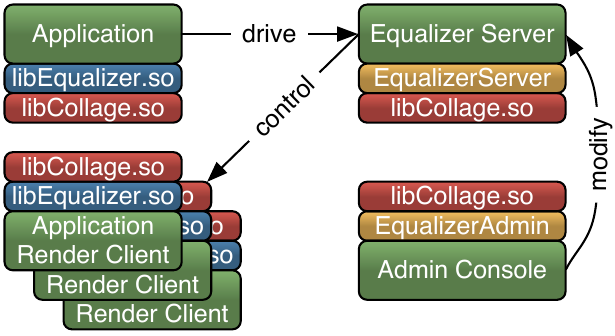}
 {\caption{\label{fProcessing}Parallel Rendering Entities}}
\end{wrapfloat}

Equalizer is a framework to facilitate the development of distributed and
multi-threaded parallel rendering applications. The programming interface is
based on a set of C++ classes, modelled closely to the resource hierarchy of a
graphics rendering system. The application subclasses these objects and
overrides C++ task methods, similar to C callbacks. These task methods will be
called in parallel by the framework, depending on the current configuration.
This parallel rendering interface is significantly different from Chromium
\cite{HHNFAKK:02} and more similar to VRJuggler \cite{BJHMBC:01} and OpenGL
Multipipe SDK \cite{BRE:05}.

To separate the responsibilities in a parallel rendering application, different
entities are responsible for different aspects of the runtime system: the
application process driving a rendering session, the server controlling the
parallel rendering configuration, render clients executing the rendering
tasks, and an administrative API to reconfigure the rendering session at
runtime. All processes communicate with each other through a common network
library (Collage) and a client library implementing the Equalizer API, as shown
in \fig{fProcessing}.

The administrative API connects to a server, and allows some changes to the
running configuration,, e.g., to create new output channels. Its description is
outside of the scope of this thesis, and is mentioned here for completeness.

\subsection{Application}

The main application thread in Equalizer drives the rendering, that is, it
carries out the main event loop, but does not actually execute any rendering.
Depending on the configuration, the application process often hosts one or more
render client threads. These application render threads are identical in
behaviour and implementation to render threads on the render client nodes. When
a configuration has no additional nodes besides the application node, we have a
single-process, multi-threaded rendering application: all application code is
executed in the same process, and no network data distribution has to be
implemented.

The main rendering loop is simple: The application requests a new frame
to be rendered, synchronises on the completion of a frame and processes events
received from the render clients. It may perform idle processing between the
start and synchronisation of a frame. \fig{FIG_execution} shows a simplified
execution model of an Equalizer application.

\subsection{Server}

The Equalizer server manages the parallel rendering session. It is an
asynchronous execution thread or process, which receives requests from the
application and serves these requests using the current configuration, launching
and stopping rendering client processes on nodes, determining the rendering
tasks for a frame, and synchronising the completion of tasks.

\subsection{Render Client}

During initialisation, the application provides a rendering client executable.
The rendering client is often, especially for simple applications, the same
executable as the application. However, in more sophisticated implementations,
the rendering client can be another executable which only contains the
application-specific rendering code. The server deploys this rendering client
on all nodes specified in the configuration. Render clients may run on a
different architecture or operating system from the main application, the
underlying network library ensures type safety and endian ordering.

In contrast to the application process, the rendering client main loop is
completely controlled by Equalizer, based on application commands. A render
client consists of the following threads: The node main thread, one network
receive thread, one thread for each graphic card (GPU) to execute rendering
tasks, and optionally one thread for asynchronous readback per GPU. If a
configuration also uses the application node for rendering, then the
application process uses one or more render threads, consistent with render
client processes. The Equalizer client library implements the main loop, which
receives network commands, processes them, and invokes the necessary task
methods provided by the developer.

The task methods clear the frame buffer as necessary, execute the OpenGL
rendering commands as well as readback, and assemble partial frame results for
scalable rendering. All tasks have default implementations so that only the
application specific methods have to be implemented, which at least involves
the {\tt frameDraw()} method executing a rendering task. For example, the
default callbacks for frame recomposition during scalable rendering implement
tile-based assembly for sort-first and stereo decompositions, and unordered
$z$-buffer compositing for sort-last rendering.

\subsubsection{Render Context}

The render context is the core entity abstracting the application-specific
rendering algorithm from the system-specific configuration. It specifies:

\begin{compactdesc}

 \item[Buffer] OpenGL-style read and draw buffer as well as colour mask. These
 parameters are influenced by the current eye pass, eye separation and
 anaglyphic stereo settings.

 \item[Viewport] Two-dimensional pixel viewport restricting the rendering area.
 The pixel viewport is influenced by the destination viewport definition and
 viewports set for sort-first decompositions.

 \item[Frustum] Frustum parameters as defined by {\em glFrustum}. Typically
 the frustum is used to set up the OpenGL projection matrix. The frustum is
 influenced by the destinations view definition, sort-first decomposition,
 tracking head matrix and the current eye pass.

 \item[Head Transformation] A transformation matrix positioning the frustum. For
 planar views this is an identity matrix and is used in immersive rendering.
 It is usually used to set up the `view' part of the modelview matrix, before
 static light sources are defined.
 \item[Range] A one-dimensional range within the interval [0..1]. This parameter is
 optional and should be used by the application to render only the appropriate
 subset of its data for sort-last rendering.

\item[View] The view object from the logical rendering rendering configuration,
as introduced below. Holds view-specific data, such as camera, model or any other
application state.

\end{compactdesc}

\subsubsection{Event Handling}

Event handling routes events from the source window in the rendering thread to
the application main thread for consumption. At each step, events can be
observed, transformed or dropped. Events are received from the operating system
in the rendering thread, transformed there into a generic representation, and
sent to the application main thread. The application processes them in the main
loop and modifies its internal state accordingly. This follows the natural data
flow for most windowing systems and has natural semantics for thread safe
event handling. For Qt, Equalizer internally dispatches events from the process
main thread to the render threads to ensure consistent behaviour.

\subsection{NUMA Aware Thread Scheduling}

Non-Unified Memory Access (NUMA) is a common hardware architecture for
high-performance visualisation clusters. Modern multi-socket render nodes use a
NUMA architecture, where each CPU socket has a number of locally-attached
memory buses, GPU and network devices, and CPU sockets are linked with an
interconnect to each other. Accessing a memory address located on another
processor has a performance penalty for both bandwidth and latency, and
accessing a GPU or network interface from a remote processor is slower than a
local access.

\begin{wrapfloat}{figure}{O}{.618\textwidth}
 \includegraphics[width=.618\textwidth]{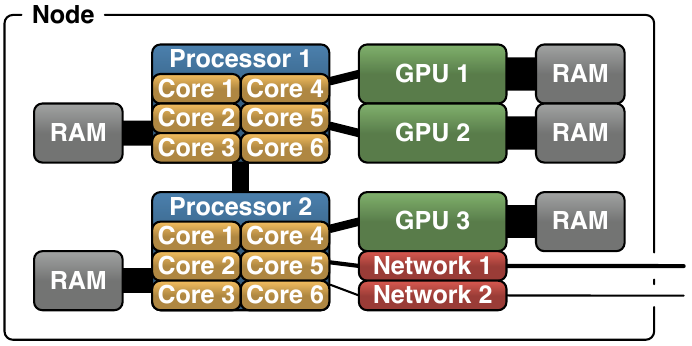}
 {\caption{\label{fNumaNode}Exemplary Dual-Socket NUMA Node}}
\end{wrapfloat}

\fig{fNumaNode} shows one such NUMA visualisation node, used in the experiments
of \cite{EBAHMP:12}. It has two CPU sockets with six cores each, three GPUs
connected to the two sockets, and two network cards (10 Gigabit Ethernet and
InfiniBand) connected to one socket.

In our parallel rendering system, a number of threads are used to drive a
single process in the cluster: the main thread (main), one rendering thread for
each GPU (draw) and one thread to finish asynchronous downloads (read), one
thread for receiving network data (recv), one command processing thread (cmd),
and one thread for image transmission to other nodes (xmit). We have
implemented automatic thread placement by extending and using the hwloc library
in Equalizer. We restrict all node threads (main, recv, cmd, xmit) to the cores
of the processor local to the network card, and all GPU threads (draw, read) to
the cores of the processor closest to the respective GPU.

\begin{wrapfloat}{figure}{O}{.618\textwidth}
 \includegraphics[width=.618\textwidth]{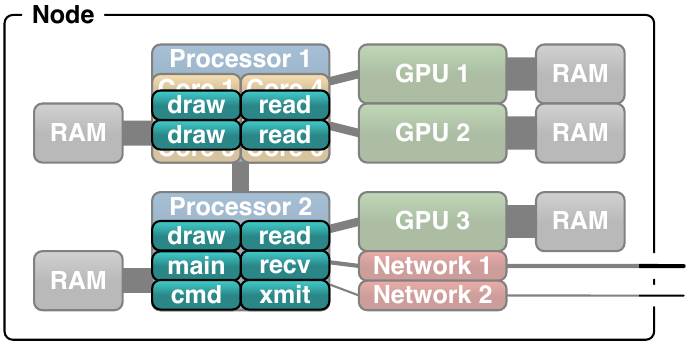}
 {\caption{\label{fNumaThreads}Thread Placement on a NUMA Node}}
\end{wrapfloat}

\fig{fNumaThreads} shows the thread placement for the node used in
\fig{fNumaNode}. Threads are bound to all cores of the respective socket, and
the ratio of cores to threads varies with the used hardware and software
configuration. Many of the threads do not occupy a full core at runtime,
especially node threads are mostly idle on a rendering client.

When using the default {\em first-touch} memory placement strategy, memory
is allocated on the processor where it is first accessed. All GPU-specific
memory allocations are done by the render threads executing the rendering code,
therefore placing the CPU-side buffers onto the same socket as the
corresponding GPU. Similarly, network buffers are allocated and used from the
one of the node threads.

\begin{wrapfloat}{benchmark}{O}{.618\textwidth}
 \includegraphics[width=.618\textwidth]{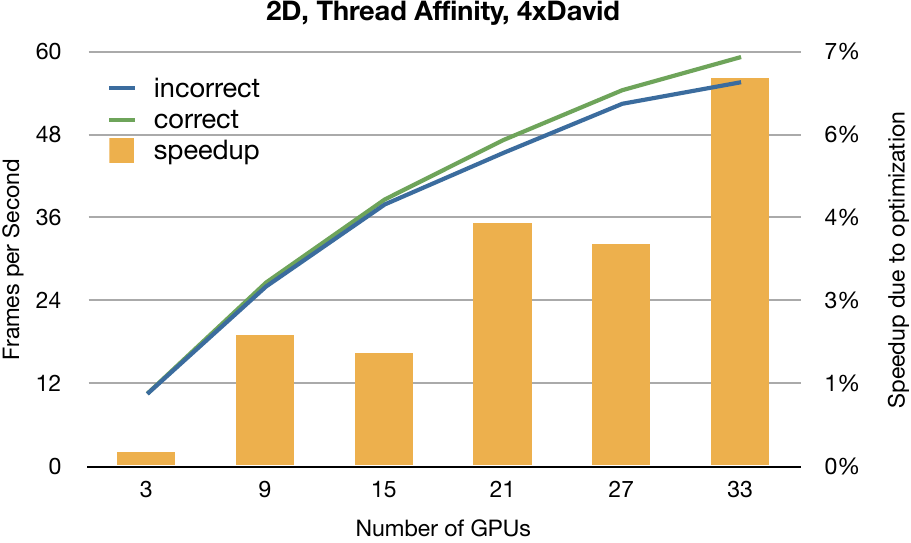}
 {\caption{\label{rNuma}Thread Affinity on NUMA Hardware}}
\end{wrapfloat}

We tested the influence of thread placement by explicitly placing the threads
either on the correct or incorrect processor. A low-level memory bandwidth test
shows a $2\times$ performance difference between these two settings. We found
that this leads to a performance improvement of more than 6\% in real-world
rendering loads, as shown in \bench{rNuma}. This benchmark uses the
aforementioned cluster nodes, and renders polygonal data using sort-first
scalable rendering. The exact experiment setup is described in
\cite{EBAHMP:12}. While this is a relatively small influence, it becomes more
important with higher frame rates as the relative draw time decreases, and the
memory-intensive compositing step importance increases. Thread placement is therefore one
of the components to achieve scalability on larger visualisation clusters with
NUMA nodes.

\section{Configuration}

A configuration consists of the declaration of the rendering resources, the
physical and logical description of the projection system, and the
configuration on how the aforementioned resources are used for parallel and
scalable rendering. A configuration is an instantiated class hierarchy in
memory used by the server to compute rendering tasks, and has a serialised text
file format to read and write configuration files.

The rendering resources are represented in a hierarchical tree structure
which corresponds to the physical and logical resources found in a 3D
rendering environment: nodes (computers), pipes (graphic cards),
windows, and channels (2D rendering area in a window).

Physical layouts of display systems are configured using canvases with
segments, which represent 2D rendering areas composed of multiple displays or
projectors. Logical layouts are applied to canvases and define the views on a
canvas. Observers observe multiple views and represent a head-tracked user in a
visualisation application.

Scalable resource usage is configured using a compound tree, which is a
hierarchical representation of the rendering decomposition and result
recomposition across the resources.

\subsection{Rendering Resources}

The first part of the configuration is a hierarchical structure of node
$\rightarrow$ pipes $\rightarrow$ win\-dows $\rightarrow$ channels describing
the rendering resources. The developer will use instances of these classes to
implement application logic and manage data.

The {\em node} is the representation of a single computer in a cluster. One operating
system process of the render client executable will be used for each node. Each
configuration might also use an application node, in which case the application
process is also used for rendering.

The {\em pipe} is the abstraction of a graphics card (GPU), and uses an
operating system thread for rendering. All pipe, window and channel task methods
are executed from the pipe thread. The pipe maintains the information about the
GPU to be used by the windows for rendering.

The {\em window} encapsulates a drawable and an OpenGL context. The drawable
can be an on-screen window or an off-screen pbuffer or framebuffer object
(FBO). Windows on the same pipe share their OpenGL rendering resources. They
execute their rendering tasks sequentially on the pipe's execution thread, in
the order they are defined in the configuration.

The {\em channel} abstracts an OpenGL viewport within its parent
window. It is the entity executing the actual rendering. The channel's
rendering context is overwritten when it is rendering for another channel
during scalable rendering. Multiple channels in application windows may be used
to view the model from different viewports. Sometimes, a single window is split
across multiple projectors, e.g., by using an external splitter such as the
Matrox TripleHead2Go.

\subsection{Display Resources}

Display resources are the second part of the configuration. They describe the
physical display setup (canvases $\rightarrow$ segments), logical display
(layouts $\rightarrow$ views) and head tracking of users within the
visualisation installation (observers).

A {\em canvas} represents one physical projection surface, e.g., a
PowerWall, a curved screen, an immersive installation, or a window on a
workstation. Canvases provide a convenient way to configure projection
surfaces. They group a set of segments (displays or projectors) into a 2D
projection surface. A canvas uses layouts describing logical views.
Typically, a desktop window uses one canvas, one segment, one layout and one
view. One configuration might drive multiple canvases, for example a projection
wall with an operator station. Planar surfaces, e.g., a display wall, configure
a frustum for the respective canvas. For non-planar surfaces, the frustum will
be configured on each display segment. The application rendering code has
access to the 2D area being updated, for example to draw 2D menus on top of the
3D rendering.

\begin{wrapfloat}{figure}{O}{.618\textwidth}
 \includegraphics[width=.618\textwidth]{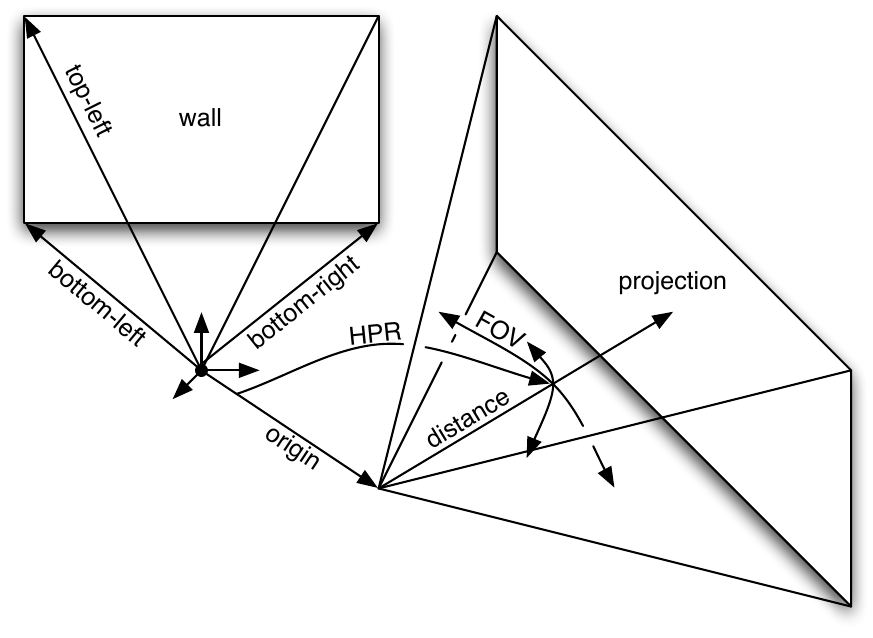}
 {\caption{\label{fFrusta}Wall and Projection Parameters}}
\end{wrapfloat}

The frustum can be specified as a wall or projection description in the global
reference system, which is shared with the head-tracking matrix of the
application. A wall is defined by the bottom-left, bottom-right and top-left
coordinates relative to the origin. A projection is defined by the position and
head-pitch-roll orientation of the projector, as well as the horizontal and
vertical field-of-view and distance to the projection wall. \fig{fFrusta}
illustrates the wall and projection frustum parameters. All size units are in
meters.

A canvas consists of one or more segments. A planar canvas typically has a
frustum description, which initialises the segment frustum based on the 2D area
covered by it. Non-planar frusta are configured by overriding the default
segment frusta. These frusta typically describe a physically correct display
setup for Virtual Reality installations.

A canvas has one or more layouts. One of the layouts is the active layout, that
is, this set of views is currently used for rendering. It is possible to
specify {\em OFF} as a layout, which deactivates the canvas. It is supported
to use the same layout on different canvases, for example to mirror a display
wall layout on a control station window.

A {\em segment} represents one output channel of the canvas, e.g., a
projector or a display. A segment has an output channel, which references the
channel to which the display device is connected. To synchronise the video
output, a swap barrier is configured to synchronise the respective window
buffer swaps. Swap barriers can use network-based software synchronisation or
hardware synchronisation based on NVidia's G-Sync hardware.

\begin{wrapfloat}{figure}{O}{.618\textwidth}
  \includegraphics[width=.618\textwidth]{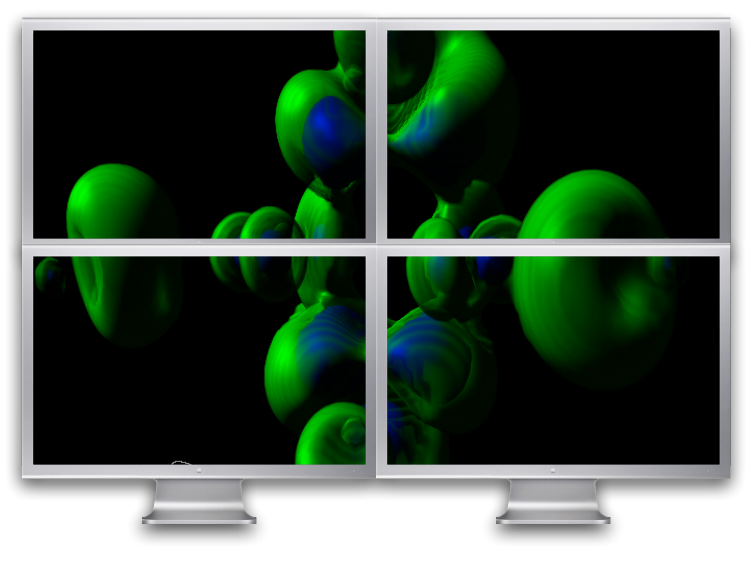}
  {\caption{\label{fCanvas}A Canvas using four Segments}}
\end{wrapfloat}

A segment covers a two-dimensional region of its parent canvas, configured by
the segment viewport. The viewport is in normalised coordinates relative to the
canvas. Segments might overlap (edge-blended projectors) or have gaps between
each other (display walls, \fig{fCanvas}\footnote{Dataset courtesy of VolVis
distribution of SUNY Stony Brook, NY, USA.}). The viewport is used to configure
the segment's default frustum from the canvas frustum description, and to place
logical views correctly.

A {\em layout} is the grouping of logical views. It is used by one or more
canvases. For all given layout/canvas combinations, Equalizer creates
destination channels when the configuration is loaded. These destination
channels may later be referenced by compounds to configure scalable rendering.
Layouts can be switched at runtime by the application. Switching a layout will
activate different destination channels for rendering.

A {\em view} is a logical view of the application data, in the sense used by
the Model-View-Controller pattern. It can configure a scene, viewing mode,
viewing position, or any other representation of the application's data. The
view object is accessible to the application thread and all render threads
contributing to its rendering. This allows the application to manage
view-specific data by attaching it as a distributed object to the view, which
will be synchronised from the application main thread to the render clients at the beginning of each frame.

A view has a fractional viewport relative to its layout. A layout is usually
fully covered by its views. Each view can have a frustum description. The
view's logical frustum overrides physical frusta specified at the canvas or
segment level. This is typically used for non-physically correct rendering,
e.g., to compare two models side-by-side on a canvas. If the view does not
specify a frustum, it will use the sub-frustum resulting from the covered area
on the canvas. A view might reference an observer, in which case its frustum is
head-tracked.

\begin{wrapfloat}{figure}{O}{.618\textwidth}
 \includegraphics[width=.618\textwidth]{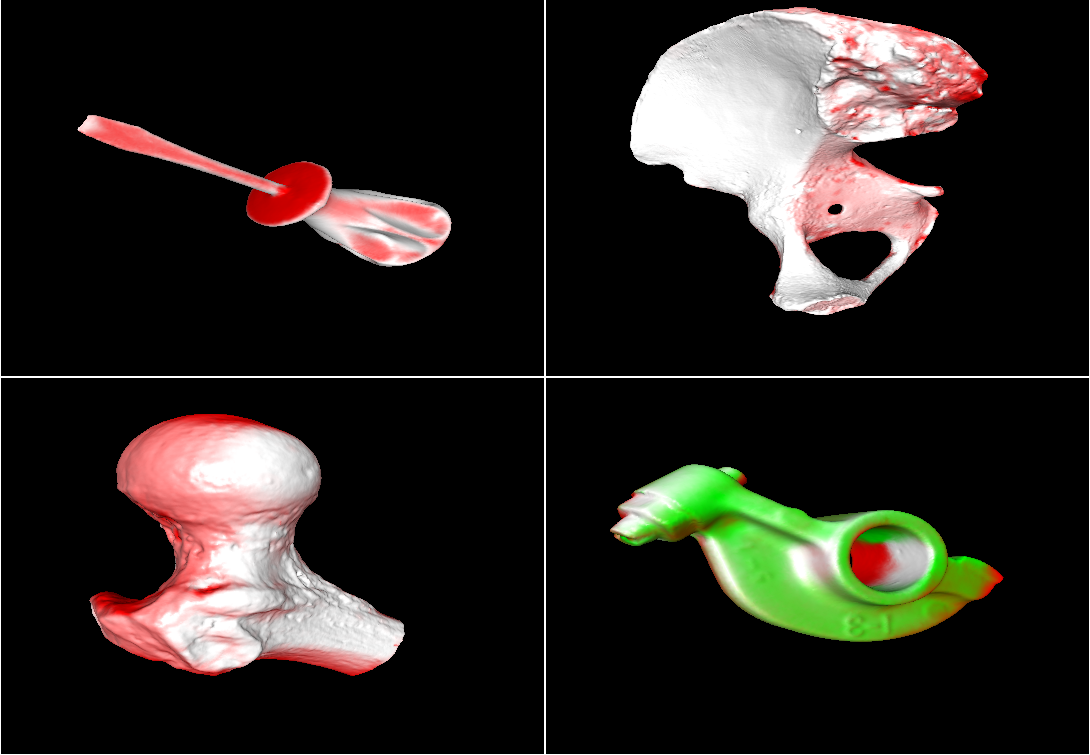}
 {\caption{\label{fLayout}Layout with four Views}}\vspace{-0.5ex}
\end{wrapfloat}

\begin{figure}[t]\center
 \includegraphics[width=\textwidth]{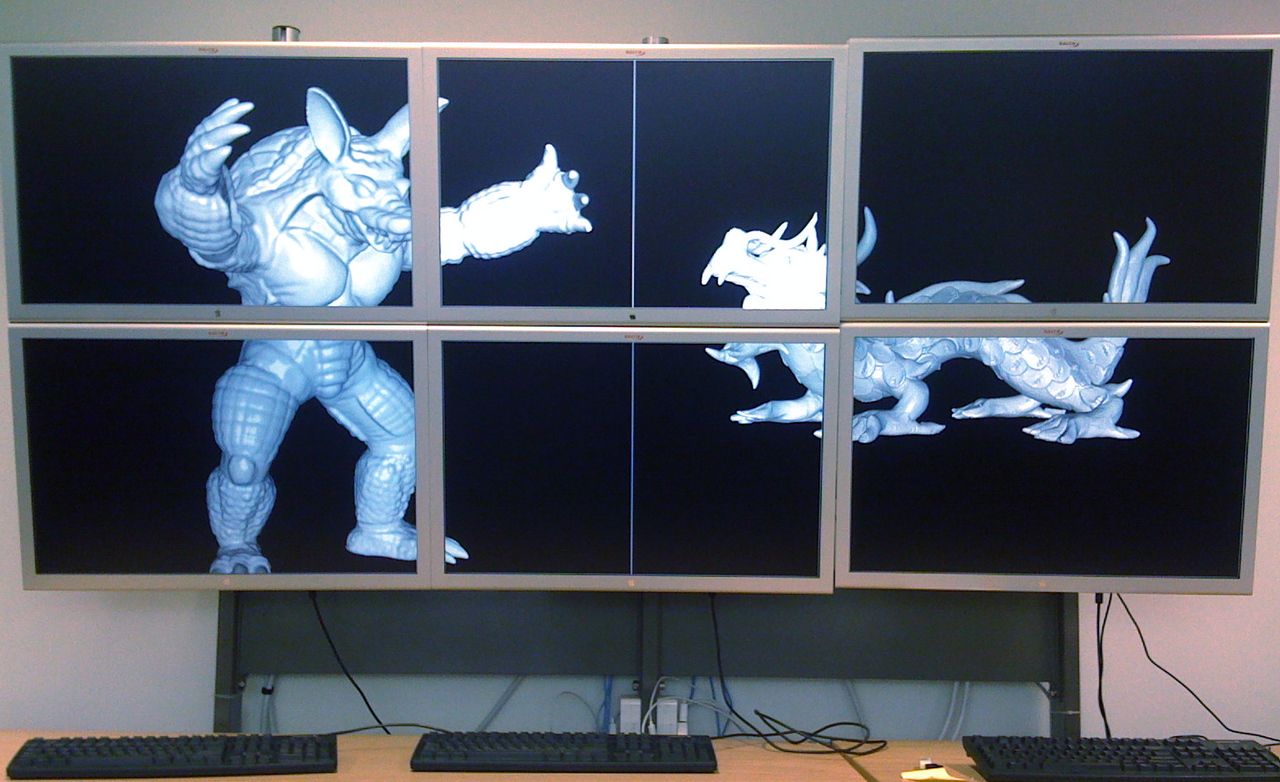}
 {\caption{\label{fDisplay}Tiled Display Wall using one Canvas with six Segments and a Layout with two Views}}
\end{figure}

\fig{fLayout} shows an example layout using four views on a single segment.
\fig{fDisplay} shows a real-world setup of a single canvas with six segments
using underlap for the display bezels, with a two-view layout. This
configuration generates eight destination channels.

An {\em observer} represents an actor looking at one or multiple views. It
has a head matrix, defining its position and orientation within the world, eye
offsets and focus distance parameters. Typically, a configuration has one
observer. Configurations with multiple observers are used if multiple,
head-tracked users are in the same configuration session, e.g., a non-tracked
control host with two tracked head-mounted displays.

\subsection{Compounds}

Compound trees describe how multiple rendering resources are combined to
produce the desired output, especially how multiple GPUs are aggregated to
increase rendering performance. They are one of the core innovations, enabling
a flexible resource configuration. Compounds are modified at runtime by
{\em equalizers} to implement dynamic behaviour, e.g., for load balancing.

Compounds are a data structure to describe the execution of rendering tasks in
the form of a tree. Each compound corresponds to some rendering tasks (clear,
draw, assemble, readback) and references a channel from the resource
description executing the tasks. The allocation of channels on pipes and
nodes determines which resources execute the task, and what can be executed in
parallel. A compound may provide output frames from the readback task to
others, and can request input frames from others for its own assembly task.
Output frames are linked to input frames by name.

Compound trees are a logical description of the rendering pipeline, and only
reference the actual physical resources through their channels. This allows
mapping a compound tree to different physical configurations by simply
replacing the channel references. For example, one can test the functionality
of a sort-last configuration by using channels of different windows on a
single-GPU workstation before deploying it to multiple physical GPUs.

A simple leaf compound description for rendering a part of the data set, given
by the data {\em range}, into a particular region of the {\em viewport} is
shown in \fig{FIG_leaf_compound}. The data range is a logical mapping of the
data set onto the unit interval and is left to the application to interpret
appropriately. Hence, the range [0 $\frac{1}{2}$] indicates that the first half
of the data set should be rendered, for example the first $\frac{n}{2}$
triangles of a polygonal mesh with $n$ faces. The viewport is indicated by the
parameters [x y width height] as fraction of the parent's viewport, and in the
example the data is thus rendered into the left half of the viewport. The
resulting framebuffer data -- including per-pixel colour and depth -- of the
rendering executed on this channel is read back and is made available to other
compounds by the name left\_half.

\begin{figure}[h!t]\center
 {\begin{tabbing} \ \ \ \ \ \=\ \ \ \=\ \ \ \=\ \ \ \=\ \ \ \=\ \ \ \= \kill
   \> {\bf compound} \{						\\
   \>\> {\bf channel} "draw"					\\
   \>\> buffer  [ COLOR DEPTH ]				\\
   \>\> range [0 $\frac{1}{2}$]				\\
   \>\> viewport [ 0 0 $\frac{1}{2}$ 1 ]                 \\
   \>\> outputframe \{name "left\_half" \}	\\
   \> \}
  \end{tabbing}}
 \vspace{-2mm}
 \caption{Compound Rendering half of the Data Set into half of the Viewport\label{FIG_leaf_compound}}
\end{figure}

A non-leaf compound performing image assembly and compositing task is
provided in \fig{FIG_inner_compound}. Framebuffer data is read from two other
compounds, which did execute rendering for part\_a and part\_b of the data
set in parallel. The compound itself executes by default $z$-depth visibility
compositing of the two input images on its channel and returns the resulting
colour framebuffer in the output frame named {\em frame.display}.

\begin{figure}[h!t]\center {\begin{tabbing} \ \ \ \ \ \=\ \ \ \=\ \ \ \=\ \ \
\=\ \ \ \=\ \ \ \= \kill \> {\bf compound} \{ \\ \>\> {\bf channel} "display"
\\ \>\> inputframe \{ name "part\_a" \} \\ \>\> inputframe \{ name "part\_b" \}
\\ \>\> outputframe \{ buffer [ COLOR ] \} \\ \> \} \end{tabbing}}
\vspace{-2mm} \caption{Compound Performing Image
Compositing\label{FIG_inner_compound}} \end{figure}

Leaf compounds execute all tasks by default, but the focus is often on the draw
task with a default assemble and standard readback task used to pass the
resulting image data on to other compounds for further compositing. While
leaf compounds execute the rendering in parallel, non-leaf compounds often
correspond to, but are not restricted to, the (parallel) image compositing and
assembly part. The readback or assemble tasks are only active if output or input
frames have been specified, respectively. Otherwise the rendered image frame is
left in-place for further processing in a parent compound sharing the same
channel.

Note that non-leaf nodes in the compound tree structure traverse their children
first before performing their default assemble and readback tasks. Furthermore,
compounds only define the logical task decomposition structure, while its
execution is actually performed on the referenced channels. Therefore, since
compounds can share channels, as often done between a parent and one of its
child compounds, rendered image data can sometimes be left in place, avoiding
readback and transfer to another node.

All attributes, as well as the channel, are inherited from the parent compound if
not specified otherwise. The {\em viewport}, {\em data range} and {\em eye}
attributes are used to describe the decomposition of the parents' 2D viewport,
database range, temporal, pixel, subpixel and eye passes, respectively.

A more formal classification of compound entities is:

\begin{compactdesc}
 \item [Root compound] is the top-level compound of a compound tree. It might
 also be a destination compound, or can be empty (not referencing a channel)
 when synchronising multiple destination channels.

\item [Destination compound(s)] are the top-most compounds referencing a
channel, which becomes the destination channel. This destination channel
determines the rendering context for the whole subtree, that is, compounds and
their channels lower in the hierarchy contribute to the rendering of the
destination channel by executing part of the destination render context and
providing output frames which will eventually be composited onto the
destination channel.

 \item [Source compounds] are the leaf nodes in a compound tree. They typically
 use a different channel from the destination channel and configure scalability
 by overriding render context parameters. This decomposes the rendering of the
 destination channel. By adding output and input frames, the partial results are
 collected and composited:
 \begin{compactdesc}
  \item[Decomposition] On each child compound the rendering task of that
  child can be limited by setting the {\em viewport}, {\em range},
  {\em period} and {\em phase}, {\em pixel}, {\em subpixel},
  {\em eye} or {\em zoom} as desired.

  \item[Compositing] Source compounds define an {\em output\_frame} to read
  back the result. This output frame is used as an {\em input\_frame} on the
  destination compound receiving the pixels. The frames are connected with each
  other by their name, that has to be unique within the root compound tree.
  For parallel compositing, the algorithm is described by defining multiple
  input and output frames across all source compounds and restricting the task
  to assemble and readback.

 \end{compactdesc}
 \item[Intermediate compounds] may be used to simplify the task decomposition or
 to configure parallel compositing.
\end{compactdesc}

\section{Virtual Reality}

Virtual Reality is an important field for parallel rendering. It requires
special attention to support it as a first-class citizen in a generic parallel
rendering framework. {\em Equalizer} has been used in many virtual reality
installations, such as the Cave2 \cite{FNTTL:13}, the high-resolution C6 CAVE
at the KAUST visualisation laboratory, and head-mounted displays
(\fig{FIG_teaser}). In the following we lay out the features needed to support
these installations, motivated by application use cases.

\subsection{Head Tracking}

Head tracking is the minimal feature needed to support immersive installations.
{\em Equalizer} does support multiple, independent tracked views through
observer abstraction. Built-in VRPN support enables the direct,
application-transparent configuration of a VRPN tracker device. Alternatively,
applications can provide a $4\times 4$ tracking matrix. Both CAVE-style tracking
with fixed projection surfaces, and HMD tracking with moving displays are
implemented.

\subsection{Dynamic Focus Distance}

To our knowledge all parallel rendering systems have the focal plane coincide
with the physical display surface. For better viewing comfort, we introduce a
new dynamic focus mode, where the application defines the distance of the focal
plane from the observer, based on the current \textit{lookat} distance.

\begin{figure}[h!t]\center
 \includegraphics[width=\textwidth]{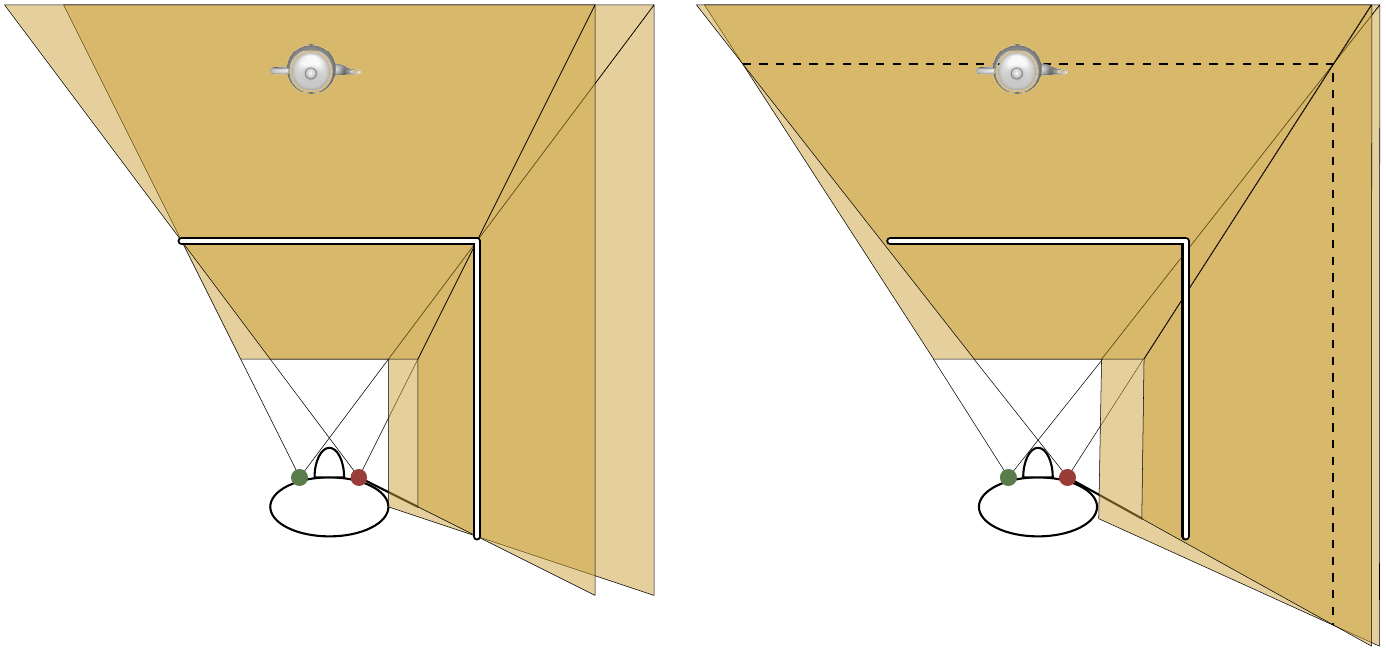}
 {\caption{\label{fFocus}Dynamic Focus in a Cave}}
\end{figure}

\fig{fFocus} illustrates this feature in a top-down view of a Cave. The
observed teapot is significantly behind the front projection wall in the
virtual world. In a standard implementation (left side), the focal plane
coincides with the projection surface. In our implementation, the application
configures a focus distance to coincide with the observed teapot (right side).
The dotted line shows the focal plane for both projection walls. Initial
experiments show that this provides better viewing comfort, in particular for
objects placed in front of the physical displays.

\subsection{Asymmetric Eye Position}

Traditional head tracking computes the left and right eye positions by using an
interocular distance. However, since human heads are not symmetric, we support
an optional configuration of individual, measured 3D eye translations relative
to the tracking matrix.

\subsection{Model Unit}

This model unit allows applications to specify a scaling factor between the
model and the real world, allowing exploration of macroscopic or microscopic
worlds in virtual reality. The unit is per view, allowing different scale
factors within the same application. It scales both the specified projection
surface, as well as the eye position (and therefore eye separation) to achieve
the necessary effect.

\subsection{Runtime Stereo Switch}

Applications can switch each view between mono and stereo rendering at runtime,
and run both monoscopic and stereoscopic views concurrently. This switch does
potentially involve the start and stop of resources and processes for passive
stereo or stereo-dependent task decompositions.

\section{Tiled Display Walls}\label{sTIDE}

Simulations performed on today’s high performance supercomputers produce
massive amounts of data, which are often too expensive to move to another
system. Tiled display walls have proven to help understand complex data due
to their size, resolution and collaborative usage. Often the two systems are
not located in the same facility because of power constraints or other factors.

\begin{wrapfloat}{figure}{O}{.618\textwidth}
  \includegraphics[width=.618\textwidth]{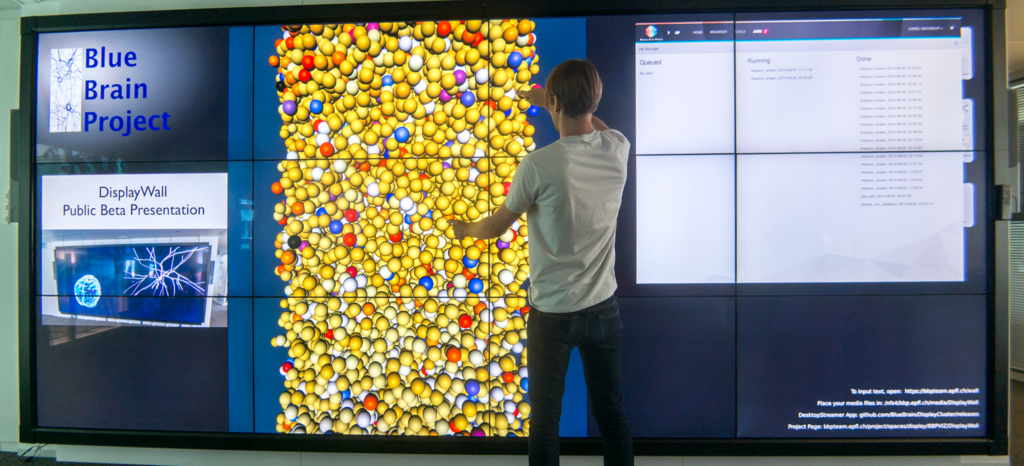}
  {\caption{\label{fTide}Tiled Display Wall with Remote Rendering of the Equalizer-based RTNeuron Application}}
\end{wrapfloat}

Software for driving tiled display walls has converged on the collaborative
aspect of these installations. Sage, Sage~2, DisplayCluster and Omegalib
implement a multi-window environment around a shared framebuffer concept.
DisplayCluster provides a dynamic, desktop-like windowing system with built-in
media viewing capability, that supports ultra high-resolution imagery and video
content, as well as remote streaming allowing arbitrary applications from
remote sources to be shown. \fig{fTide} shows our evolution of DisplayCluster
called Tide~\cite{tide} running on a 24~megapixel, $4\times 3$ tiled display
wall.

Streaming to a Tide wall is implemented using the Deflect~\cite{deflect} client
library. The application provides an image buffer to Deflect, which will be
compressed using libjpeg-turbo, and sent asynchronously and in parallel by the
stream library. Multiple stream sources from multiple processes can provide
content to a single wall window, enabling parallel streaming for parallel
rendering applications. Deflect also implements an event model, where the
application registers to receive keyboard, mouse and window management events
from the wall.

We integrated the stream library into Equalizer to send the framebuffer of each
destination channel of a view to DisplayCluster, using a direct FBO download
(if possible) or a texture download. We use asynchronous transmission to
pipeline compression, streaming, and rendering. Received events from
DisplayCluster are converted and forwarded to Equalizer’s event system. This
integration allows all Equalizer applications to benefit from streaming without
code changes, configured by specifying the DisplayCluster hostname on all views
to be streamed.

\begin{wrapfloat}{figure}{O}{.618\textwidth}
  \includegraphics[width=.618\textwidth]{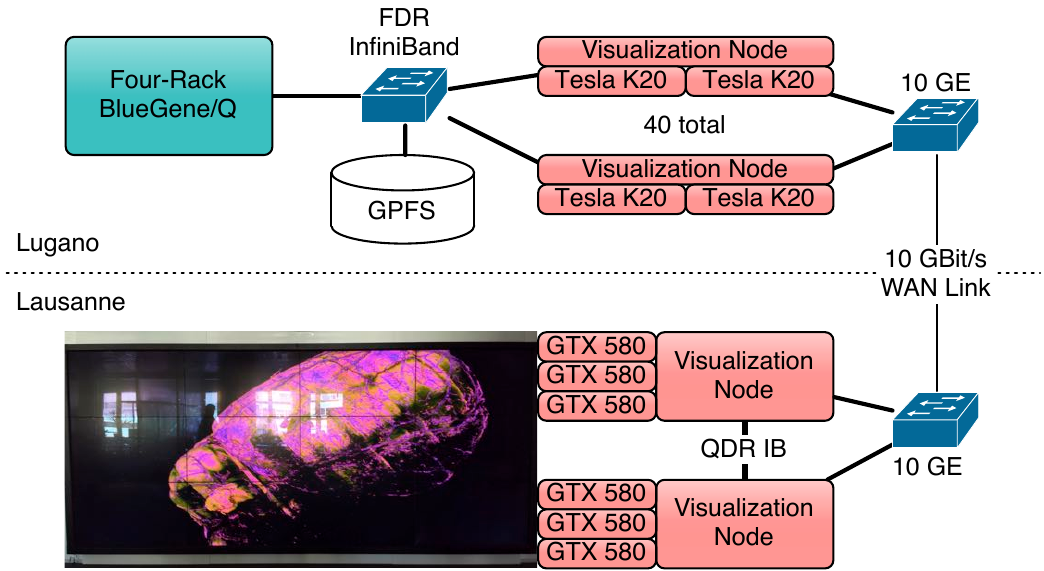}
  {\caption{\label{fTideSetup}Remote Streaming Scenario}}
\end{wrapfloat}

We evaluated the overall system performance using the Blue Brain Project setup
shown in \fig{fTideSetup}. The supercomputer and data is located in a remote
supercomputing centre in Lugano, whereas the tiled display wall is at the
project's main office in Lausanne. Both locations are linked using a high-speed
WAN link. The HPC installation has a colocated visualisation cluster for remote
rendering scenarios.

\begin{benchmark}[h!t]\center
  \includegraphics[width=\columnwidth]{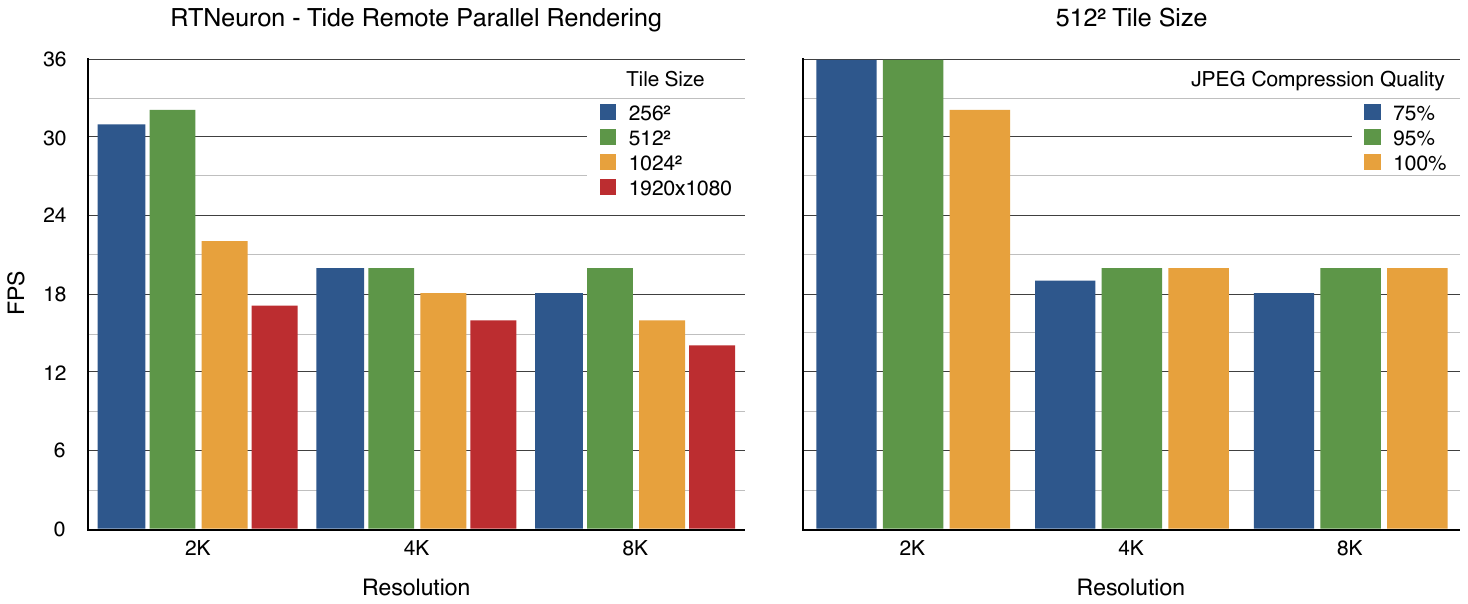}
  \caption{\label{rTide}Remote RTNeuron - Tide Parallel Rendering}
\end{benchmark}

\bench{rTide} shows the performance of streaming RTNeuron rendering from the
Lugano cluster to the remote 24 Megapixel wall. We tested three resolutions
($1920\times 1080$, $3840\times 2160$ and $7680\times 3240$), and four
different tile sizes ($256^2$, $512^2$, $1024^2$ and $1920\times 1080$). Due to
a configuration issue, the WAN link delivered only 1 GBit/s throughput during
the benchmark. RTNeuron is an Equalizer-based application used in the Blue
Brain Project to analyse results from detailed simulations of neuronal
simulations.

The results show that interactive frame rates are available even at the full
native resolution, that a $512^2$ tile size is the best option, and that 95\%
compression delivers the best performance in most cases. Based on the
experiments we settled on a $512^2$ tile size and 100\% compression quality to
avoid artefacts as the default settings in Equalizer.

Decoupling the display system and software from the rendering system has many
benefits. It increases robustness, provides reliably performance on a shared,
collaborative device, facilitates media and device inteagration, and minimises
data movement.

\section{Compositing}

In contrast to most other parallel rendering frameworks, Equalizer decouples the
compositing algorithm from the task decomposition. This is a key aspect of our
architecture, allowing a flexible configuration, often in many unforeseen
ways. The compound tree with its task decomposition, input and output frames, is
a specialised description to ``program''  scalable rendering across parallel
resources.

Compositing is configured using output frames connected to input frames,
compound tasks and eye passes, as well as frame parameters. In its simplest
form, a sort-first source compound provides an output frame, which is routed
to an input frame using the same name on the destination compound. The source
viewport decomposes the task, and the output frame collects this partial result
from the source channel to composite it using the correct offset onto the
destination channel.

Frame parameters customise pixel handling. They include the transfer buffers
(colour, depth), partial channel viewport, pixel zooming (upscale and
downscale), and transport method (on-GPU texture or CPU memory). An output
frame may be connected to multiple input frames. Frame parameters are used
together with compound tasks for parallel compositing, and advanced features
such as monitoring and dynamic frame resolution, introduced in
\cref{sLoadBalancing}.

\begin{wrapfloat}{figure}{O}{.618\textwidth}
  \includegraphics[width=.618\textwidth]{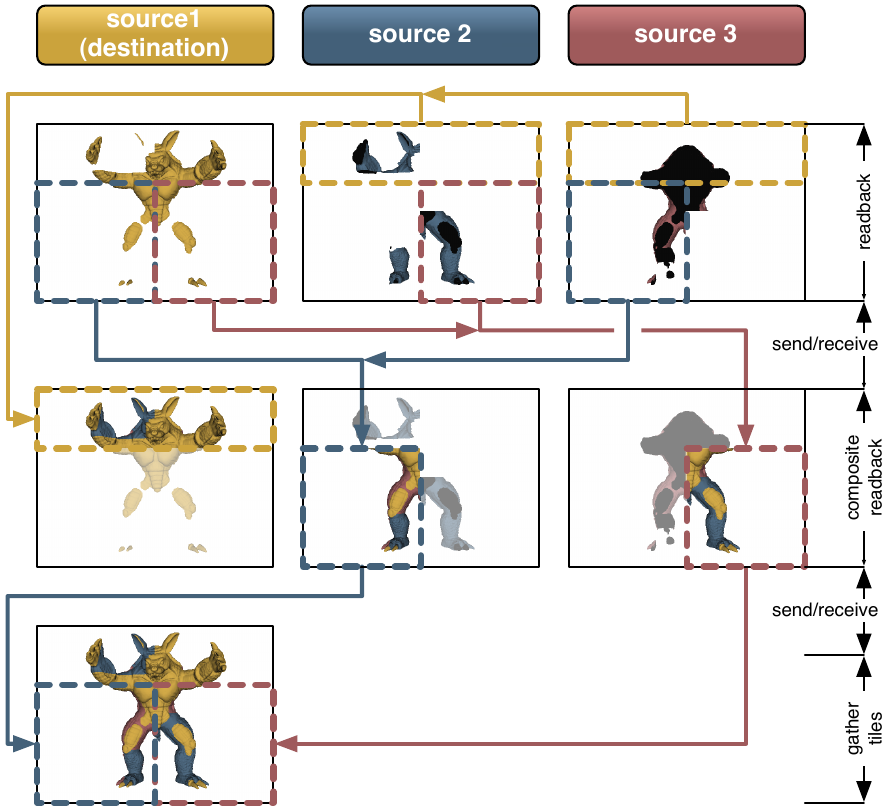}
  {\caption{\label{fDirectSend}Direct Send Compositing}}
\end{wrapfloat}

\fig{fDirectSend} shows the pixel flow of direct send compositing for a three-way
sort-last decomposition, where the destination channel also contributes to the
rendering. In the first step, each channel exchanges two colour$+$depth tiles with
its neighbours, and then $z$-composites its own tile. This yields one complete
tile on each channel, of which two colour tiles are then assembled on the
destination channel, where the third tile is already in place.

The corresponding compound tree is shown schematically in \fig{fDirectSendCmp}.
Each of the three channels has a child compound to execute the rendering and
read back two incomplete tiles for sort-last compositing on the corresponding
two sibling compounds. These three leaf compounds represent the first step in
\fig{fDirectSend}. One level up, each channel receives two tiles and assembles
them onto its partially rendered result, creating a complete tile (middle step
in \fig{fDirectSend}). For the two source-only channels, a final colour-only
output image is connected to the destination channel. The arrows illustrate the
pixel flow for one of the tiles.

\begin{wrapfloat}{figure}{O}{.618\textwidth}
  \includegraphics[width=.618\textwidth]{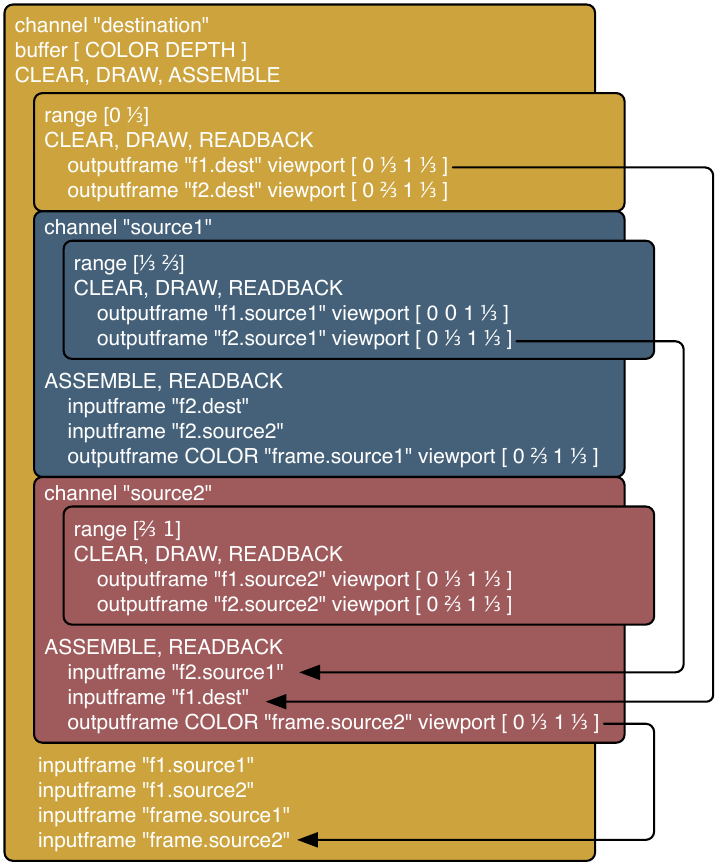}
  {\caption{\label{fDirectSendCmp}Direct Send Compound}}
\end{wrapfloat}

For most rendering applications even a relatively complex setup such as this
example requires very little programmer involvement. The abstractions provided
by the resource description, render context and compounds enable Equalizer to
reconfigure the application almost transparently. For polygonal rendering, it
suffices that the application honours the {\em range} parameter of the render
context to decompose its rendering. All other tasks, in particular the parallel
compositing and pixel transfers, are fully handled by Equalizer. Applications
which require ordered compositing, for example volume rendering, overwrite the
assemble callback to reorder the input frames correctly, before passing them on
to the compositing code.

The abstraction through frames is flexible, but still allows many architectural
optimisations:
\begin{compactdesc}
\item[Unordered Compositing:] Unless overwritten by the application, Equalizer
will composite all input frames by default in the order they become available, not in the
order they are configured. In the example in  \fig{fDirectSend}, the
destination channel will assemble its four input frames one by one as the output
frame is received. Due to asynchronous execution and resulting pipelining of
operations, the availability changes each frame depending on the runtime of
other tasks.

\item[Parallel Compression, Downloads and Network Transfers:] Compressing,
transmitting and receiving frames between nodes is handled by threads
independent from the render thread. GPU transfers are handled by asynchronous
PBO transfers. Pipelining all these operations with the actual rendering
significantly minimises the compositing overhead.

\item[On-GPU Transfers:] Occasionally the source and destination channel are on
the same GPU. Using textures as pixel buffers minimises overhead for the
output to input frame transfer.
\end{compactdesc}

In \cref{sCompositing} we provide detailed information on our compositing advances.

\section{Load Balancing}

Compounds provide only a static configuration of the parallel rendering setup.
{\em Equalizers} are algorithms hooking into a compound and modify
parameters of their respective subtree at runtime, to dynamically optimise the
resource usage. Each equalizer focuses on tuning one aspect of the
decomposition, allowing them to be composited in a configuration. Due to their
nature, they are transparent to application developers, but might have
application-accessible parameters to tune their behaviour. Resource
equalisation is the critical component for scalable parallel rendering, and
therefore the eponym for the {\em Equalizer} project name.

Compounds are a static snapshot of a configuration, and equalizers provide
dynamic configuration on top. This separation of responsibilities is an
important architectural component of our parallel rendering framework. In
\cref{sLoadBalancing} we provide an extensive overview over the available
equalizers.

\section{Runtime Reconfiguration}

Supported by the distributed execution layer, introduced in the next section,
Equalizer implements dynamic reconfiguration of a running visualisation
application. This functionality is used by runtime layout switches and runtime
reliability.

Runtime reconfiguration is designed to be a side effect of the internal resource
management algorithm, that is, the initialization and exit of a configuration
uses the same code path as the runtime addition of a single channel. Rendering
resources in Equalizer are reference counted by the compounds using them, and the
state change from inactivated to activated triggers a launch or stop of the
associated resource. These resource counters are propagated up the resource
hierarchy: A channel will (de)activate its window, pipe and node.

Channel and window (de)activation are relatively lightweight and only incur the
creation and initialisation of the class instance (and associated OpenGL
resources) on the client. A pipe (de)activation incurs additionally a new (or
removed) operating system thread, and a node (de)activation is tied to a
process. Depending on the application logic, at some level of the resource
hierarchy application data has to be distributed to the rendering client. An
application may use pre-launched rendering clients which run even when not
active, and can use this to cache application state for faster reconfiguration.

Layout switches are caused by the activation of a different layout on a canvas
by the application code. A typical layout switch will only (de)active channels,
which is a very lightweight operation. Since each combination of a layout and a
canvas creates a unique set of destination channels, the destination compounds
of these channels may use a different set of source channels for rendering,
which may reside on different GPUs or even nodes in the cluster. Some
configurations use a different number of rendering GPUs or even nodes,
 causing the startup or exit of new rendering processes in the cluster.

Runtime reliability detects failed nodes in a visualisation cluster,
independent of the cause (hardware or software failure). The server tracks a
`last seen' time stamp for each node. When waiting for a task to finish, the
server uses this time stamp to detect failures. Potentially failed nodes are
pinged with a special command, which is processed even if all application
threads are busy. Nodes still answering this command are considered alive for a
longer period, after which they are considered failed, likely due to an
infinite loop in the application code. Failed nodes are removed from the
configuration, and their associated compounds are deactivated. In the case of
load balanced source channels, the load equalizer will simply reassign the work
to other source channels. For static configurations, the source channel
contribution will be missing from the final image. For destination channels,
the corresponding output display will disappear from the configuration.

\section{Distributed Execution Layer}

\begin{figure}[ht]\center
  \includegraphics[width=\columnwidth]{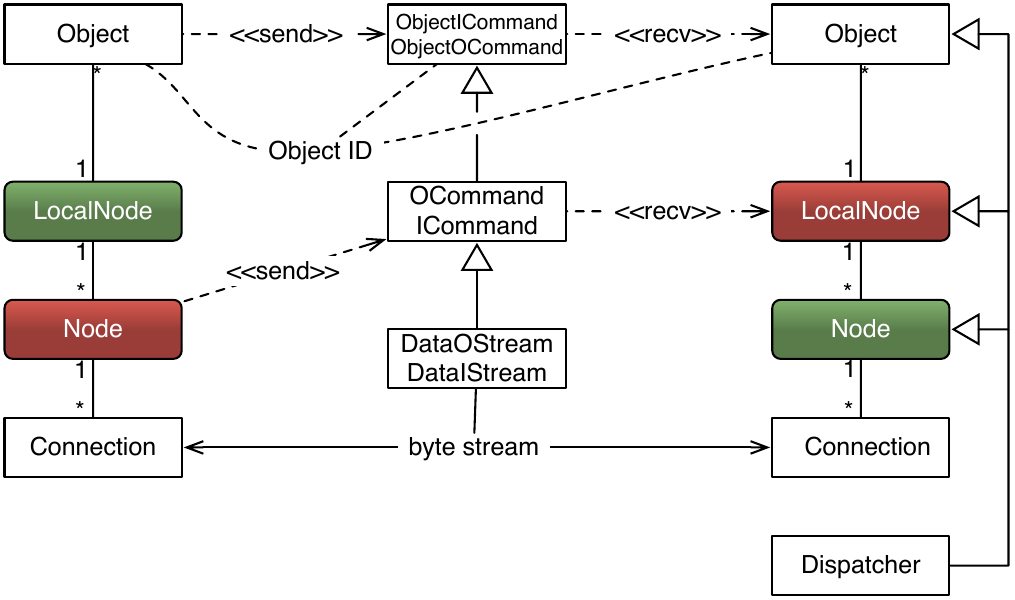}
  \caption{\label{fNetObject}Communication between two Collage Objects}
\end{figure}

An important part of writing a parallel rendering application is the
communication layer between the individual processes. Equalizer relies on the
Collage network library for its internal operation. Collage provides networking
functionality of different abstraction layers, gradually providing higher level
functionality for the programmer. \fig{fNetObject} shows the main primitives in
Collage:

\begin{compactdesc}
\item[Connection] A stream-oriented point-to-point communication
  line. Connections
  transmit raw data reliably between two endpoints for unicast connections, and
  between a set of endpoints for multicast connections. For unicast,
  process-local pipes, TCP and InfiniBand RDMA are implemented. For multicast,
  a reliable, UDP-based protocol is discussed in \sref{sec:RSP}.
\item[DataI/OStream] Abstracts the marshalling of C++ classes from or to
  a set of connections by implementing output stream operators. Uses buffering
  to aggregate data for network transmission. Performs byte swapping during
  input if the endianness differs between the remote and local node.
\item[Node and LocalNode] The abstraction of a process in the cluster. Nodes
  communicate with each other using connections. A {\em LocalNode} listens on various
  connections and processes requests for a given process. Received data is
  wrapped in {\em ICommand}s and dispatched to command handler methods. A Node is a
  proxy for a remote LocalNode.
\item[Object] Provides object-oriented and versioned data distribution of C++
  objects between nodes. Objects are registered or mapped on a Local\-Node.
\end{compactdesc}

{\em Collage} implements a few generic distributed objects, which are used by
{\em Equalizer} and other applications. A barrier is a distributed primitive
used for software swap synchronisation. Its implementation follows a simple
master-slave approach, which has shown to by sufficient for this use case.
Queues are distributed, single producer, multiple consumer FIFO queues are used for
tile and chunk compounds (\sref{sTile}). To hide network latencies, consumers
prefetch items.

An object map facilitates distribution and synchronisation of a collection of
distributed objects. Master versions can be registered on a central node, e.g.,
the application node in {\em Equalizer}. Consumers, e.g., {\em Equalizer}
render clients, can selectively map the objects they are interested in.
Committing the object map will commit all registered objects and sync their new
version to the slaves. Syncing the map on the slaves will synchronise all mapped
instances to the new version recorded in the object map. This effective design
allows data distribution with minimal application logic.

\cref{sCollage} contains more information on our network library.

\chapter{Scalable Rendering}\label{sScalable}

\section{Overview}

Scalable rendering is a subset of parallel rendering which aims to improve the
framerate of a rendering work load by decomposing it over multiple rendering
resources. Parallel rendering includes other use cases, for example to use
multiple GPUs to drive individual displays of a tiled display wall. This
chapter addresses the research question on how we can improve the rendering
performance of visualisation applications to enable users to explore more data.

Scalable rendering research has put a lot of focus on two of the three
architectures classified by \cite{Molnar92}: sort-first and sort-last rendering.
Sort-middle rendering is still largely confined to hardware implementations due
to its high communication cost of sorting and distributing fragments to
processing units.

In this chapter, we present new parallelisation variants of sort-first
rendering, and other decompositions which break out of this standard
classification. For each mode, we introduce its algorithm and implementation,
potential impact on the application code, as well as its strengths and
weaknesses. Due to the flexible architecture of our parallel rendering system,
these modes are largely usable with any Equalizer application and can be
combined with all other modes.

Most of these rendering modes are similar to sort-first rendering, in that they
decompose the final view spatially or temporally, while computing complete
pixels on each source channel. Stereo compounds decompose per eye pass, DPlex
compounds temporally, pixel and subpixel compounds use equal spatial
decompositions. Finally, tile and chunk compounds implement implicit
load balancing for sort-first and sort-last rendering using queueing of work
items.

This wide set of decomposition modes for scalable rendering, embedded in our
generalised compound structure, enables applications and researchers to
decompose the rendering task in, as far as we know, any way possible for a
rendering pipeline. To our knowledge no other implementation provides this
breath and flexibility, and some algorithms appear for the first time in
Equalizer.

\section{Sort-First}

Sort-first rendering decomposes the rendering task in screen space. It has many
variants: tiled display walls and similar installations perform sort-first
parallel rendering naturally by using multiple GPUs to drive the output
displays, and classic sort-first scalable rendering assigns one screen-space
region to each rendering resource, often combined with load balancing.
Equalizer supports these classic sort-first modes. In the following subsections
we present other variants of sort-first rendering, each tailored to a certain
use case.

\subsection{Stereo}

\begin{wrapfloat}{figure}{O}{.618\textwidth}
 \includegraphics[width=.618\textwidth]{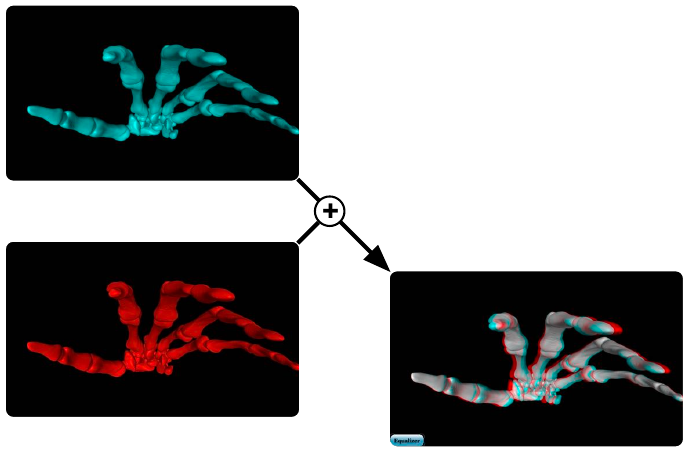}
 {\caption{\label{fStereo}Anaglyphic Stereo Compound}}
\end{wrapfloat}

Stereo, or eye decomposition, is a specialised version of sort-first rendering.
Two GPUs get assigned each one of the eye passes during stereo rendering. For
passive stereo, there is no compositing step needed, whereas for active and
anaglyphic stereo the frame buffer for one eye pass has to be copied to the
destination channel. Due to the strong similarity between both eye passes, this
mode provides close to perfect load balance. \fig{fStereo} shows an ana\-gly\-phic stereo compound.

While many visualisation applications provide passive or active stereo rendering and
sometimes decomposition using two GPUs, our implementation within our flexible
compound structure allows to fully exploit stereo decomposition.  Stereoscopic
and monoscopic rendering is no special case in the architecture, but rather a
configuration of the rendering resources. Among other things, this allows
extending a two-way stereo decomposition with further resources and any other
scalable rendering mode. One can also easily set up an application with mixed
rendering, e.g., to render a monoscopic view on a control workstation while
rendering stereoscopic on a larger immersive installation.

Stereo compounds are configured by restricting each source to a single eye
pass. Typically, one of the channels also configures the {\em cyclop} eye
pass, which gets activated when the view is switched to monoscopic rendering.

\subsection{Pixel}

\begin{wrapfloat}{figure}{O}{.618\textwidth}
 \includegraphics[width=.618\textwidth]{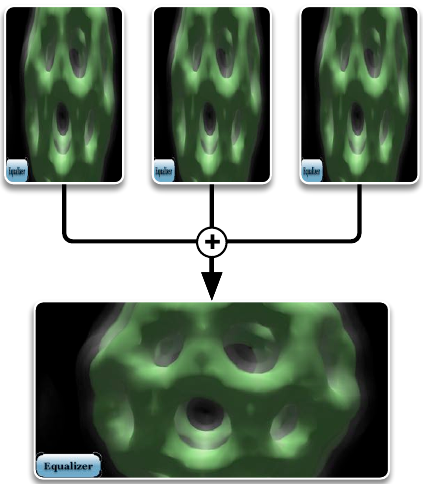}
 {\caption{\label{fPixel}Pixel Compound}}
\end{wrapfloat}

Pixel compounds (\fig{fPixel}) decompose the destination channel by
interleaving pixels in image space. They are a variant of sort-first rendering
well suited for fill-limited applications which are not geometry bound, for
example direct volume rendering. Source channels cannot reduce geometry load
through view frustum culling, since each source channel has almost the same
frustum as the destination channel. However, the fragment load on all source
channels is reduced linearly and well load balanced due to the interleaved
distribution of pixels. This functionality is transparent to {\em Equalizer}
applications, and the default compositing uses the stencil buffer to blit
pixels onto the destination channel.

Pixel compounds are configured by restricting each source compound with a pixel
kernel. The kernel describes the size of the decomposition in 2D space, and the
2D pixel offset within this region. This follows our design philosophy of
enabling features by generalising the underlying algorithm rather then
hardcoding them.

\subsection{Subpixel}

\begin{wrapfloat}{figure}{O}{.618\textwidth}
 \includegraphics[width=.618\textwidth]{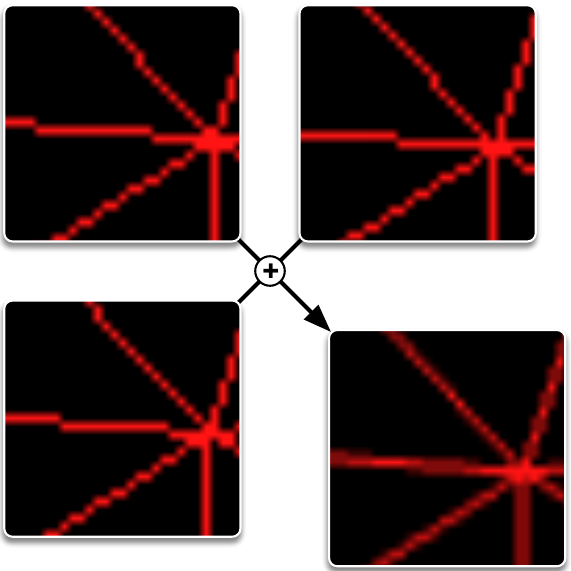}
 {\caption{\label{fSubpixel}Subpixel Compound}}
\end{wrapfloat}

Subpixel compounds (\fig{fSubpixel}) are similar to pixel compounds, but they
decompose the work for a single pixel, for example during Monte-Carlo ray
tracing, FSAA or depth of field rendering. The default compositing algorithm
uses accumulation and averaging of all computed fragments for a pixel. Like
Pixel compounds, this mode is naturally load balanced on the fragment
processing stage but cannot scale geometry processing. This feature is not
fully transparent to the application, since it decomposes rendering algorithms
which render multiple samples per pixel. Applications needs to adapt their
rendering code, for example to jitter or tilt the frustum based on the subpixel
executed in the current subpixel rendering pass.

Subpixel compounds increase the amount of pixels to be composited linearly with
the number of source channels. They can use the same parallel compositing
algorithms as sort-last rendering. Since the compositing logic is decoupled
from the task decomposition, it reuses the same code as sort-last parallel
compositing except for the combination step on each GPU.

Subpixel compounds are configured on each source compound with a subpixel
kernel. This kernel describes the number of contributing sources, and the
offset for each source in this range.

\subsection{Tiles \label{sTile}}

\begin{wrapfloat}{figure}{O}{.618\textwidth}
 \includegraphics[width=.618\textwidth]{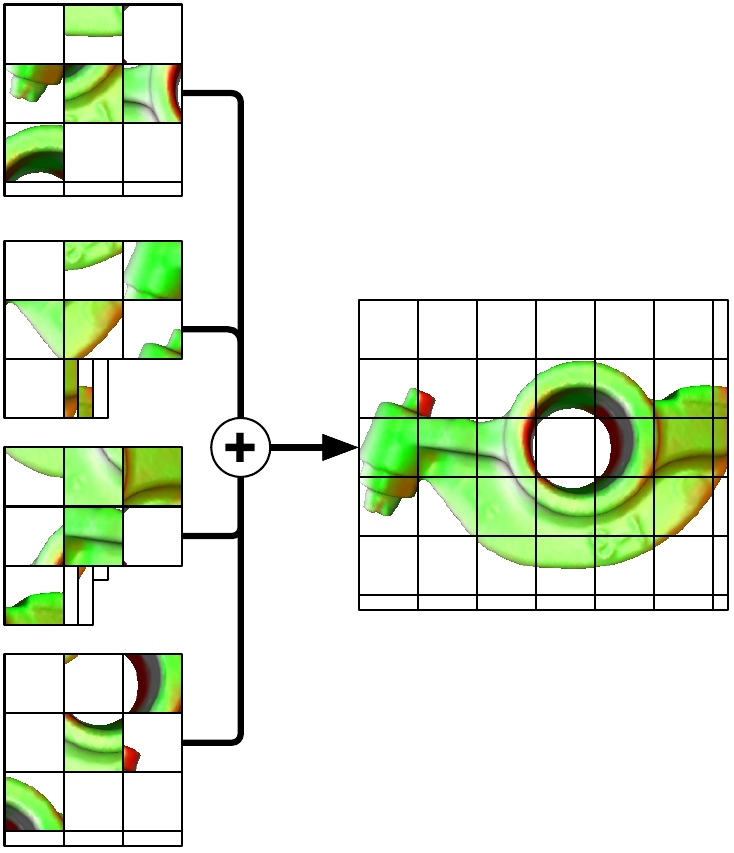}
 {\caption{\label{fTile}Tile Compound}}
\end{wrapfloat}

Tile (\fig{fTile}) decompositions are a variant of sort-first rendering. They
decompose the scene into a set of fixed-size image tiles. These tasks, or work
packages, are queued and processed by all source channels by polling a
server-central queue. Prefetching ensures that the task communication overlaps
with rendering.

As shown in~\cite{SPEP:16}, work packages can provide better performance due to
being implicitly load balanced, as long as there is an insignificant overhead
for the render task setup. This mode is transparent to {\em Equalizer}
applications. We have used a tile compound to scale an interactive ray tracing
application to hundreds of rendering nodes using RTT Deltagen.

Tile compounds are configured using output and input queues. The destination
channel has an output queue, which configures the tile size and represents the
server-side end of the task queue. Each source compound has an input queue of
the same name, which represents the client-side queue end polling tasks from
the server. Output frames from tile sources are automatically configured with
the current tile offset and size for correct assembly on the destination
channel.

\section{Sort-Last}

Sort-last rendering decomposes the rendering task in object space, that is,
each rendering resource produces an incomplete full-resolution image. To our
knowledge, sort-last rendering always requires a compositing step, which is the
challenging part for this decomposition mode. It is often addressed using
parallel compositing, which we discuss in \sref{sparcomp}.

Equalizer does support classical sort-last rendering with or without load
balancing, where each resource renders one part of the applications database.
Furthermore, we also implement chunk compounds, which are similar to tile
compounds (\sref{sTile}), with which they share a lot of the infrastructure.
Chunk compounds also produce work packages, although using a fixed-size
subrange of data for each package instead of the tile coordinates used for tile
compounds.

\section{Time-Multiplex}

Time-multiplexing distributes full frames over the available resources, such
that each resource only renders a subset of the visible frames (\fig{fDPlex}).
This mode is also called alternate frame rate or DPlex, was first implemented
in \cite{BRE:05} for shared memory machines. The algorithm is however much
better suited for distributed memory systems, since the separate memory space
makes concurrent rendering of different frames much easier to implement. While
it increases the framerate almost linearly, it cannot improve the latency
between user input and the corresponding output. At best, it can achieve the
same latency compared to the single-GPU case, when perfect linear scalability
is achieved. Consequently, this decomposition mode is mostly useful for
non-interactive film generation.

\begin{wrapfloat}{figure}{O}{.618\textwidth}
\includegraphics[width=.618\textwidth]{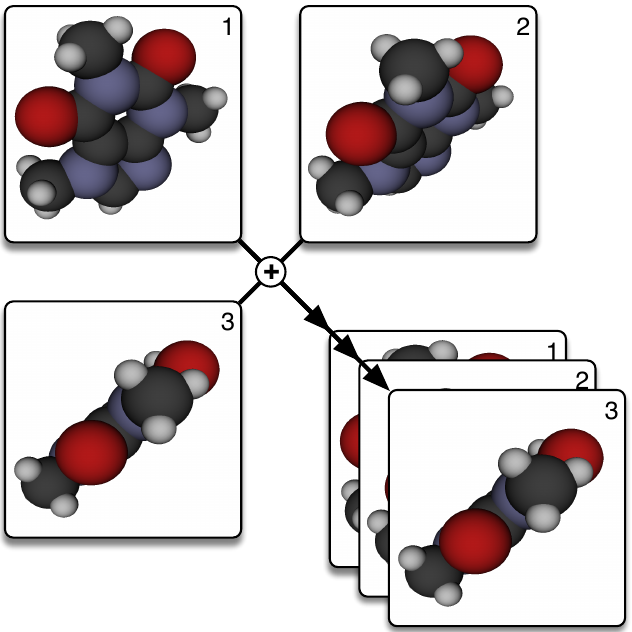}
{\caption{\label{fDPlex}Time-Multiplex Compound}}
\end{wrapfloat}

DPlex is very well load balanced, since most applications observe a strong
frame-to-frame coherence with respect to the rendering load. This decomposition
mode has the peculiarity that small imbalances tend to accumulate such that the
concurrent frames all finish simultaneously. To provide a smooth framerate, if
so desired, a framerate equalizer can be installed on the destination compound.
\sref{sFramerateEq} covers this functionality. It is transparent to
{\em Equalizer} applications, but does require the configuration latency to
be greater or equal to the number of source channels.

DPlex rendering is not hard-coded into our framework, but configured by
restricting the rendering task temporally on each source compound. This is
achieved by setting a {\em period} and {\em phase} parameters, which
configure the number of frames skipped and starting offset on the given source
compound. A simple DPlex compound would have a destination compound with $n$
source compounds, where each source has a period of $n$ and one phase from
$0..n-1$. While this generalization may seem artificial, it opens up different
use cases, for example giving a fast GPU a smaller period, thus giving it more
work.

\section{Stereo-Selective Compounds}

\begin{figure}[h!t]\center
 \includegraphics[width=\columnwidth]{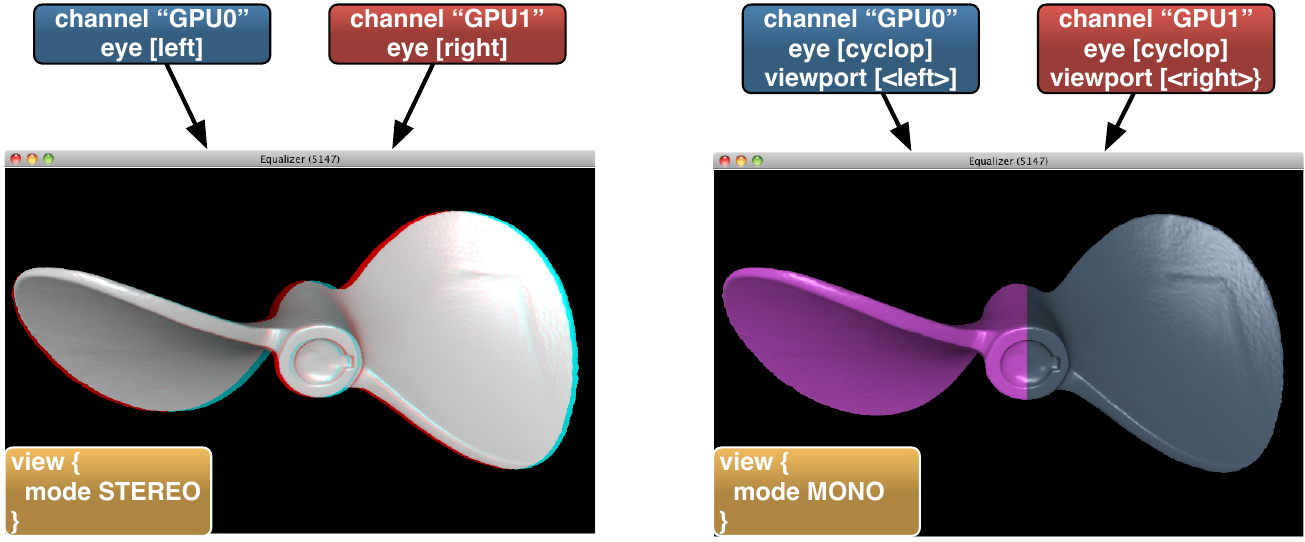}
 {\caption{\label{fStereoSwitch}Stereo-Selective Compound}}
\end{figure}

Stereo-selective compounds have different configurations, depending on the
current rendering mode. Each compound sub-tree can restrict the eye passes it
renders from the default left, right, cyclop passes. Depending on the active
stereo mode (stereo or mono), restricted compound trees may be skipped or
activated. This is used on one hand to configure stereo compounds, but may also
be used to configure different decompositions depending on the stereo mode.
\fig{fStereoSwitch} shows a simple example: A dual-GPU setup is used with
eye-parallel rendering during stereo rendering, and a standard sort-first
parallel rendering during monoscopic rendering. Note that the rendering mode is
runtime-configurable, that is, the application can switch the view from
monoscopic to stereoscopic rendering at any time, activating and deactivating
the configured compounds and attached resources. It is also possible to
configure a different set of resources (nodes and GPUs) per stereo mode,
triggering the launch and exit of render client processes during the stereo
switch.

\section{Mixed Mode Compounds}

A major contribution of our parallel rendering system is the flexible system
architecture. While many applications and frameworks implement a subset of the
features mentioned above, most of them hardcode the algorithms, predetermining
the number of possible configurations. In Equalizer, both the decomposition and
the recomposition of the rendering task are derived through a number of
orthogonal parameters, which are easily combined to configure common scalable
rendering modes. For advanced usage, they can also be configured for many other
use cases. During deployment of Equalizer, we have seen many interesting and
unforeseen configurations:

\begin{compactitem}

\item Reusing the {\em period} parameter used to configure number of frames
in a DPlex compound, an underpowered control workstation for a large tiled
display wall was configured to render only every other frame using a period of
two. Due to the standard latency of one frame, this meant that the display wall
rendering became the bottleneck. It could now render at a substantially higher
framerate than before, when the control host was the bottleneck.

\item Rerouting one of the eye passes of a head-mounted display to a large
display using an output and input frame, external users could observe the
interaction and view of the person using the HMD. The same can be achieved by
mirroring the video signal by other means, but this was not available on the
given setup.

\item Using combined stereo and sort-first decomposition on the central tiles of
a tiled display wall. Often times the central tiles of a tiled display wall
receive a higher rendering load then the outer tiles. In this particular
configuration, each tile was driven by a dual-GPU node using active stereo
compounds, and the middle segments where given an additional machine setting up
a two-way sort-first decomposition under each node of the two-way stereo
compound.

\item Combined sort-last and sort-first decomposition: Sort-first rendering is
typically limited in the scalability of the decomposition step, where geometry
overlap between resources often yields diminishing returns after about ten
GPUs. Sort-last rendering on the other hand is often limited by the overhead of
the compositing step. Combining both modes enables to balance these constraints
for better scalability.

\end{compactitem}

\section{Benchmarks}

Benchmarks for static compound configurations are relatively rare, since most
practical settings use some type of load balancing. They are however
interesting in that they show how well different rendering algorithms are
naturally load-balanced. In \cite{ESP:18}, we collected some data for polygonal
and volume rendering. \bench{rStatic} provides a strong scalability benchmark
for both types of rendering and a set of compounds. The linear scaling graph
provides a theoretical limit for perfect scalability compared to the
single-threaded, single-GPU rendering performance.

\begin{benchmark}[h!t]\center
 \includegraphics[width=\columnwidth]{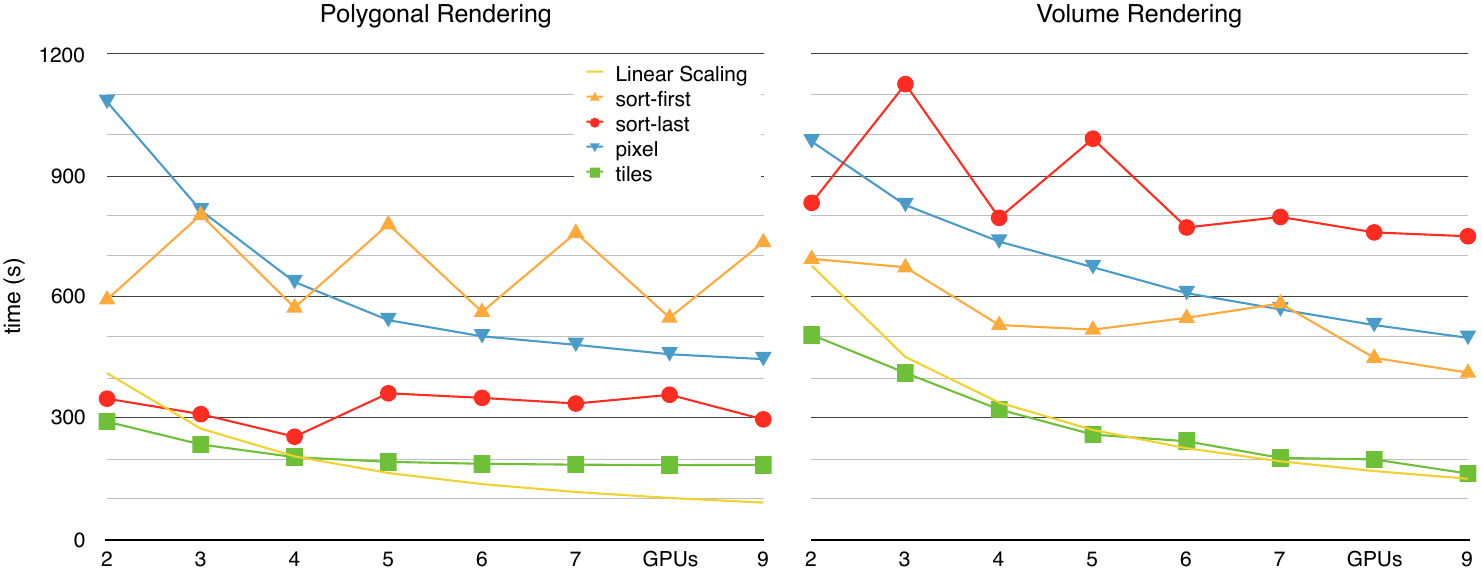}
 {\caption{\label{rStatic}Compound Scalability}}
\end{benchmark}

For static task decomposition, polygonal rendering performs better with
sort-last compared to sort-first. Sort-last performs a static decomposition in
data space, which reduces the geometry processing load per GPU, which is the
dominant factor in our polygonal rendering code. Since this decomposition can
be computed easily, even a static decomposition is relatively balanced. A
sort-first decomposition can reduce the geometry processing through view
frustum culling, but the remaining visible set will be relatively unbalanced on
each GPU, depending on the current camera position.

For volume rendering, this balance is reversed and sort-first performs better.
Typically, a volume renderer is bound by fragment processing. Consequently,
sort-first rendering scales better than sort-last, since the screen-space is
equally divided. For both rendering algorithms, one can observe static
imbalances in the zig-zag graphs, where odd number of resources coincidentally
split the rendering load less balanced than even numbers.

Tile compounds provide close to linear scalability, and in some cases
super-linear scaling. Compared to the other compounds, tile compounds are
naturally load balanced, providing this excellent scalability relative to
static sort-first and sort-last. Super-linear scaling is due to their small
work package size, which makes rendering more cache-friendly. Polygonal
rendering has a higher static overhead per tile due to the CPU-side view
frustum culling, and therefore scales less well compared to volume rendering.

Pixel compounds provide predictable scalability, but fail to approach ideal
linear scaling due to their increased compositing cost and constant geometry
load.

Benchmarks \ref{rCompounds} \ref{rEqualizers}, \ref{rCSLBFPS} and \ref{rCSLB}
provide more realistic scalability data when using these compounds with load
balancers.

\chapter{Compositing}\label{sCompositing}

\section{Overview}

Compositing collects and combines partial results from multiple resources
during scalable rendering onto one or more destination channels. While
significant characteristics of the decomposition step, discussed in the
previous chapter, are dependent on the application rendering code, compositing
is largely a generic problem and can be implemented and optimised in a parallel
rendering framework. Consequently, this area of parallel rendering research has
received significant attention. By integrating many state-of-the art
optimisations into our parallel rendering framework, we provide a generic
solution that scales well on modern visualisation cluster architectures.

We present new insight into the behaviour of known sort-last parallel
compositing algorithms on mid-size visualisation clusters (compared to high-end
HPC systems), the importance of streaming sort-last compositing and spatial
sort-last polygonal rendering, the impact of state of the art optimisations
such as region of interest and asynchronous compositing, as well as image
compression algorithms for high-speed interconnects. This chapter addresses the
research question which new algorithms will decrease the time needed to
composite rendering results, in particular for sort-last rendering.

\section{Parallel Compositing\label{sparcomp}}

Parallel compositing leverages multiple compute resources, memory bandwidth and
network bandwidth within a visualisation cluster to accelerate the compositing
step in parallel rendering. During sort-last and subpixel decompositions, each
rendering resource produces an output which needs to be combined with the
result of other resources on a per-pixel level. This compositing step reduces
the amount of information, either through depth-sorting or blending multiple
input fragments into a single pixel. This loss of information in the
compositing step can be exploited by distributing the work over multiple
resources, and then collecting the reduced image tiles, commonly called parallel
compositing.

\subsection{Spatial and Depth-Sorted Sort-Last Compositing}

For sort-last rendering, two approaches to combine the partial
results exist: $z$-sorting using the depth buffer, and spatial rendering
decomposition with ordered compositing.

\begin{figure}[h!t]\center
  \subfigure[]{
 \includegraphics[height=4.5cm]{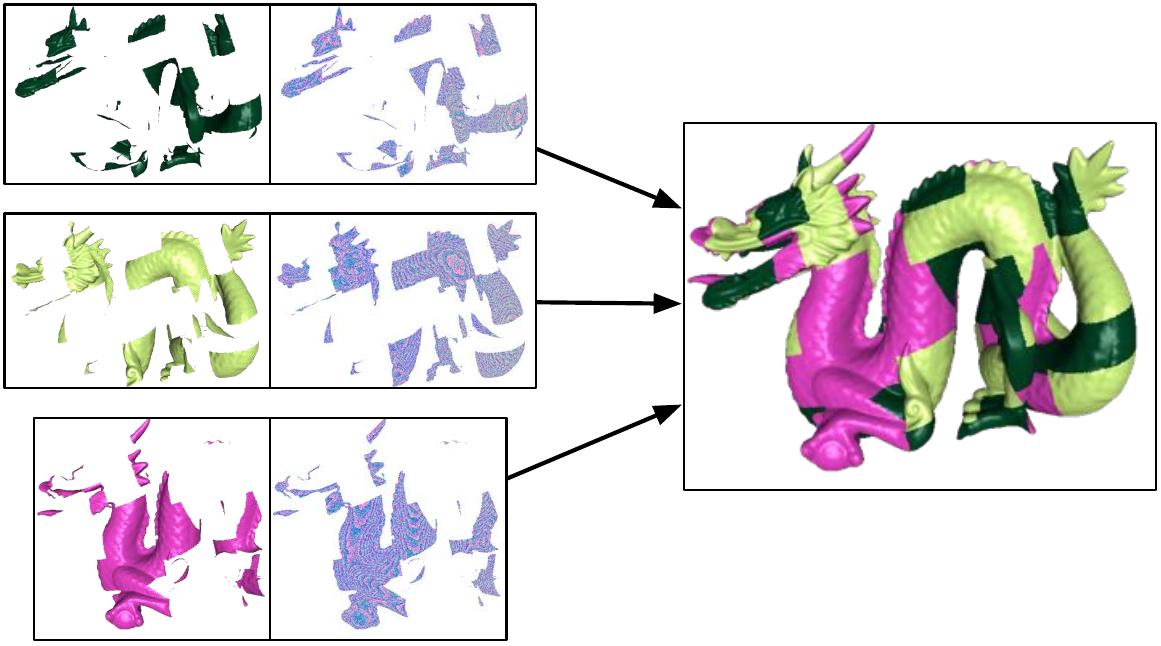}
    \label{fDepth}
  }\hfil
  \subfigure[]{
    \includegraphics[height=4.5cm]{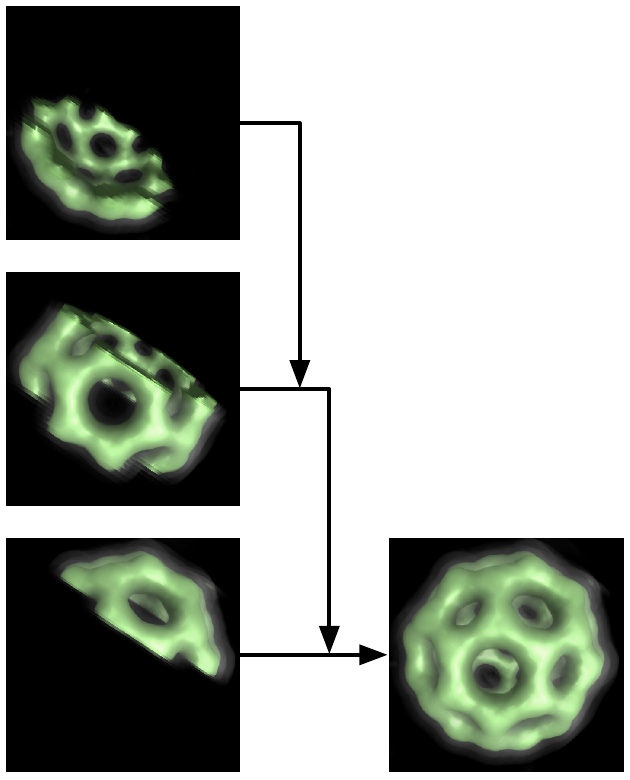}
    \label{fSorted}
  }
  \caption{Depth-Sorted and Back-to-Front Sort-Last Compositing}
\end{figure}

The first algorithm uses both the colour and depth buffer, and assigns the
final pixel to the colour of the source with the front-most depth buffer
values. It requires no spatial ordering of the data during rendering. It does
not correctly composite pixels with transparent geometry, since there is no
guarantee of the blending order. Owing to the use of the depth buffer, it is
also more expensive, since both color and depth data needs to be processed.
Furthermore, depth buffer readback tends to be slower, and compression
algorithms for depth buffer data do not to perform as well as colour buffer
compression. Depth-sorted compositing is often used for polygonal data, as
shown in \fig{fDepth}, since these applications often do not sort their
geometry into convex spatial regions.

Spatial compositing is often used for direct volume rendering and requires the
application to render convex regions of data on each source, and then
depth-sort the partial images produced by each source. The partial images are
composited in order, typically with alpha-blending. Since the sorting happens
at the image level, rather then the fragment level as in the first algorithm,
it can operate using only the colour buffer, as shown in \fig{fSorted}. This
algorithm can produce correct transparency, since the convex regions allow
ordered blending.

Spatial compositing provides better performance, and better scalability when
used with other optimisations, such as region of interest and load balancing
due to compact regions produced by the spatial sorting. Typically used for
volume rendering, we have applied spatial sort-last rendering and compositing
to polygonal data, by sorting the data spatially and using clipping planes to
generate perfectly convex rendering subsets.

\begin{wrapfloat}{benchmark}{O}{.618\textwidth}
 \includegraphics[width=.618\textwidth]{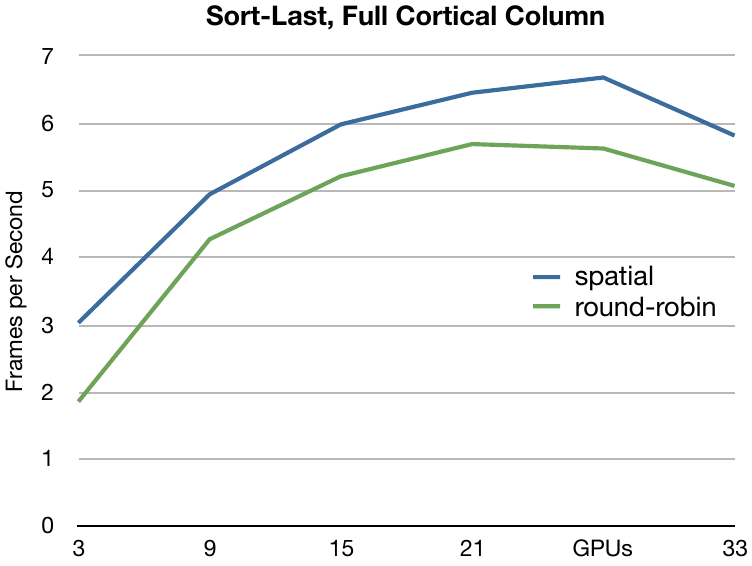}
 {\caption{\label{rDBSpatial}Spatial versus Depth-Sorted Sort-Last Rendering}}
\end{wrapfloat}

\bench{rDBSpatial} shows the difference between spatial and depth-sorted
sort-last rendering in RTNeuron (\sref{sRTNeuron}). Due to complexities in the
application data model and the disadvantageous geometrical structure of
neurons, the spatial rendering in RTNeuron scales less than the
round-robin allocation used for the depth-sorted mode. Still, owing to the
significantly reduced compositing load of this mode, both due to a smaller
region of interest and no depth buffer transfers, spatial sort-last rendering
has a significantly better framerate. The exact experiment setup can be found
in \cite{EBAHMP:12}.

\subsection{Direct Send and Binary Swap}

Contrary to most other implementations, parallel compositing algorithms in
Equalizer are not hardcoded, but rather configured explicitly. The transport
of pixel data for compositing is expressed through
connected output and input frames. Output and input frames are connected by
name; they do not need to follow the compound hierarchy, and a single output
frame may be consumed by multiple input frames. Output frame parameters
configure a subset of the rendering (viewport and framebuffer attachments), and
are read back after rendering and assembly. Furthermore, every step of the
compositing pipeline is implemented in Equalizer and transparent for the
application developer. Some steps may be replaced with application code, for
example ordering frames during compositing.

Two commonly used parallel compositing algorithms are direct send and binary
swap. Both distribute the compositing task equally over all available
resources, then collect the composited tiles on the destination channel.

\begin{figure}[h!t]\center
 \includegraphics[width=\columnwidth]{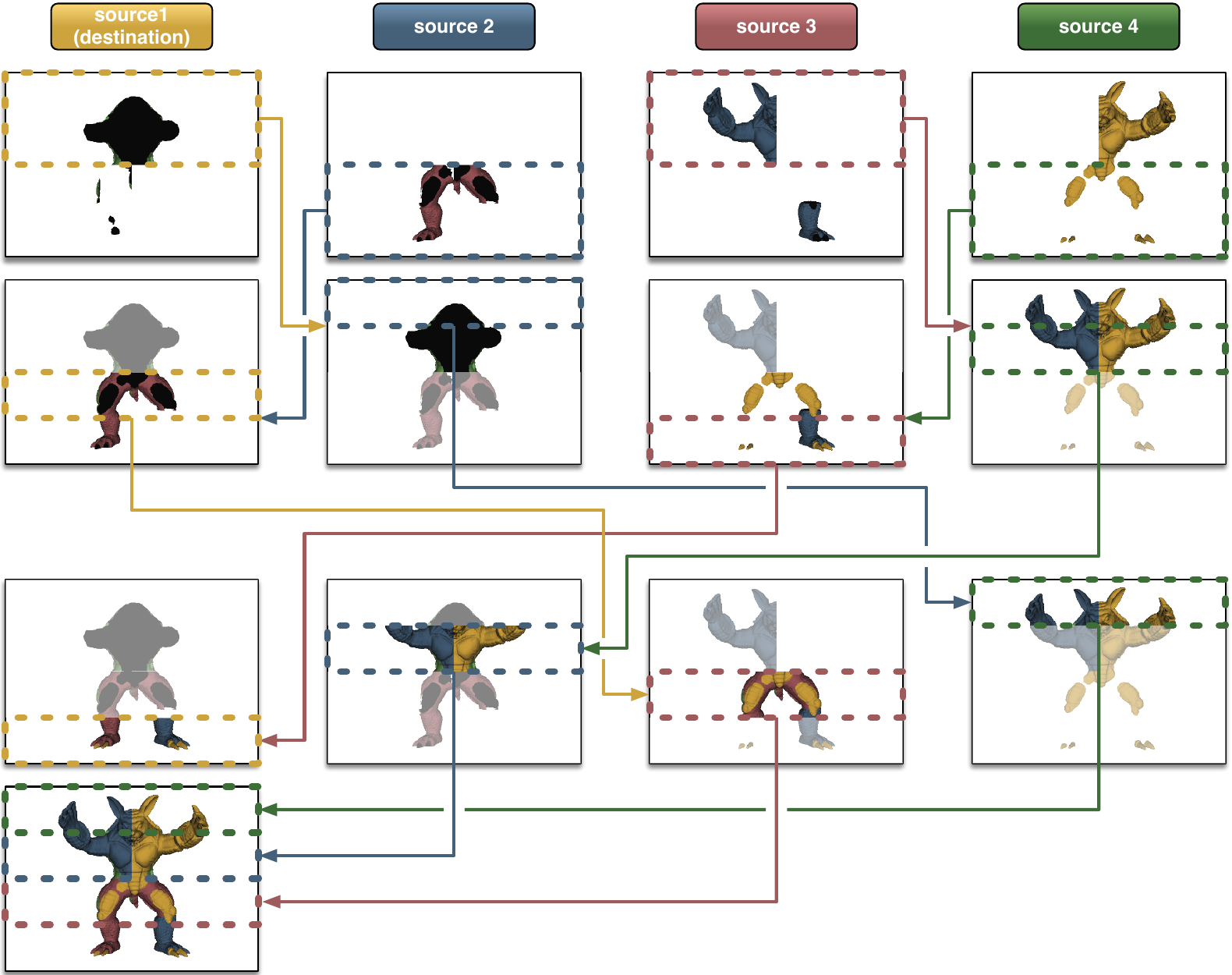}
 {\caption{\label{fBS}Binary Swap Sort-Last Compositing}}
\end{figure}

Direct send, shown in \fig{fDirectSend}, uses one assemble operation on each
resource to fully composite a single tile. Binary swap, shown in \fig{fBS},
exchanges pixels between pairs of nodes using a binary compositing tree which
gradually assembles a tile on each resource. Both use a sort-first-like
assembly operation to collect the fully assembled tiles on the destination
channel. 2-3 swap \cite{Yu:2008:MPV:1413370.1413419} is an extension to binary
swap, which overcomes the power-of-two source channel requirement by exchanging
compositions between groups of two or three nodes in the compositing tree.

In \cite{EP:07} we have shown that on commodity clusters, direct send
compositing provides better performance over binary swap commonly used on HPC
systems. While it uses more messages in total, direct send has fewer
synchronisation points than binary swap. Moreover, as a result of the early
assembly optimisation (\sref{sEarlyAss}), direct send can handle imbalances
between nodes better, since a late channel has a smaller penalty on the overall
execution time.

\subsection{Streaming Sort-Last Compositing}

As a result of the asynchronous architecture of our framework, streaming sort-last
compositing is a viable alternative to more involved parallel compositing
algorithms in smaller sort-last configurations.

\begin{figure}[h!t]\center
  \subfigure[]{
    \includegraphics[height=7cm]{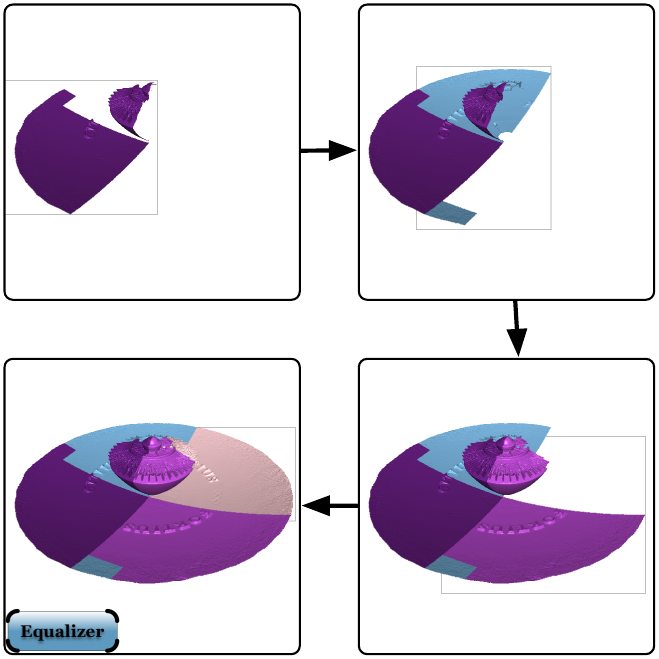}\label{fDBStream}
  }\hfil\subfigure[]{
    \includegraphics[height=7cm]{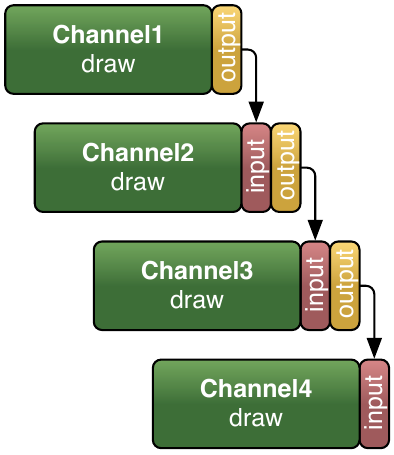}\label{fDBStreamFlow}
  }
  {\caption{Streaming Sort-Last Compound}}
\end{figure}

\fig{fDBStream} shows a streaming sort-last compound. The output of one source
channel is copied to the next channel in the chain, which then composites it on
top of its own rendering, streaming the combined frame on to the next source.
At the end of the chain, the destination channel completes the input frame by
compositing it with his rendering.

Equalizer only synchronises the input to the output frames, therefore this
configuration creates a pipelined configuration, where the compositing
operations form the ``critical path''. Each channel has its draw pass
delayed by the time taken by all preceding compositing operations, as shown
schematically in \fig{fDBStreamFlow}. This pipelining emerges naturally due to
the synchronisation points introduced by the compositing configuration.

The total system latency for sort-last stream compounds is $t_{draw} + (n -
1)\times (t_{readback} + t_{assemble})$. Note that the readback and assemble
times are usually an order of magnitude smaller than the render time, which
makes this compound attractive for small-to-medium sized decompositions, since
it has minimal compositing overhead and less synchronization compared to
parallel compositing algorithms.

\section{Region of Interest}

During scalable rendering pixel data has to be copied from the source channel
framebuffer to either the destination channel framebuffer, or to an
intermediate channel during parallel compositing. The associated distributed
image compositing cost is directly dependent on how much data has to be sent
over the network, which in turn is related to how much screen space is actively
covered. For sort-last rendering every node potentially renders into the entire
frame buffer, resulting in a linear increase in the amount of pixels composited
for an increasing number of nodes. Depending on the data set and
viewpoint, only a subset of the framebuffer shows pixels generated from the
data. With an increasing number of nodes, the set of affected pixels typically
decreases, leaving blank areas that can be omitted for transmission and
compositing.

Equalizer provides an API for the programmer to provide the region of interest
(ROI). The ROI is the screen-space 2D bounding box fully enclosing the data
rendered by a single resource, which can be easily computed by calculating the
screen-space projection of the model's bounding volume. We have extended the
core parallel rendering framework to use this application-provided ROI to
optimise the {\em load\_equalizer} and {\em tree\_equalizer}, as well as image compositing.

\fig{fDBStream} outlines the region of interest of each source. The compositing
code uses the ROI to minimise image readback size, and consequently network
transmission. The ROI is an output frame parameter, and is transmitted to all
input frames together with the pixel data. On the input frame, the compositing
code respects this parameter to place the pixel data in the right position.
Further, the ROI of the rendering pass is automatically merged with the ROI
of the composited frames for readback. The usage of ROI for load balancing is
described in \sref{sLoadEqualizer}.

Applying ROI for sort-first rendering provides a small improvement for the
rendering performance, as shown in \bench{rROI}(a) from \cite{EBAHMP:12}. As the
number of resources increases, the ROI becomes more important since the
relative amount of time spent in compositing increases with the rendering load
decreasing. With ROI enabled we observed performance improvements between
5-20\%, reaching 60~Hz when using 33 GPUs. Without ROI, the framerate peaked at
less than 50~Hz when using 27 GPUs.

\begin{benchmark}[h!t]\center
    \includegraphics[width=.48\textwidth]{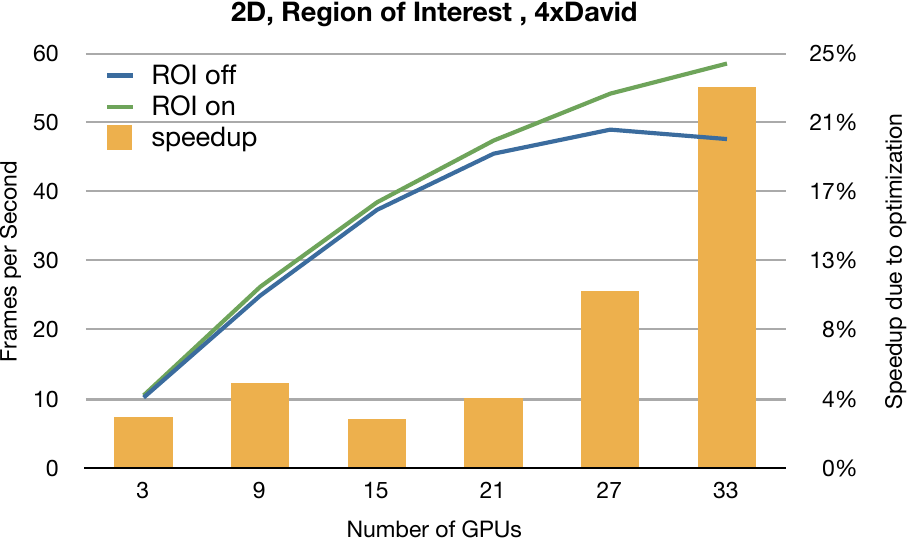}\hfil
    \includegraphics[width=.48\textwidth]{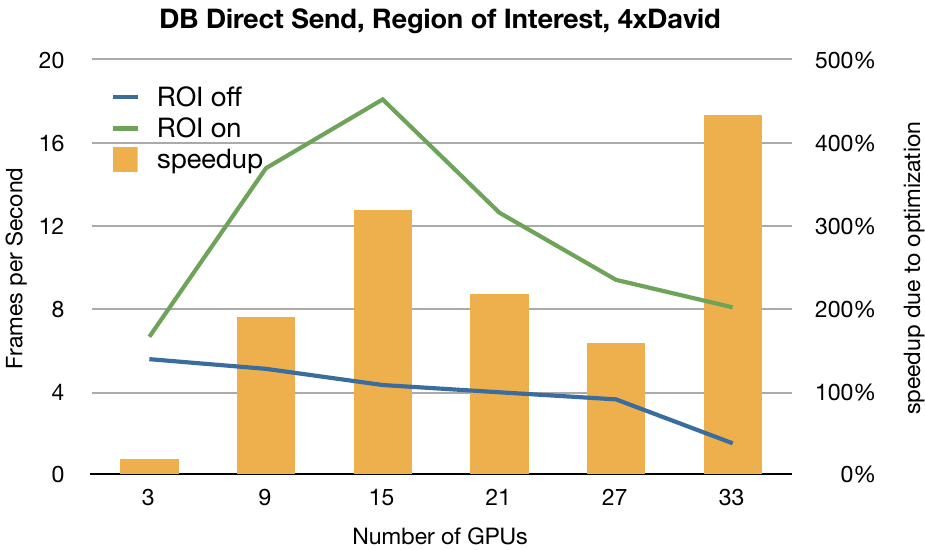}\\{\small(a)\hspace{.5\textwidth}(b)}\\
 {\caption{\label{rROI}Region of Interest for Sort-First and Sort-Last Rendering}}
\end{benchmark}

ROI is crucial for sort-last rendering performance. In our experiments in
\cite{EBAHMP:12}, we used a polygonal renderer creating relatively compact
regions during sort-last decomposition, while still using depth-sorted
compositing (cf. \fig{fDBStream}). This is a relatively common use case for
sort-last rendering. In this mode, we can observe significant speedups with ROI
(up to 4x), as shown in \bench{rROI}(b). In \cite{MEP:10} this
application-provided ROI was extended by an algorithm which automatically
computes the ROI by analysing the framebuffer. This algorithm has the advantage
of simplifying the application developers' life, and can also conveniently
detect holes in the rendered framebuffer.

\section{Asynchronous Compositing}

Asynchronous compositing pipelines pixel transfers with rendering operations by
moving them to a separate thread. Compositing in a distributed parallel
rendering system is decomposed into readback of the produced pixel data (1),
optional compression of this pixel data (2), transmission to the destination
node consisting of send (3) and receive (4), optional decompression (5) and
composition consisting of upload (6) and assembly (7) in the destination
framebuffer.

\begin{wrapfloat}{benchmark}{O}{.618\textwidth}
 \includegraphics[width=.618\textwidth]{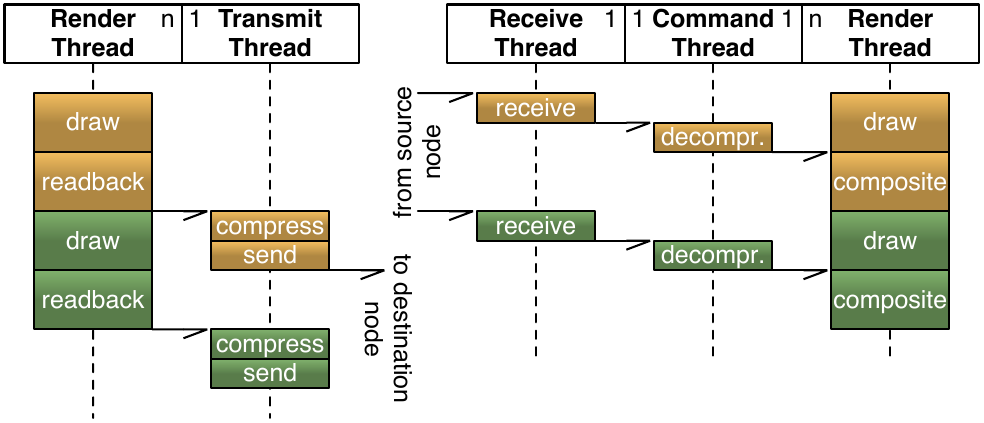}
 {\caption{\label{fSyncRB}Synchronous Readback and Upload}}
\end{wrapfloat}

\begin{figure}[t]\center
  \includegraphics[scale=0.6]{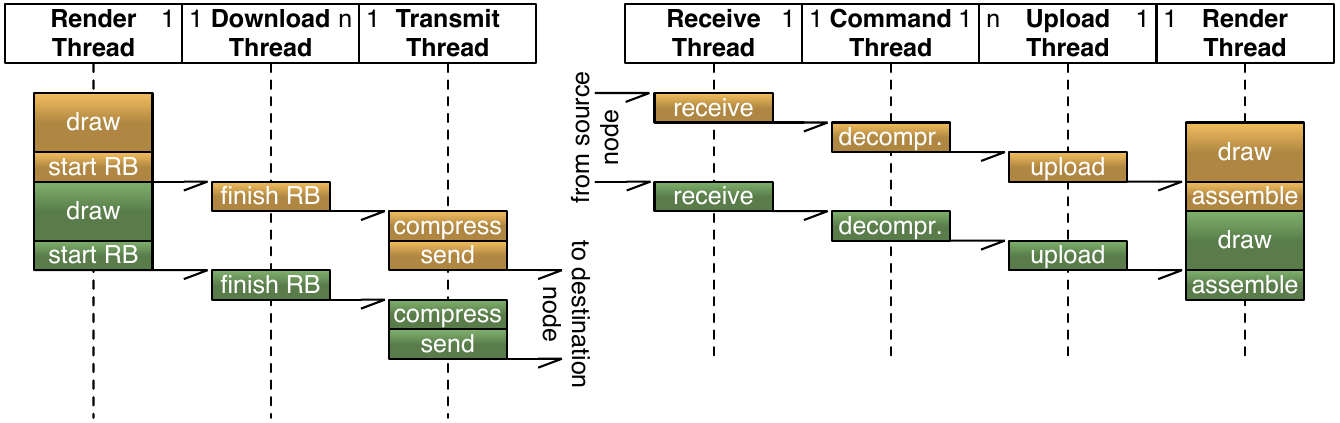}
  \caption{Asynchronous Readback and Upload\label{fAsyncRB}}
\end{figure}

In a naive implementation operations 1 to 3 are executed serially on one
core, 4 to 7 on another. In our parallel rendering system, operations 2
to 5 are executed asynchronously to the rendering operations 1, 6 and 7.
Furthermore, we use a latency of one frame which means that two rendering frames
are always in execution, allowing the pipelining of these operations, as shown
in \fig{fSyncRB}. We have implemented asynchronous readback using OpenGL pixel
buffer objects, further increasing the parallelism by pipelining the rendering
and pixel transfers, as shown in \fig{fAsyncRB}.

In the asynchronous case, the rendering thread performs only
application-specific rendering operations, since the overhead of starting an
asynchronous readback becomes negligible. Equalizer uses a plugin system to
implement GPU-CPU transfer modules that are runtime loadable. We extended this
plugin API to allow the creation of asynchronous transfer plugins, and
implemented such a plugin using OpenGL pixel buffer objects (PBO). At runtime,
one rendering thread and one download thread are used for each GPU, as well as
one transmit thread per process. The download threads are created lazy when
needed.

\begin{wrapfloat}{benchmark}{O}{.618\textwidth}
 \includegraphics[width=.618\textwidth]{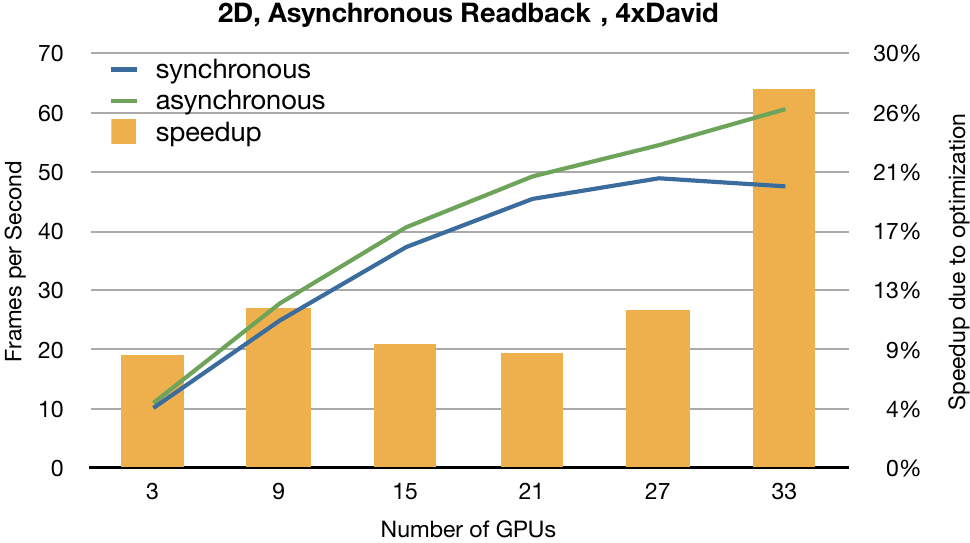}
 {\caption{\label{rAsync}Asynchronous Compositing Sort-First Rendering}}
\end{wrapfloat}

Asynchronous compositing is, together with region of interest, one of the most
influential optimisations for scaling rendering to large cluster sizes. For
sort-first rendering, shown in \bench{rAsync} from \cite{EBAHMP:12}, pipelining
the readback with the rendering yields a performance gain of about 10\%. At
higher frame rates, when the rendering time of a single resource decreases,
asynchronous readback has an even higher impact of over 25\%.

\section{Compression for Image Compositing}

The image compositing stages in distributed rendering are fundamentally limited
by the GPU-to-node and node-to-node image data throughput. Efficient
image coding, compression and transmission must be considered to minimise that
bottleneck.

Basic run-length encoding (RLE) has been used as a fast algorithm to improve
network throughput for interactive image transmission. However, it only gives
sufficient results in specific rendering contexts and fails to provide a
general improvement as shown in \cite{MEP:10}. RLE only works in compacting large
empty or uniform colour areas, but is often useless for non-trivial full frame
colour results. We developed two enhancements to improve RLE: per-component RLE
compression and reordering of colour bits. These preconditioning steps exploit
typical characteristics of image data for run-length encoding.

Equalizer also integrates more complex compression algorithms such as {\em
lib-jpeg-turbo}, which are of little practical use on modern cluster
interconnects. Their compression overhead is often too high to be amortised by
the decreased network transmission time on 10 GBit/s or faster interconnects.
For remote image streaming, as discussed in \sref{sTIDE}, they remain a
viable compression algorithm.

Based on our work, \cite{MEP:10} implemented GPU-based YUV subsampling before
the image download, which has negligible overhead, reasonable compression
artefacts, and a good compression ratio.

\subsection{Enhanced RLE Compression}

Run-length encoding (RLE) is a simple compression scheme and is on modern
architectures purely constrained by the available memory bandwidth. For image
compression in visualisation applications, we can exploit some characteristics
of the data to improve the compression ratio over the standard RLE compression.

Our basic RLE implementation is a fast 64-bit version comparing two pixels
at the same time (8 bit per channel RGBA format). This choice is motivated by
the fact that modern processors have 64 bit registers, thus using 64 bit tokens
optimises throughput. While this method is very fast, it shows poor compression
results in most practical settings since it can only compress adjacent pixels
of the same colour. We have observed a compression rate of up to 10\% in
practical scenarios.

\begin{wrapfloat}{figure}{O}{.618\textwidth}
  \includegraphics[width=.618\textwidth]{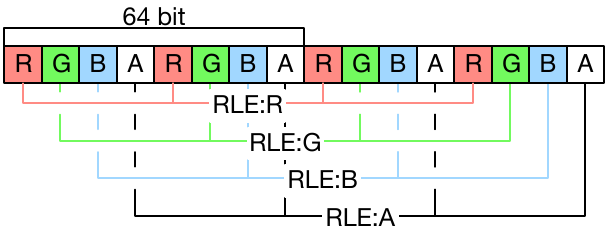}
  \caption{64 bit and Per-Component RLE Compression}
  \label{fRLE}
\end{wrapfloat}

The first improvement is to treat each colour component separately by producing
four independent RLE-compressed output streams as illustrated in \fig{fRLE}.
This per-component RLE improves the compression rate from 10\% to about 25\%,
as individual colour components change less often than full pixels.

\begin{figure}[h!t]
  \centering
  \includegraphics[width=\textwidth]{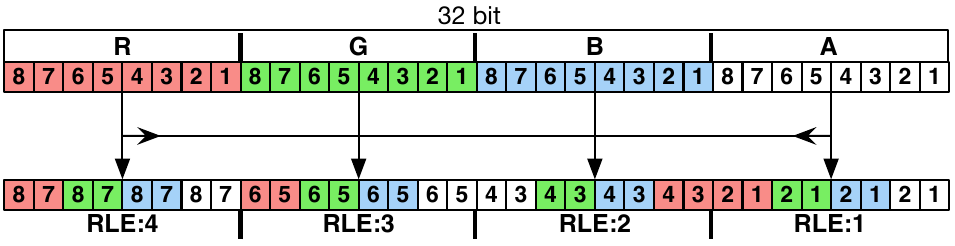}
  \caption{Swizzling Preconditioning of 32-bit RGBA Values}
  \label{fSwizzle}
\end{figure}

The second improvement is {\em bit-swizzling} of colour values before
per-component compression. This swizzling step is a data pre-conditioner, which
reorders and interleaves the per-component bits as shown in \fig{fSwizzle} by
grouping them by significance. Now the per-component
RLE compression separately compresses the higher, medium and lower order bits
in separate streams, thus achieving stronger compression for smoothly changing
colour values, since high-order bits change less often.

Swizzling improves the compression rate to up to 40\% for the same scenario as
above. The preconditioning step only requires bit shift and mask
operations, is entirely executed in registers and has no measurable
impact on performance, since the whole algorithm is memory bound on modern CPUs.

All RLE compressors perform a data decomposition on the input image, and
parallelise the compression of the resulting sub-images across multiple
threads. This parallel execution improves the performance by saturating
multiple memory channels compared to a single-threaded implementation.

\begin{wrapfloat}{benchmark}{O}{.618\textwidth}
  \includegraphics[width=.618\textwidth]{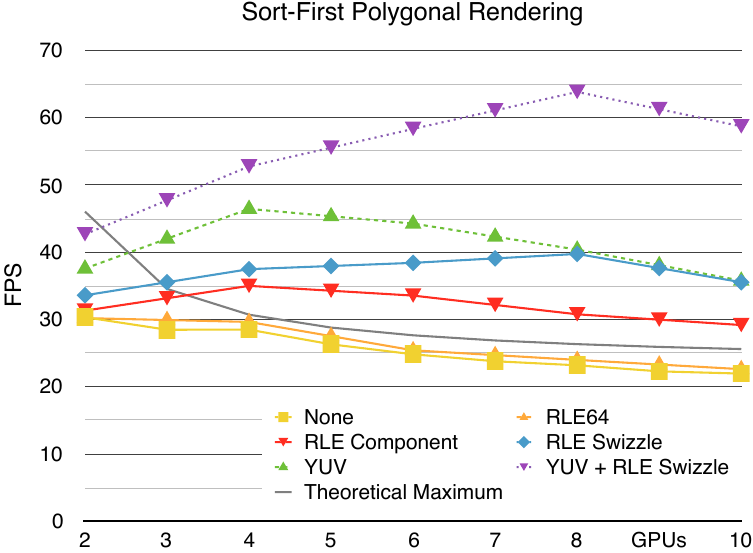}
  \caption{Image Compression in Sort-First Polygonal Rendering}
  \label{rRLE}
\end{wrapfloat}

\bench{rRLE} summarises the compression results from \cite{MEP:10}. We have
chosen sort-first rendering, since this highlights the results for the RLE
compressor which is optimised for colour image data. This benchmark did run on a
visualisation cluster with Gigabit Ethernet at a resolution of $1280\times
1024$ pixels. It rendered the David statuette at 1mm resolution, resulting
in a rendering time of about 28 ms on a single GPU. The theoretical maximum
line shows the upper limit for sort-first compositing with uncompressed image
data and no rendering time. It decreases as the destination channel
contributes to the sort-first rendering and does not require a pixel transfer.
With an increasing number of remote source channels its size decreases,
requiring more pixels to be transferred.

The graph shows how various incremental improvements add up to significant
performance gains. Even in a relatively difficult scenario with a fast
rendering time, and, by modern standards, slow network interconnect, we were
able to more than double the performance to above 60~Hz.

The basic 8-byte RLE Compressor performs just minimally better than no
compression. Both stay relatively close the theoretical maximum, but can't
quite reach it due to load imbalances and non-zero rendering time. The
swizzling pre-conditioner can significantly reduce the compositing time, and
even improve the overall framerate.

The YUV is compressor is an on-GPU compression plugin which performs a
color space conversion and lossy chroma subsampling. It can be combined with the
CPU-based RLE compressor, which then interleaves and compresses the Y, U and V
channels, resulting in major performance improvements. Both compression steps
have virtually no computational overhead and are memory bandwidth bound. Since
the YUV compressor runs on the GPU, it reduces the costly GPU to CPU transfer
time over the PCI Express link.

\subsection{GPU Transfer and CPU Compression Plugins}

Equalizer uses runtime-loadable plugins to transfer pixel data from and to the
GPU, as well as plugins to compress and decompress pixel data for network
compression. This separation allows different code paths for multi-GPU machines
where no CPU-based compression is used, and for distributed execution where data
is compressed before network transfer.

The GPU transfer might also apply compression. This is typically done on the
GPU to reduce the amount of memory transferred over the GPU-CPU interconnect.
One example is YUV subsampling, where a shader implements the RGB to YUV colour
space conversion and subsequent chroma subsampling. Furthermore, a GPU transfer
plugin may implement asynchronous downloads, where the download is started
from the render thread an finished in a separate download thread as shown in
\fig{fAsyncRB}. CPU compression plugins are always executed from asynchronous
threads concurrently to the rendering threads.

The implementation of these steps in plugins provides a clean separation and
interface for users and researchers interested in experimenting with image
compression for interactive parallel rendering.

\section{Out-of-Order Assembly}\label{sEarlyAss}

Early assembly provides better pipelining when the frame assembly order is not
important, for example for sort-first rendering and sort-last rendering with
$z$-compositing. Our default compositing code uses a signal on all input
frames, which is triggered for each input frame arrival. The compositing code
then picks and composites this image, assembling images early and out of order as
they become available. This decreases the time to solution, since the assemble
operation finishes once the last frame arrives, plus the time to assemble this
last frame. In-order assembly would have a statistical probability of
$\frac{n}{2}$ frames to assemble after the last frame arrives (unless other
constraints make the arrival not fully random or network-constrained).

\chapter{Load Balancing}\label{sLoadBalancing}

\section{Overview}

\begin{figure}[h!t]
  \includegraphics[width=\textwidth]{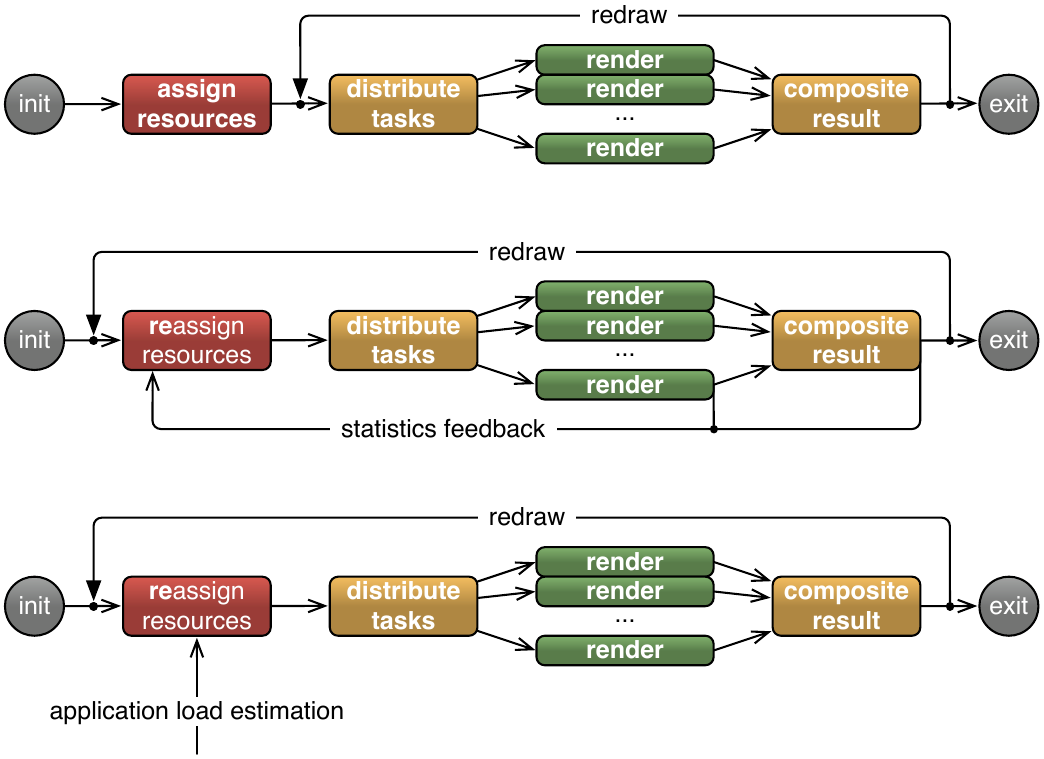}
  \caption{\label{floadbalancing}Static, Reactive and Predictive Load Balancing}
\end{figure}

Load balancing performs resource assignment per source channel based on
workload, with the goal of equalising resource utilisation. Static load
balancing, shown in \fig{floadbalancing} top, performs this assignment once
during initialisation. Dynamic load balancing can either be reactive or
predictive (middle and bottom of \fig{floadbalancing}). Reactive load balancing
utilises statistics from previous frames to estimate future load distribution.
Predictive load balancing uses an application-provided load estimate (also
called cost function) to predict the load distribution for the current frame.
Both approaches reassign resources dynamically, typically for each rendered
frame. Implicitly load balanced algorithms achieve a good load balance by other
means, for example by work stealing between resources.

In our framework load balancing is implemented by {\em Equalizers}, which are
an addition to compound trees. They modify parameters of their respective
compound subtree at runtime to dynamically optimise the resource usage, by
tuning one aspect of the decomposition or recomposition. Due to their nature,
they are transparent to application developers, but might have
application-accessible parameters to tune their behaviour. Resource
equalisation is the critical component for scalable rendering, and therefore
the eponym for the {\em Equalizer} project name.

In this section we present various {\em equalizer} implementations: two
variants for reactive load balancing for sort-first and sort-last rendering,
implicitly load balanced work packages for sort-last and sort-first rendering,
cross-segment load balancing for multi-display installations, constant frame
rate rendering using dynamic frame resolution, and monitoring of large-scale
visualisation systems. These equalizers address the research question on how we
can improve load-balancing for sort-first rendering, in particular for large
display systems.

\section{Sort-First and Sort-Last Load Balancing}\label{sLoadEqualizer}

\begin{wrapfloat}{figure}{O}{.382\textwidth}
  \includegraphics[width=.382\textwidth]{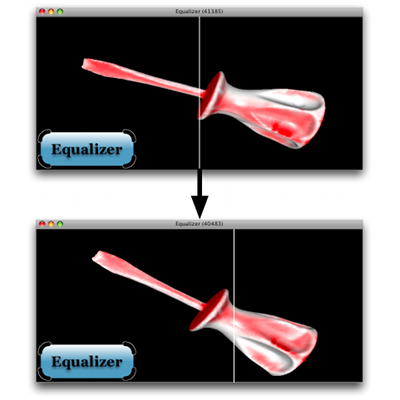}
  \caption{\label{floadeq}Load Balancing}
\end{wrapfloat}

Sort-first (\fig{floadeq}) and sort-last load balancing are the most obvious
optimisations for these parallel rendering modes. Our load equalizers are fully
transparent for application developers; they use a reactive approach
based on past rendering times. This requires a good frame-to-frame
coherence for optimal results, which is the case with most rendering applications.
Equalizer implements two different algorithms: A {\em load\_equalizer} and a
{\em tree\_equalizer}, which have shown advantageous
for different types of rendering load.

Both equalizers extract their load metrics from statistics collected by the
rendering clients, which are sent asynchronously from the clients to the
server, where the equalizers subscribe to them for operation. At the beginning
of each frame, the server triggers all equalizers on all compound trees, which
enables them to set new decomposition parameters before the rendering tasks are
computed.

The {\em load\_equalizer} builds a model of the rendering load in screen space
or data space. It stores a 2D (for sort-first) or 1D (for sort-last) grid of
the load, mapping the load of each channel. The load is stored in normalised
2D/1D coordinates using $\frac{time}{area}$ as its measure. The contributing
source channels are organised in a binary kD-tree.  The algorithm then
balances the two branches of each level by equalising the integral over the
cost area map on each side. This algorithm is similar to \cite{ACCC:04}, which
uses a dual-level tree. Our binary split tree provides more compact tiles for
larger cluster configurations, since the split direction alternates on each
level.

\begin{wrapfloat}{figure}{O}{.618\textwidth}
  \includegraphics[width=.618\textwidth]{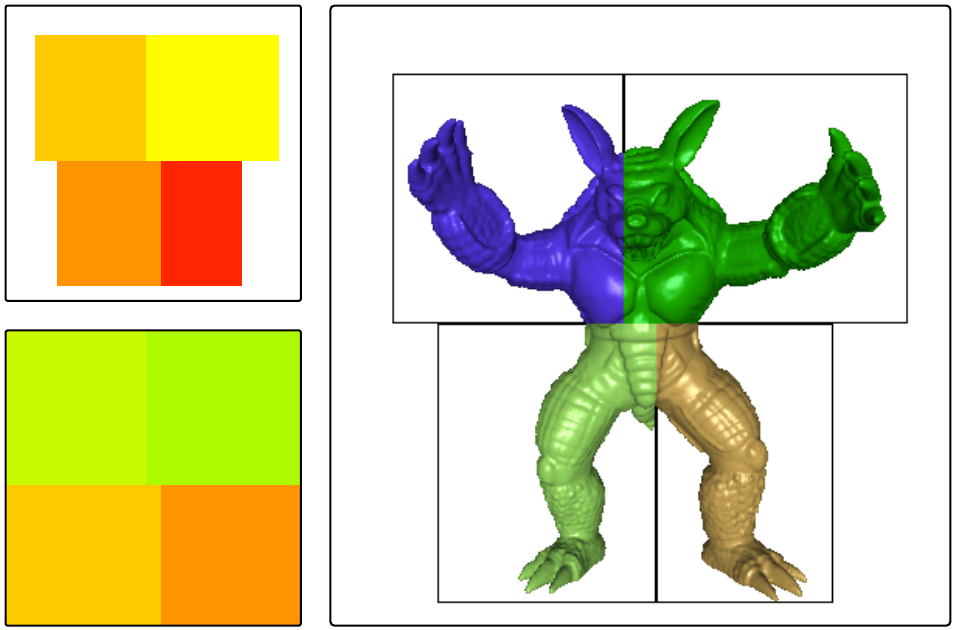}
  \caption{Load Cost Area Map with (top) and without (bottom) using Region of Interest Information}
  \label{fROI}
\end{wrapfloat}

The load balancer has to operate on the assumption that the load is uniform
within one load grid tile. Naturally this leads to estimation errors, since in
reality the load is not uniformly distributed, it tends to increase towards the
centre of the screen in \fig{fROI}. We reuse the Region of Interest (ROI) from
compositing of each source channel to automatically refine the load grid as
shown in \fig{fROI}, top left. In cases where the rendered data projects only
to a limited screen area, this ROI refinement provides the load balancer with a
more accurate load estimation, leading to a better load prediction during the
balancing step.

The {\em tree\_equalizer} uses the same binary kD-tree structure as the
{\em load\_equalizer} for recursive load balancing. It computes the
accumulated render time of all children for each node of the tree and uses
the result to allocate an equal render time to each subtree. It makes no assumption
of the load distribution in 2D or 1D space, it only tries to correct the
imbalance in render time.

Both equalizers implement tunable parameters allowing application developers
to optimise the load balancing based on the characteristics of their rendering
algorithm. These parameters are accessible through an API from the application
main thread:

\begin{compactdesc}
\item[Split Mode] configures the tile layout: horizontal stripes, vertical stripes,
or 2D, a binary tree split alternating the split axis on each level, resulting
in compact 2D tiles.
\item[Damping] reduces frame-to-frame oscillations. The equal load distribution
within the region of interest assumed by the load balancers is in reality not
equal, causing the load balancing to overshoot. Damping is a normalised scalar
defining how much of the computed delta from the previous position is applied to
the new split.

\item[Resistance] eliminates small deltas in the load balancing step, i.e., it
only changes the viewport or range if the change is over the configured limit.
This can help the application to cache visibility computations, since the
frustum does not change with each frame.

\item[Boundaries] define the modulo factor in pixels onto which a load split may
fall. Some rendering algorithms produce artefacts related to the OpenGL raster
position, e.g., screen door transparency. It can be eliminated by aligning
the boundary to the pixel repetition. Furthermore, some rendering algorithms are
sensitive to cache alignments, which can again be exploited by choosing the
corresponding boundary.

\item[Usage] is a per-child normalised resource utilisation coefficient. The
equalizer will assign proportional work to this resource, and deactivate it if
the usage is 0. This parameter is primarily used by the cross-segment load
balancer to reassign resources between destination channels. It can also be
used to configure heterogeneous GPU resources more efficiently.

\end{compactdesc}

\subsection{Dynamic Work Packages}

Load balancing can be classified into explicit and implicit approaches.
Explicit methods centrally compute a task decomposition up-front, before a new
frame is rendered, while implicit methods decompose the workload into task
units that can be dynamically assigned to the resources during rendering, based
on the work progress of the individual resources. Explicit load balancing
typically assigns a single task to each resource to minimise static per-task
cost. The aforementioned load and tree equalizers implement explicit reactive load
balancing.

Implicit load balancing uses a finer granularity of significantly more rendering tasks
to resources. These tasks are assigned using central task distribution, or
 task stealing between resources. Implicit algorithms are more
commonly used in offline raytracing compared to real-time rasterisation,
 because of practically non-existent per-tile cost in raytracing.
Since each rendered task directly sends its result for
compositing, work packages exhibit a better pipelining of rendering and
compositing operations.

Our work package implementation uses a task pulling mechanism, an approach that
has been employed before in distributed computing. Rather than having the
server push tasks to the rendering clients, our dynamic work packages approach
manages fine grained tasks on a server-side queue, while the clients
request and execute the tasks as they become available. Every rendering client
employs a small local, prefetched queue of work packages to hide the round-trip
latency to fetch new packages. During rendering, a client first works on
packages from its local queue and concurrently requests packages from the
server whenever the amount of available packages sinks below a threshold.
In~\cite{SPEP:16} this basic, random task assignment was extended with
client-affinity models.

At the beginning of each frame, the server generates tiles or database ranges
of a configurable size for each compound with an output queue. Compounds with
an input queue matching the name of the output queue generate a special draw
task, which causes the render client to set up its input queue, and to call
{\em frameDraw} and {\em frameReadback} for each received work package.

\subsection{Benchmarks}

\bench{rCompounds} shows the performance of static and dynamic sort-first
and sort-last rendering. The experiment setup is described in \cite{ESP:18}.
The results show that, as expected, load balancing improves the performance
over a static task decomposition significantly. The simpler
{\em tree\_equalizer} outperforms the load grid-driven
{\em load\_equalizer} in most cases, except for sort-first volume rendering,
where the load in the region of interest is relatively uniform. This result is
counter-intuitive, since the {\em load\_equalizer} operates with more
information and should be able to produce better results. It seems to confirm
that simple algorithms often outperform theoretically better, but more complex
implementations. The decoupling of the load balancing algorithm from the rest
of the system enabled by the compound architecture opens the possibility for
more research in this area.

\begin{benchmark}[h!t]
  \includegraphics[width=\textwidth]{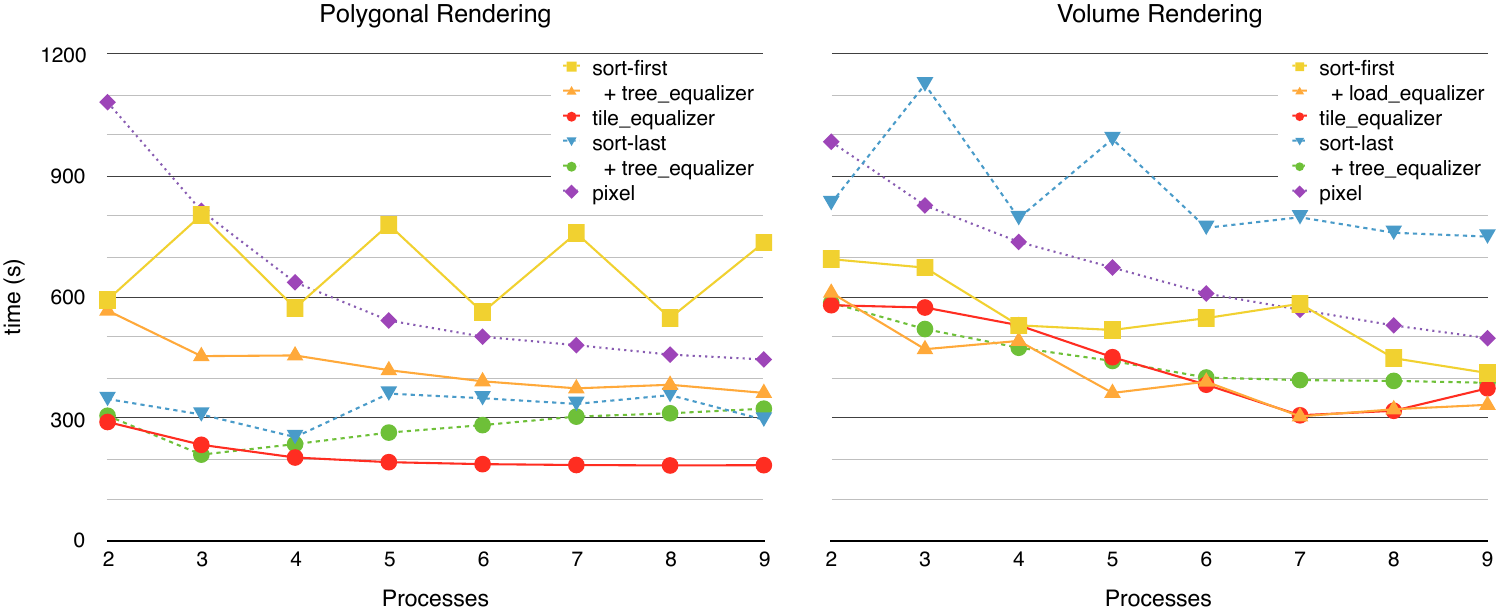}
  \caption{\label{rCompounds}Sort-First and Sort-Last Scalability}
\end{benchmark}

\bench{rEqualizers} provides a more detailed analysis of the different
equalizers. Using volume rendering, we measure the performance of decomposition
modes under heterogeneous load, which was achieved by varying the
number of volume samples used for each fragment (1-7) while rendering. This
allowed for a consistent linear scaling of rendering load, which was randomly
varied either per frame, or per node. The linear scaling of load per node
corresponds to a scaling of resources. Doubling the rendering load on a
specific node corresponds to halving its available rendering resources. To the
system this node would then contribute the value 0.5 in terms of normalised
compute resources.

\begin{benchmark}[h!t]
  \includegraphics[width=\textwidth]{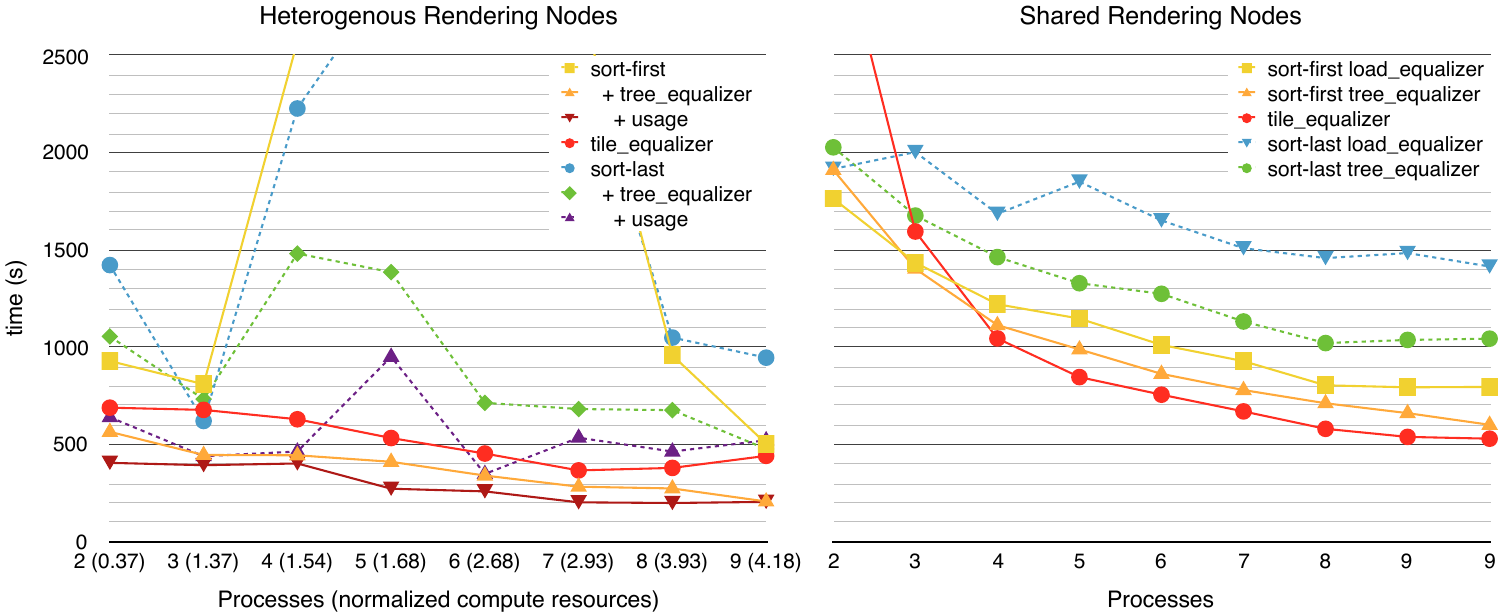}
  \caption{\label{rEqualizers}Sort-First and Sort-Last Equalizer Behaviour}
\end{benchmark}

\bench{rEqualizers} (left) models how individual modes perform on heterogeneous
systems. In this case the tree equalizer performs best, as it allows us to a
priori define how much usage it should make of individual nodes, i.e., bias the
allocation of rendering time in accordance with the (simulated) compute
resources. \bench{rEqualizers} (right), on the other hand, illustrates how the
decomposition modes perform on a system where compute resources fluctuate
randomly every frame, as can be the case for shared rendering nodes in
virtualised environments. For this scenario tile equalizer seems best suited,
as it load balances implicitly and does not assume coherence of
available resources between frames. The simpler tree equalizer outperforms the
load equalizer in this experiment.

The {\em tile\_equalizer} often outperforms the {\em tree\_equalizer}. This
suggests that the underlying implicit load balancing of task queues can be
superior to the explicit methods of {\em load\_equalizer} and
{\em tree\_equalizer} in high load situations, where the additional overhead
of tile generation and distribution is more justified. The relatively simple
nature of our rendering algorithms is also favouring
work packages, since they have a near-zero static overhead per rendering pass.
\cite{SPEP:16} contains additional experiments.

\section{Cross-Segment Load Balancing}

\begin{figure}[h!t]
  \includegraphics[width=\textwidth]{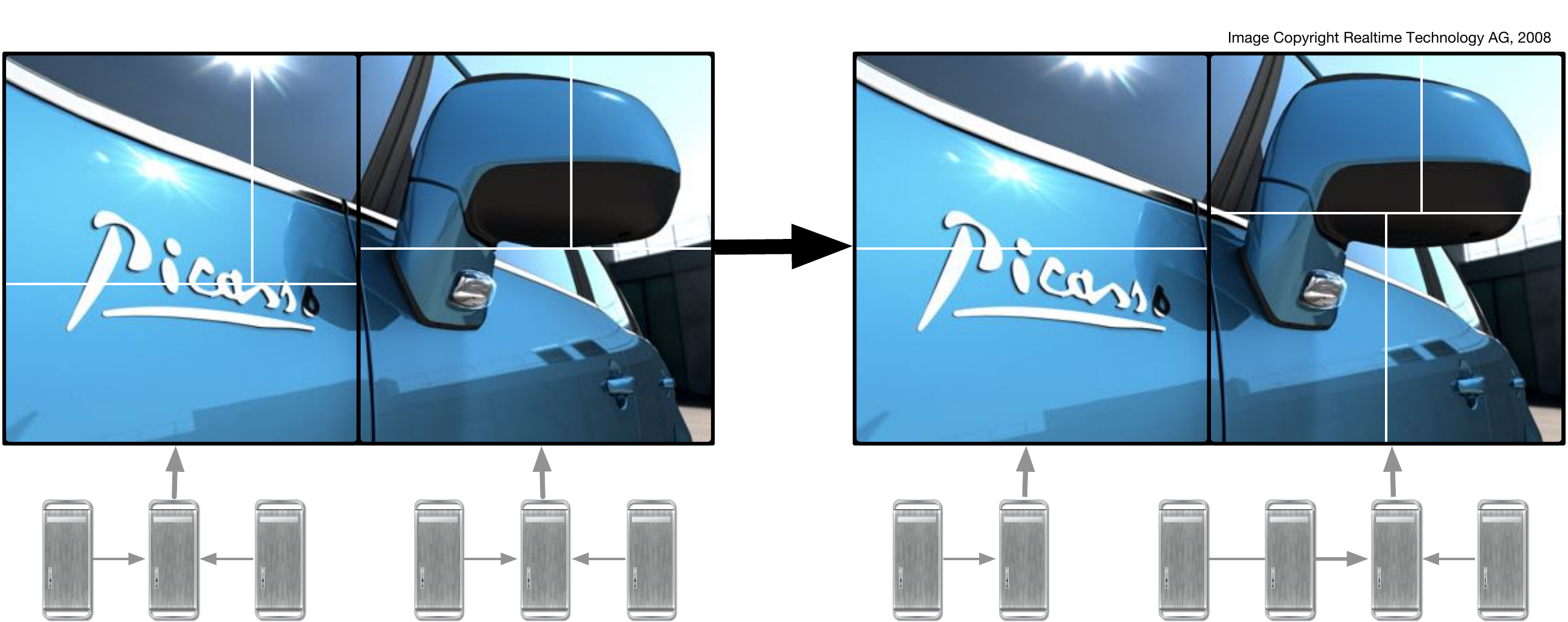}
  \caption{\label{fvieweq}Six GPU to two Display Cross-Segment Load Balancing}
\end{figure}

A serious challenge for all distributed rendering systems driving large
multi-display installations is dealing with the varying rendering load per
display, and therefore the graphics load on its driving GPUs. Cross-segment load
balancing (CSLB) is a novel dynamic load balancing approach to dynamically
allocate $n$ rendering resources to $m$ output channels (with $n\geq m$), as
shown in \fig{fvieweq}.

The $m$ output channels each drive a display or projector of a multi-display
system. Commonly, each destination channel is solely responsible for rendering
and potentially compositing of its corresponding display segment.

A key element of CSLB is that the $m$ GPUs physically driving the $m$ display
segments are not restricted to a one-to-one mapping of rendering tasks to the
corresponding display segment. CSLB performs a dynamic assignment of $n$
graphics resources from a pool to drive $m$ different destination display
segments, where the $m$ destination channel GPUs themselves may also be part of
the pool of graphics resources. CSLB also does not require a planar display
surface, that is, the algorithm works equally for tiled display walls and
immersive installations.

Dynamic resource assignment is performed through load balancing components that
exploit statistical data from previous frames for the decision of optimal GPU
usage for each segment, as well as optimal distribution of work among them.
The algorithm is also compatible with predictive load balancing based on a load
estimation given by the application.

CSLB is implemented in Equalizer as two layers of hierarchically organised
components specified in the configuration. The first level globally assigns
fractions of resources to each destination channel, and the second level
consist of {\em load\_equalizer}s or {\em tree\_equalizer}s balancing the
assigned resources for each destination segment.

\begin{figure}[h!t] \center
  \subfigure[CSLB resources setup.]{\includegraphics[width=0.47\textwidth]{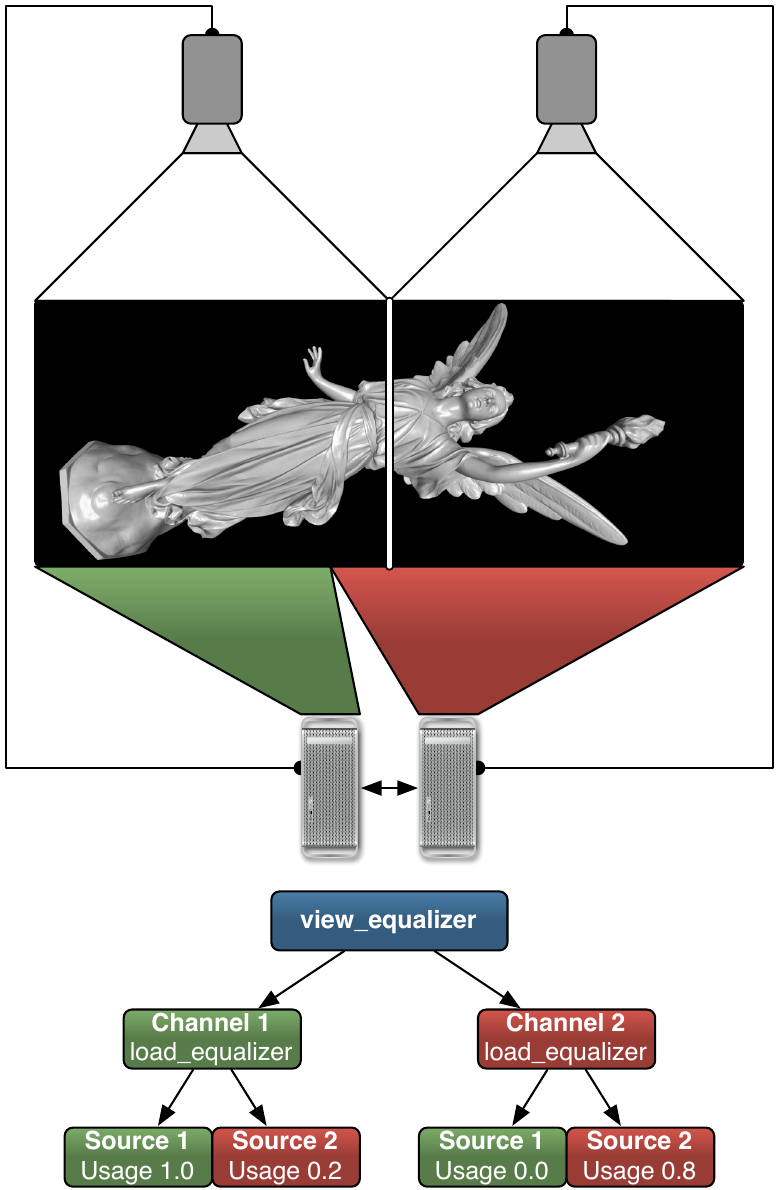}
  \label{fViewEqualizerSetup}}
  \hfill
  \subfigure[CSLB configuration file format.]{
	\begin{minipage}[b]{0.46\textwidth}
	\setbox0=\vbox to 290pt{\tt\scriptsize
compound \\
\{ \\
\quad view\_equalizer \{\} \\
\quad compound \\
\quad \{ \\
\quad \quad channel "Channel1" \\
\quad \quad load\_equalizer\{\} \\
\quad \quad compound \{\} \# self \\
\quad \quad compound \\
\quad \quad \{ \\
\quad \quad \quad channel "Source2" \\
\quad \quad \quad outputframe \{\} \\
\quad \quad \} \\
\quad \quad inputframe\{\} \\
\quad \quad ... \\
\quad \} \\
\quad compound \\
\quad \{ \\
\quad \quad channel "Channel2" \\
\quad \quad load\_equalizer\{\} \\
\quad \quad compound \{\} \# self \\
\quad \quad compound \\
\quad \quad \{  \\
\quad \quad \quad channel "Source1"  \\
\quad \quad \quad outputframe \{\}  \\
\quad \quad \} \\
\quad \quad inputframe\{\} \\
\quad \} \\
\quad ... \\
\} \\
	\label{fViewEqualizerConfig}}
	\fbox{\box0}
	\end{minipage}
   }
\caption{Dual-GPU, Dual-Display Cross-Segment Load Balancing Setup}
\label{fViewEqualizer}
\end{figure}

\fig{fViewEqualizer} depicts a snapshot of the simplest CSLB setup along with
its configuration file. Two destination channels, {\em Channel1} and {\em
Channel2}, each connected to a projector, create the final output for a
multi-projector view. Each projector is driven by a distinct GPU, constituting
the source channels {\em Source1} and {\em Source2}. Each source channel GPU
can contribute to the rendering of the other destination channel segment. For
each destination channel, a set of potential resources are allocated. A
top-level {\em view\_equalizer} assigns the usage to each resource, based on
which per-segment {\em load\_equalizer}s compute the 2D split to balance the
assigned resources within the display. The left segment of the display has a
higher workload, so, both {\em Source1} and {\em Source2} are used to render
for {\em Channel1}, whereas {\em Channel2} uses only {\em Source2} to render
the image for the right segment. The schematic also shows the current usage of
the four potential source compounds, where only three have an active draw pass
at this point in time.

CSLB uses a two-stage approach, where a {\em view\_equalizer} at the top level
of the compound hierarchy handles the resource assignment. Each child of this
root compound has one destination (segment) channel, corresponding to one of
the $m$ display segments, using a {\em load\_equalizer} or {\em
tree\_equalizer}. The view equalizer supervises the different destination
channels of a multi-display setup; the load equalizers on the other hand are
responsible for the partitioning of the rendering task of each segment among
its child compounds. They use the precomputed usage of each child to allocate a
corresponding amount of work for the child. Therefore, each destination channel
of a display segment has its source channel leaf nodes sharing the actual
rendering load. One physical GPU assigned to a source channel can be referenced
in multiple leaf nodes, and thus contribute to different displays.

For performance reasons the view equalizer assigns each resource to at most two
rendering tasks, e.g., to update itself and to contribute to another display.
Furthermore, it gives priority to the source compound using the same channel as
the output channel of each segment to minimise pixel transfers.

Cross-segment load balancing allows for optimal resource usage of multiple
GPUs used for driving the display segments themselves, as well as any additional
source GPUs for rendering. It combines multi-display parallel rendering with
scalable rendering for optimal performance.

\begin{wrapfloat}{benchmark}{O}{.618\textwidth}
  \includegraphics[width=.618\textwidth]{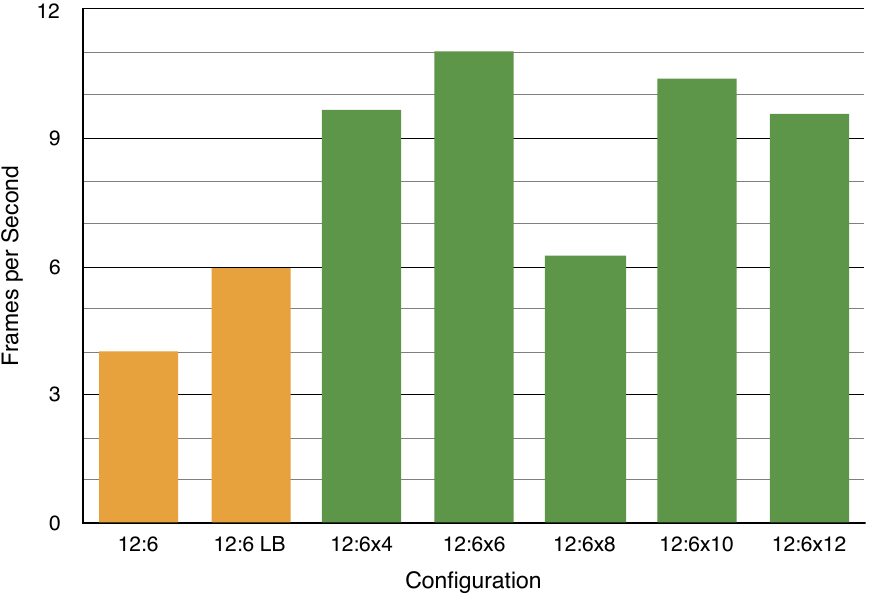}
  \caption{\label{rCSLBFPS}Cross-Segment Load Balancing}
\end{wrapfloat}

\cite{EEP:11} provides experimental results for a six-monitor tiled display wall,
driven by twelve GPUs. \bench{rCSLBFPS} shows an overview for the achievable
performance improvements. The first two configurations use a static assignment
of two GPUs to one output channel, where the first one statically assigns one
half to each GPU, and the second uses a load balancer to dynamically split the
work between the two GPUs. Already this 2D load balancing improves the
framerate by almost 50\%. The remaining configurations add a view equalizer on
top of the per-segment 2D load equalizer. The configuration assigns up to 4, 6,
8, 10 or all GPUs to each segment, that is, any segment may use up to $n$ GPUs,
and the GPUs are shared evenly across multiple segments. While theoretically
the all-to-all ($12:6\times 12$) configuration should provide the best
performance, mispredictions of the equalizers lead to a sweet spot of GPU
sharing between segments. In our $12:6$ setup, assigning up to six GPUs per
segment almost doubles the performance over the state-of-the-art sort-first
load balanced setup.

\bench{rCSLB} shows the rendering time over a fixed camera path of 540 frames.
In the static case two GPUs are responsible for each of the six outputs of the
tiled display wall used. For the CSLB graph, up to eight GPUs were dynamically
reassigned each frame to each of the six output channels, depending on the
current load distribution. Except for a few camera positions, where the model is
positioned evenly over all outputs, CSLB outperforms the fixed assignment.

\begin{benchmark}[h!t]
  \centering
  \includegraphics[width=\textwidth]{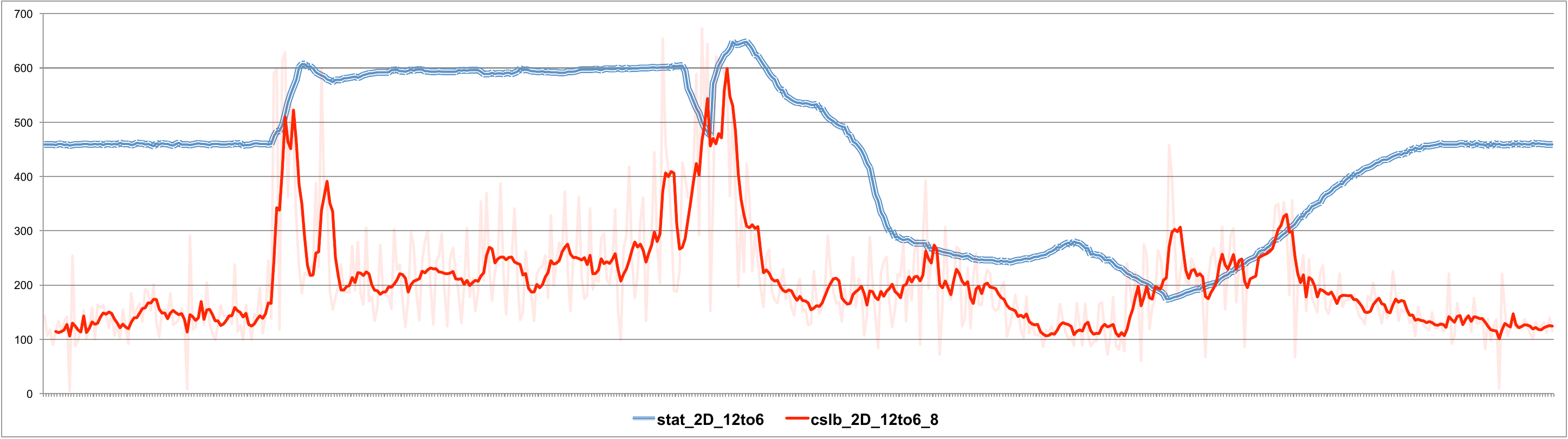}
  \caption{Cross-Segment Load Balancing for six Displays and 12 GPUs compared to a static two-to-one six Display Sort-First Rendering}
  \label{rCSLB}
\end{benchmark}

A strength of this algorithm lies in its flexibility. On one hand, it can
perform dynamic resource assignment not only for a planar display system, as
some approaches which built a single virtual framebuffer, but also for curved
displays and CAVE installations. On the other hand, it allows a flexible
assignment of potential contributing GPUs to each output channel individually.
Each output may have a different, potentially overlapping, set of GPUs which
may contribute to its rendering.

\section{Dynamic Frame Resolution}

Dynamic Frame Resolution (DFR) (\fig{fdfr}) provides a functionality similar to
dynamic video resizing \cite{MBDM:97}, specifically it maintains a constant
framerate by adapting the rendering resolution of a fill-limited application.

\begin{wrapfloat}{figure}{O}{.618\textwidth}
  \includegraphics[width=.618\textwidth]{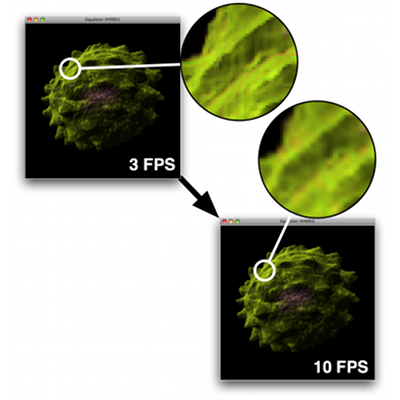}
  \caption{\label{fdfr}Dynamic Frame Resolution}
\end{wrapfloat}

While the aforementioned uses a now-obsolete hardware implementation, our
implementation works on commodity hardware and is purely implemented in
software.

DFR works by rendering into a source channel (often on a FBO) separate from
the destination channel, and then scaling the rendering during the transfer
(typically through an on-GPU texture) to the destination channel. The DFR
equalizer monitors the rendering performance and accordingly adapts the
resolution of the source channel and the zoom factor for the source to
destination transfer. If the performance and source channel resolutions allow,
this will not only subsample, but also supersample the destination channel to
reduce aliasing artefacts.

DFR can be combined with other scalability features, e.g., sort-first
rendering. It is also notable that it does not need any additional code in the
core compound logic, it simply exploits existing functionality such as
texture-based compositing frames and frame zoom with dynamic per-frame
adjustments.

\section{Frame Rate Equalizer}\label{sFramerateEq}

The framerate equalizer smooths the output frame rate of a destination channel
by instructing the corresponding window to delay its buffer swap to a minimum
time between swaps. This is regularly used for time-multiplexed decompositions,
where source channels tend to drift and finish their rendering unevenly
distributed over time. This equalizer is however fully independent of DPlex
compounds, and may be used to smooth the framerate of irregular rendering
algorithms. Due to the artificial sleep time before swap, it may incur a small
performance penalty, but it greatly improves the perceived rendering quality for
users in DPlex compounds.

\section{Monitoring}

The monitor equalizer (\fig{fmonitor}) allows reusing of the rendering from one
or more channels on another channel, typically for monitoring a larger display
setup on a control workstation.

\begin{wrapfloat}{figure}{O}{.618\textwidth}
  \includegraphics[width=.618\textwidth]{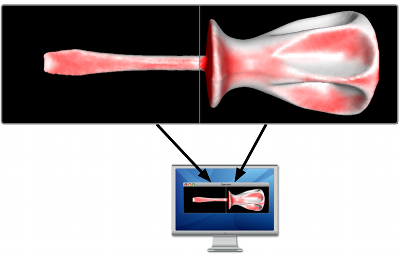}
  \caption{\label{fmonitor}Monitoring}
\end{wrapfloat}

Output frames on the display channels are connected to input frames on the
monitoring channel. The monitor equalizer changes the scaling factor and offset
between the output and input, so that the monitor channel has the same, but
typically downscaled view, as the originating segments. While this is not
strictly a scalable rendering feature, it optimises resource usage by not
needlessly rendering the same view multiple times. It reuses the zoom parameter
of compositing frames, and adapts this every time one of the channels is
resized.

\chapter{Data Distribution and Synchronisation}\label{sCollage}

\section{Overview}

Most research in parallel rendering does not look into the problem of managing
application state in a distributed rendering session. For basic parallel
rendering research this problem is trivial to solve, whereas in real-world
applications it is often one of the major challenges for using a distributed
rendering cluster. Researching and improving the system behaviour of
non-trivial applications is critical for meaningful parallel rendering
research, and therefore providing a distributed network library is a key
component of a parallel rendering system.

For this reason we have spend significant effort in researching, designing and
implementing a distributed execution layer used by Equalizer and
applications built on Equalizer. The {\em Collage} network library is an
independent open source project. In the following sections we highlight core
features and show how they are different from other distribution mechanisms,
e.g., the MPI library.

The {\em Collage} network library was conceived with the requirements of a dynamic
parallel rendering system in mind. Some of the features implemented by Collage
emerged with the growing complexity of Equalizer and its applications, and are
often layered on top of the basic primitives. The core requirements are:

\begin{compactdesc}

\item[Peer-to-peer network:] Whilst the execution model of an Equalizer
application follows a master-slave approach, and Equalizer internally uses a
client-server model, the core transport layer should be agnostic to these
higher-level abstractions. In Collage, each communicating process is equal to
all others, and no traffic prioritisation or communication pattern is enforced
by a node type. This has proven particularly useful during the implementation of
parallel compositing algorithms, where the compositing nodes form an ad-hoc
peer-to-peer sub-network.

\item[Dynamic connection management:] As a consequence of the peer-to-peer
network, all nodes in a cluster are equivalent. Due to the heterogeneous nature
of a parallel rendering application, we furthermore imposed no constraints on
the management of connections between nodes. Nodes are identified and addressed
by an universally unique identifier. The network layer lazily establishes a
connection to any given node by querying its known neighbours or a zeroconf
network for connection parameters. Connections may be established concurrently
by both sides of a node pair (e.g. during parallel compositing), which requires
a robust handshake protocol during connection establishment. For larger cluster
installations, a fully connected peer-to-peer network would be suboptimal. For
example on Windows operating systems there is a latency penalty once more than
64 connections are needed, caused by low-level implementation details. This
feature also allowed us to implement runtime configuration switches involving a
changing set of rendering resources.

\item[Transport layer abstraction:] The actual network protocol is abstracted
by an API defining byte-oriented stream semantics. While this choice of
abstraction makes it harder for RDMA-based protocols to deliver full
performance, it has proven useful in supporting a large set of transports, from
standard Ethernet sockets, SDP for InfiniBand, native Verbs for InfiniBand, UDT
to a fully-featured reliable multicast implementation. In particular, the ease
of integration of multicast transport is strong evidence for the usefulness of
this abstraction.

\item[Convenient to use for existing applications:] The history and code
structure of visualisation applications is often very different from other
distributed applications, such as simulation codes. They have been developed for
years for desktop systems, are often single-threaded and have data models and
object hierarchies built for their domain-specific problems and algorithms. The
network library needs to provide primitives which match this reality as closely
as possible by providing a modern, object-oriented C++ API.

\end{compactdesc}

\section{Architecture}

\begin{figure}[h!t]\center
  \includegraphics[width=\textwidth]{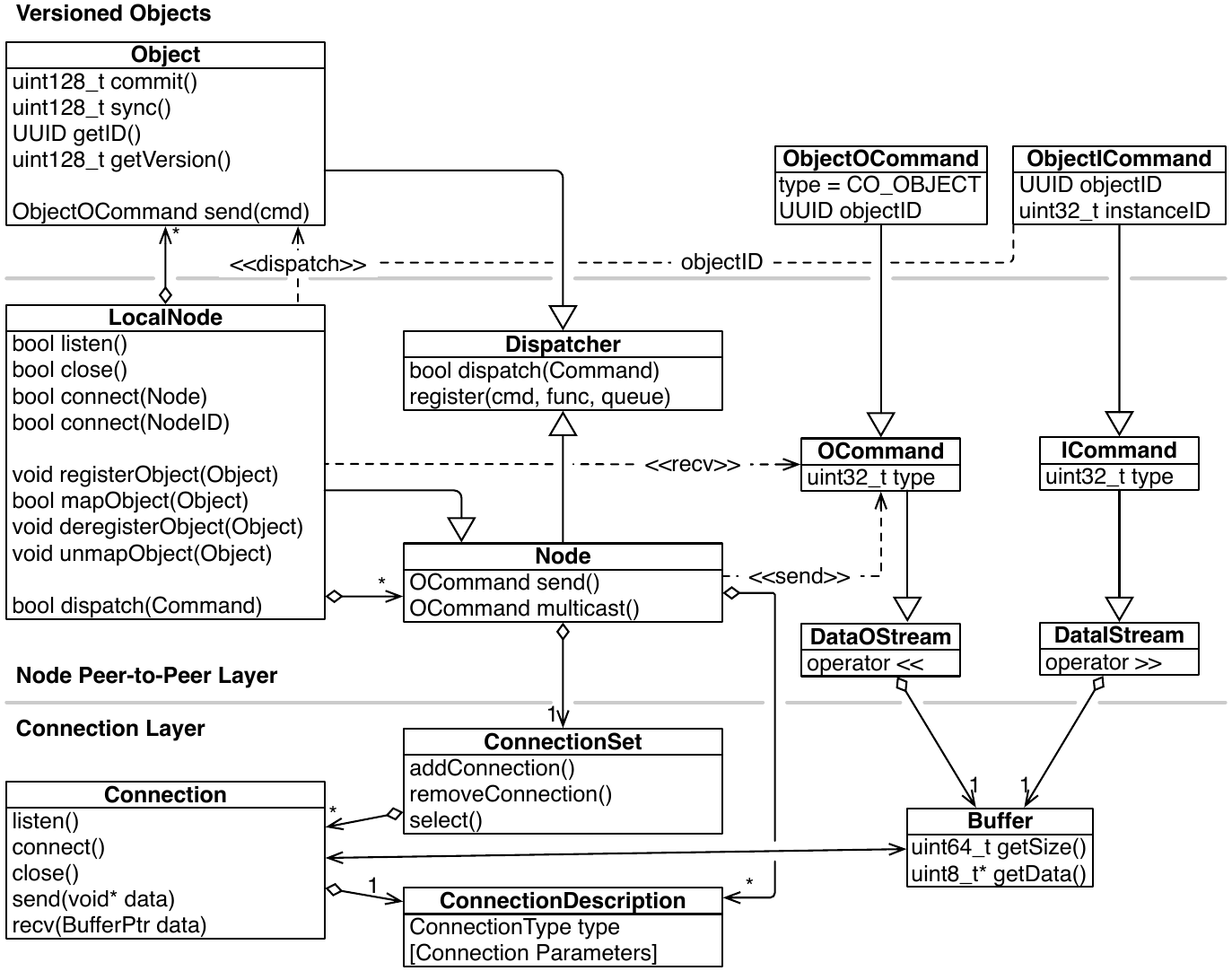}
  {\caption{\label{fCollageUML}UML class diagram of the major Collage classes}}
\end{figure}

Our Collage network library provides a peer-to-peer communication
infrastructure, offering different abstraction layers which gradually provide
higher level functionality to the programmer. Collage is used by Equalizer to
communicate between the application node, the server and the render clients.
Many resource entities described in \cref{sArchitecture} are distributed
Collage objects. \fig{fCollageUML} provides an overview of the major Collage
classes and their relationship. The main classes, in ascending abstraction
level, are:

\begin{compactdesc}

\item[Connection:] A stream-oriented point-to-point communication line.
Different implementations of a connection exist. A connection transmits a raw
byte stream reliably between two endpoints for unicast connections, and between
a set of endpoints for multicast connections.

\item[DataOStream:] Abstracts the output of C++ data types onto a set of
  connections by implementing output stream operators. Uses buffering to
  aggregate data for network transmission.
\item[OCommand:] Extends DataOStream to implement the protocol between Collage
  nodes by adding node and command type routing information to the stream.
\item[DataIStream:] Decodes a buffer of received data into C++ objects and PODs
  by implementing input stream operators. Performs endian swapping if the
  endianness differs between the originating and local node.
\item[ICommand:] The other side of OCommand, extending DataIStream.
\item[Node and LocalNode:] The abstraction of a process in the cluster. Nodes
  communicate with each other using connections. A LocalNode listens on various
  connections and processes requests for a given process. Received data is
  wrapped in ICommands and dispatched to command handler methods. A Node is a
  proxy for communicating with a remote LocalNode.
\item[Object:] Provides object-oriented, versioned data distribution of C++
  objects between nodes within a session. Objects are registered or mapped on a
  Local\-Node.
\end{compactdesc}

\subsection{Connection}

A {\em Connection} is the basic primitive used for communication between
processes in Collage. It provides a stream-oriented communication between two
endpoints. A connection is either closed, connected or listening. A closed
connection cannot be used for communication. A connected connection can be
used to read or write data to the communication peer. A listening connection
can accept connection requests leading to new, connected connections.

A {\em ConnectionSet} is used to manage multiple connections. The typical use
case is to have one or more listening connections for the local process, and a
number of connected connections for communicating with other processes. The
connection set is used to select one connection requiring some action.
This can be a connection request on a listening connection, pending data on a
connected connection, or the notification of a disconnect. It is an
encapsulation of the {\em poll} or {\em WaitForMultipleObject} system calls.

The connection and connection set can be used by applications to implement
other network-related functionality, e.g., to communicate with a sound server
on a different machine. They do not require a particular wire protocol. A
{\em LocalNode} has a connection set and uses it to manage connections
with other nodes.

\subsection{Data Streams}

Data streams implement serialisation and buffering on top of connections. They
use output and input stream operators (\textless\textless\ and
\textgreater\textgreater) with function overloads to provide serialisation for
all common data types. The input stream will perform byte swapping if the
endianness differs between the sending and receiving node. Applications can
easily provide overloads for their own classes for serialisation. All
serialised data is assembled in a memory buffer and sent over the connection
once the data is complete. An output data stream might send its data to many
connections, e.g., when an object update is sent to all subscribed slave nodes.

\subsection{Thread-aware Command Dispatch}

Collage sends commands over connections to implement remote procedure calls. A
command is identified by its type (typically the C++ class handling it) and a
command identifier. These fields are used to implement thread-aware dispatching
of received commands to handler functions.

Nodes and objects communicate using commands derived from data streams. The
basic command dispatch is implemented in the {\em Dispatcher} class, from
which {\em Node} and {\em Object} are sub-classed.

The dispatcher allows the registration of commands with a dispatch queue and an
invocation method. Each command has a type and command identifier, which is
used to identify the receiver, registered queue and method. The dispatch pushes
the packet to the registered queue. When the commands are dequeued by the
processing thread, the registered command method is invoked.

This dispatch and invocation functionality is used within Equalizer to dispatch
commands from the receiver thread to the appropriate node or pipe thread, and
then to call a specific method when it is processed by these threads. All
Equalizer task methods available to the application are triggered by this
mechanism. This dispatch provides object-oriented semantics, since C++
instances can register themselves on the dispatcher, and get automatically
invoked in the correct thread when an appropriate command arrives.

\subsection{Nodes}

The {\em Node} is the abstraction of one process in the peer-to-peer network.
Each node has a universally unique identifier. This identifier is used to
address nodes, e.g., to query connection information to connect to the node.
Nodes use connections to communicate with each other by sending
{\em OCommand}s.

The {\em LocalNode} is the specialisation of the node for the given process.
It encapsulates the communication logic for connecting remote nodes, as well as
object registration and mapping. Local nodes are set up in the listening state
during initialisation.

A remote {\em Node} can either be connected explicitly by the application,
or implicitly due to a connection from a remote node. The explicit connection
can be done by programmatically creating a node, adding the necessary
{\em ConnectionDescription}s and connecting it to the local node. It may
also be done by connecting the remote node to the local node by using its
{\em NodeID}. This will cause Collage to query connection information for
this node from the already connected nodes and zeroconf, instantiating the node
and connecting it. Both operations may fail.

Each Equalizer entity has a {\em LocalNode} for communication, and one {\em
Node} instance for each peer it communicates with.

\subsubsection{\label{sZeroconf}Zeroconf Discovery}

Each {\em LocalNode} provides a {\em Zeroconf} communicator, which allows
node and resource discovery. The zeroconf service "\_collage.\_tcp" is used to
announce the presence of a listening {\em LocalNode} using the ZeroConf
protocol to the network. The node identifier and all listening connection
descriptions are announced, used to connect unknown nodes by using the
node identifier alone.

\subsubsection{Communication between Nodes}

\begin{figure}[h!t]\center
  \includegraphics[width=\textwidth]{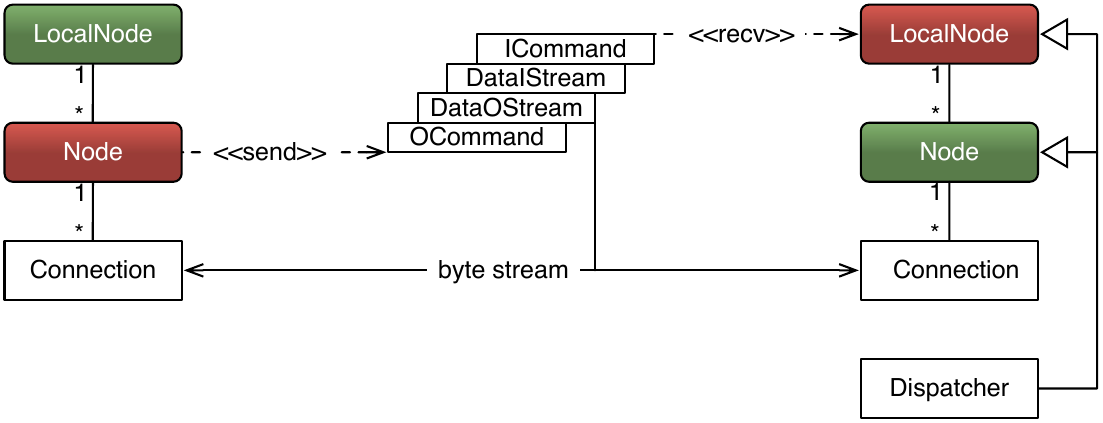}
  {\caption{\label{fNetNode}Communication between two Nodes}}
\end{figure}

\fig{fNetNode} shows the communication between two nodes. Each
{\em LocalNode} has a receiver thread, which uses a connection set to read
and dispatch incoming data from the network, and a command thread used for
higher-level functions such as object mapping. When the remote node sends a
command, the listening node receives the command and dispatches it from the
receiver thread. The dispatch will either invoke the bound function immediately,
or enqueue the command into the given queue. The queue consumer, for example the
main or command thread, will read the command of this queue and then invoke the
bound function.

\section{Reliable Stream Protocol}\label{sec:RSP}

RSP is an implementation of a reliable multicast protocol over unreliable UDP
multicast transport. RSP behaves similarly to TCP; in contrast to the underlying
UDP transport, it is not message-oriented, but implements byte stream semantics.
RSP provides full reliability and ordering of the data, and slow receivers will
eventually throttle the sender through a sliding window algorithm. This behaviour
is needed to guarantee delivery of data in all situations. Pragmatic generic
multicast (PGM~\cite{pgm}) provides full ordering, but slow clients will
disconnect from the multicast session instead of throttling the send rate. Since
we use multicast for distributing application data to all rendering clients, we
want semantics similar to TCP, expressly waiting for a client to read data is
preferable over losing this client.

RSP combines various established multicast
algorithms~\cite{adamson2004negative,Gau:2002} in an open source implementation
capable of delivering wire speed transmission rates on high-speed LAN
interfaces. The following will outline the RSP protocol and implementation,
as well as the motivation for the design decisions. Any defaults given below are for Linux
or Mac OS X, the Windows UDP stack requires different default values which can be
found in the implementation.

Our RSP implementation uses a separate protocol thread for each RSP group,
which handles all reads and writes on the multicast UDP socket. It implements
the protocol handling and communicates with the application threads through
thread-safe queues. The queues contain datagrams filled with the application
byte stream, prefixed by a header of at most eight bytes. Each connection has a
configurable number of buffers (1024 by default) of a configurable datagram
size (1470 bytes default), which are either free or in transmission. The header
contains two bytes for the datagram type (connection handshake, data,
acknowledgement, negative acknowledgement, acknowledgement request), and up two
six bytes of datagram-specific information (e.g. for acknowledgement: two bytes
read node identifier, two bytes write node identifier, two bytes sequence
number).

\begin{figure}[h!t]\center
  \includegraphics[width=\textwidth]{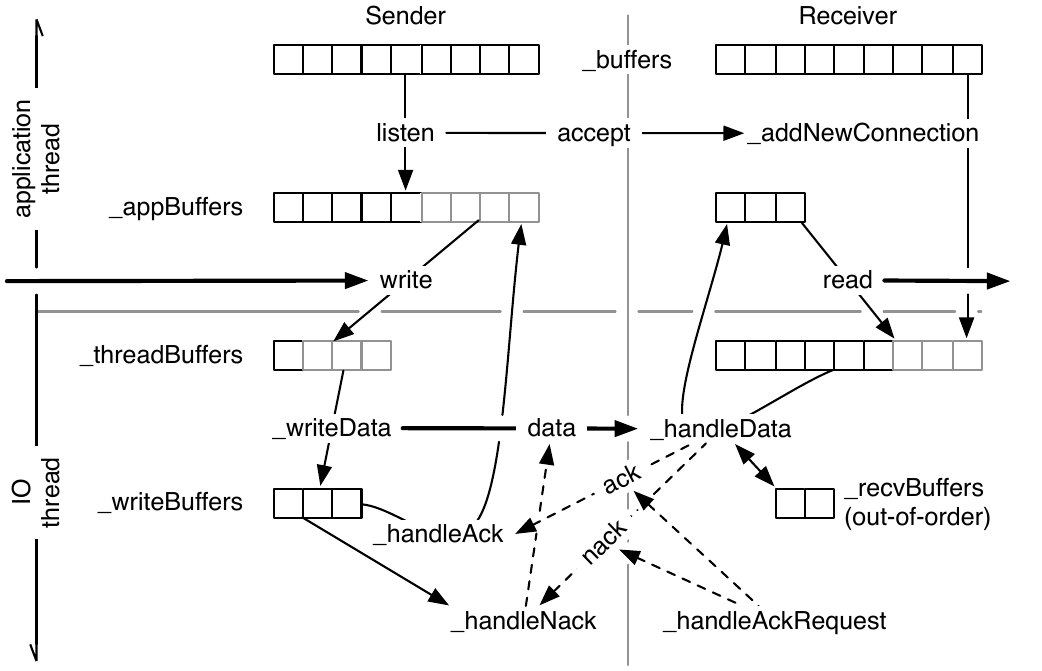}
  {\caption{\label{fRSP}RSP Data Flow}}
\end{figure}

\fig{fRSP} shows the data flow through the RSP implementation. Each member of
the multicast group opens a listening connection, which will send query
datagrams to the multicast socket. For each found member, a receiving
connection instance is created and, similar to a TCP socket, passed to the
application upon {\em accept}. Each connection instance has a fixed number
(1024 by default) of fixed-size (1470 by default) buffers, each used directly
for an UDP datagram. The listening connection uses these buffers for writing
data, and each receiving connection uses its buffers for received data. These
buffers are continuously cycled through two sets of queues: a blocking,
thread-safe queue used on the application side for reading and writing data,
and a non-blocking, lock-free and thread-safe queue on the protocol thread for
data management.

When writing data, the application thread pops empty buffers from its queue
(blocking when the data cannot be written fast enough), fills in the
{\em data} datagram header and copies the application data piece-wise into
the datagram. The datagrams are then pushed onto the protocol thread buffer
queue. The protocol thread writes the datagrams into the UDP multicast socket,
and reads and handles any incoming datagrams. On the receiver side, the protocol
receives the data, and pushes them in order to the corresponding application
thread queue. Out-of-order datagrams are stored aside and queued in order later.
Negative acknowledgements (nack) are immediately sent for missing
datagrams. The writer will repeat nack'd datagrams, recycle fully acknowledged
datagrams to the application queue, and ask for missing acknowledgements if
needed. When reading data, the application pops full buffers from the
corresponding connection queue (blocking when no data is available), copies
the data piece-wise out of the datagram into the application buffer, and recycles
the cleared buffers onto the protocol thread queue.

Handling a smooth packet flow is critical for performance. RSP uses active flow
control to advance the byte stream buffered by the implementation. Each incoming
connection actively acknowledges every $n$ (17 by default) packets fully
received. The incoming connection offset this acknowledgement by their
connection identifier to avoid ack bursts. Any missed datagram is actively
nack'd as soon as detected. Write connections continuously retransmit packets
for nack datagrams, and advance their window upon reception of all acks from the
group. The writer will explicitly request an ack or nack when it runs out of
empty buffers or finishes its write queue. Nack datagrams may contain multiple
ranges of missed datagrams, motivated by the observation that UDP
implementations often drop multiple contiguous packets.

Congestion control is necessary to optimise bandwidth usage. While TCP uses the
well-known additive increase, multiplicative decrease algorithm, we have chosen
a more aggressive congestion control algorithm of additive increase and additive
decrease. Experimentally this has proven to be more optimal: UDP is often
rate-limited by switches; packets are discarded regularly and not
occasionally. Only slowly backing off the current send rate helps to stay close
to this limit. Furthermore, our RSP traffic is limited to the local subnet,
making cooperation between multiple data stream less of an issue. Send rate
limiting uses a bucket algorithm, where over time the bucket fills with send
credits, from which send datagrams are subtracted. If there are no available credits,
the sender sleeps until sufficient credits are available.

In \cite{ESP:18} we provide experimental results showing that our
implementation can achieve above 90\% wire speed on 10~GBit/s Ethernet,
good scalability with respect to multicast group size, and is very effective
for concurrently distributing structured and unstructured application data to a large number
of rendering clients (see \bench{rNetwork}).

\section{Distributed, Versioned Objects}

\subsection{Overview}

Adapting an existing application for parallel rendering requires the
synchronisation of application data across the processes in the parallel
rendering setup. Existing parallel rendering frameworks often address this
poorly, at best they rely on MPI to distribute data. Real-world, interactive
visualisation applications are typically written in C++ and have complex data
models and class hierarchies to represent their application state. As outlined
in \cite{EMP:09}, the parallel rendering code in an {\em Equalizer}
application only requires access to the data needed for rendering, as all
application logic is centralised in the application main thread. We have
encountered two main approaches to address this distribution: Using a shared
filesystem for static data, or using data distribution for static and dynamic
data. Distributed objects are not required to build {\em Equalizer}
applications. While most developers choose to use this abstraction for
convenience, we have seen applications using other means for data distribution,
e.g., MPI.

\subsection{Object Types}

Distributed objects in {\em Collage} provide powerful, object-oriented data
distribution for C++ objects. They facilitate the implementation of data
distribution in a cluster environment. Distributed objects are created by
subclassing from {\em co::Serializable} or {\em co::Object}. The
application programmer implements serialisation and deserialisation. Distributed
objects can be static (immutable) or dynamic. Objects have a universally unique
identifier (UUID) as cluster-wide address. A master-slave model is used to
establish mapping and data synchronisation across processes. Typically, the
application main loop registers a master instance and communicates the UUID to
the render clients, which map their instance to the given identifier. The
following object types are available:

\begin{compactdesc}
\item[Static] The object is neither versioned nor buffered. The instance data is
  serialised whenever a new slave instance is mapped. No additional data is
  stored.
\item[Unbuffered] The object is versioned and unbuffered. No data is stored, and
  no previous versions can be mapped.
\item[Instance] The object is versioned and buffered. The instance and delta
  data are identical; that is, only instance data is serialised. Previous
  instance data is saved to be able to map old versions.
\item[Delta] The object is versioned and buffered. The delta data is typically
  smaller than the instance data. The delta data is transmitted to slave
  instances for synchronisation. Previous instance and delta data is saved to be
  able to map and sync old versions.
\end{compactdesc}

Instance and delta objects have a memory overhead on the master instance to
store past data. The number of old versions retained is configurable per
object. For Equalizer applications, this overhead typically occurs on the
application node holding the master instances, and is configured based on the
configurations' latency. When using unbuffered objects, applications only
observe inconsistent state during the initial mapping, when a too recent
version is used by a render client. The push-based commit-sync logic eventually
brings the object into a consistent state with respect to the rendered frame.

Serialisation is facilitated using output or input streams, which abstract the
data transmission and are used like a {\em std::stream}. The data streams
implement efficient buffering and compression, and automatically select the best
connection for data transport. Custom data type serialisers can be implemented
by providing the appropriate serialisation functions. No pointers should be
directly transmitted through the data streams. For pointers, the corresponding
object is typically also a distributed object , and its UUID and version are
transmitted in lieu of a pointer.

Dynamic objects are versioned, and on {\em commit} the delta data to the
previous version is sent, if available using multicast, to all mapped slave
instances. The data is queued on the remote node, and is applied when the
application calls {\em sync} to synchronise the object to a new version. The
{\em sync} method might block if a version has not yet been committed or is
still in transmission. All versioned objects have the following characteristics:

\begin{compactitem}

\item The master instance of the object generates new versions for all slaves.
These versions are continuous. It is possible to commit on slave instances, but
special care has to be taken to handle possible conflicts during concurrent
commits from multiple slave instances.

\item Slave instance versions can only be advanced; that is,
{\em sync(version)} with a version older than the current version will fail.

\item Newly mapped slave instances are mapped to the oldest available version
by default, or to the version specified when calling {\em mapObject}.

\end{compactitem}

Blocking commits allows limiting the number of outstanding, queued versions on
the slave nodes. A token-based protocol will block the commit on the master
instance if too many unsynchronised versions exist. This is useful to limit the
amount of memory consumed by slave instance, and too prohibit run-away
conditions of the master instance.

\subsection{Serialisable}

\label{sec:Serializable}The {\em Serializable} implements one convenient usage
pattern for object data distribution which emerged during deployment of
Equalizer in applications. The {\em Serializable} data distribution is based on
the concept of dirty bits, allowing inheritance with data distribution. Dirty
bits are a 64-bit mask tracking the parts of the object to be distributed
during the next commit. Setters of the class mark the appropriate dirty bit,
and the accumulated bits are used to compute deltas during commit.

For serialization, the application developer implements {\em serialize} or
{\em deserialize}, which are called with the bit mask specifying which data
has to be transmitted or received. During a commit or sync, the current dirty
bits are given, whereas during object mapping all dirty bits are passed to the
serialisation methods. A commit will clear the dirty mask after serialisation.

\subsection{Optimisations}

The Object API provides sufficient abstraction to implement various
optimisations for faster mapping and synchronisation of data: compression,
chunking, caching, preloading and multicast.

\subsubsection{Compression}

The most obvious optimisation is compression. Recently many new compression
algorithms have been developed, exploiting modern CPU architectures and
deliver compression rates well above one Gigabyte per second. {\em Collage}
uses the Pression library~\cite{pression}, which provides an unified interface
for a number of compression libraries, such as FastLZ~\cite{jesperfast},
Snappy~\cite{snappy} and ZStandard~\cite{zstd}. It also contains a custom,
virtually zero-cost RLE compressor. Pression parallelises the compression and
decompression using data decomposition. The compression is generic and
lossless, available transparently to the application. Applications can also
use data-specific compression.

\bench{rCompressorDetail}~(top left) shows the compression ratio and speed for
generic binary data from \cite{ESP:18}. Whilst the structure of the transmitted
data varies with each application, this micro-bench\-mark gives a reasonable
estimation of the expected performance. In our context of interactive
distributed rendering applications, it is important to use the right tradeoff
between spending time and resources for data compression, and the gained
network transmission time due data reduction.

\begin{benchmark}[h!t]\center
  \includegraphics[width=\textwidth]{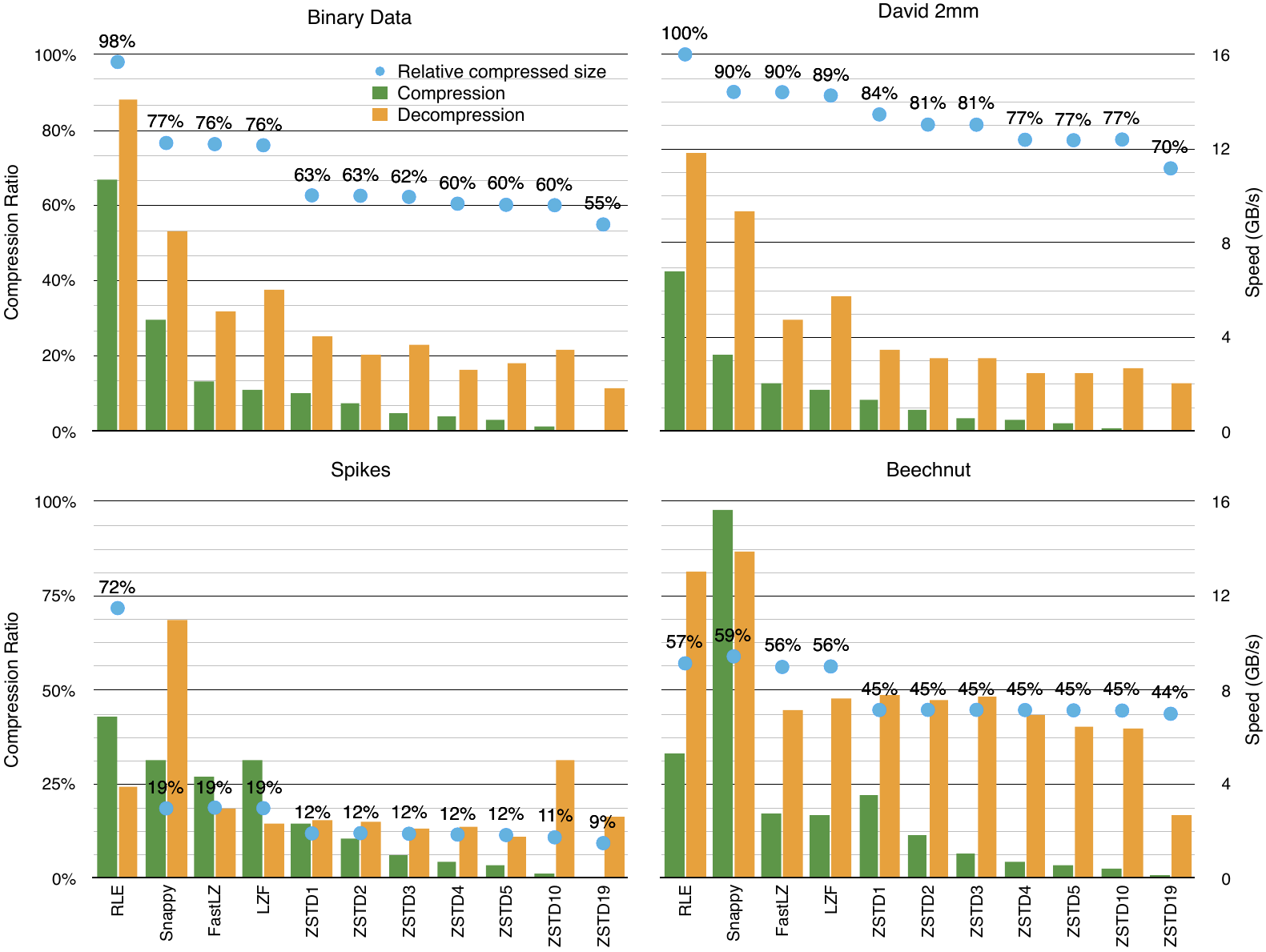}
  \caption{\label{rCompressorDetail}Compression Performance for Binary Data and the Object Data used in \bench{rMapping}}
\end{benchmark}

On current CPUs (the benchmark was executed on a 12-core node), modern
compression libraries provide performance benefits even on fast interconnects
such as 10 Gb/s Ethernet. In particular, the modern Snappy and ZStandard
libraries deliver impressive performance.

In \cite{ESP:18} we also evaluated the compression performance for concrete
application data. The results are shown in \bench{rCompressorDetail}. Polygonal
data is difficult to compress with a generic lossless compressor, due to the
floating point format used for the vertices. A data-specific compressor aware
of the data semantics can provide much better results. Volume data on the other
hand has shown to be well compressible, with typical $2:1$ compression ratios
at interactive speeds. \sref{sObjectBench} discusses how these compressors
accelerate data distribution in Equalizer applications.

\subsubsection{Chunking}

The data streaming interface implements chunking, which pipelines the
serialisation code with the network transmission. After a configurable number of
bytes has been serialised to the internal buffer, it is transmitted and
serialisation continues. This is used both for the initial mapping data, and for
commit data.

\subsubsection{Caching and Preloading}

Caching retains instance data of objects in a client-side cache, and reuses
this data to accelerate mapping of objects. The instance cache is either filled
by ``snooping'' on multicast transmissions or by an explicit preloading when
master objects are registered. Preloading sends instance data of recently
registered master objects to all connected nodes during idle time of the
corresponding node. These nodes simply enter the received data to their cache.
Preloading uses multicast when available.

\subsubsection{Multicast}

Due to the master-slave nature of data distribution, multicast is used to
optimise the transmission time of data. If the contributing nodes share a
multicast session, and more than one slave instance is mapped, {\em Collage}
automatically uses the multicast connection to send the new version
information.

\subsection{Benchmarks}\label{sObjectBench}

\bench{rMapping} analyses the performance of data distribution and
synchronisation in real-world applications. We extracted the data distribution
code from our mesh renderer (eqPly) and our volume renderer (Livre) into a
benchmark application to measure the time to initially map all the objects on
the render client nodes, and to perform a commit$+$sync of the full data set
after mapping has been established. All figures observe a noticeable
measurement jitter due to other services running on the shared cluster during
benchmarking. The details of the benchmark algorithm can be found in
\cite{ESP:18}.

\begin{benchmark}[h!t]\center
  \includegraphics[width=\textwidth]{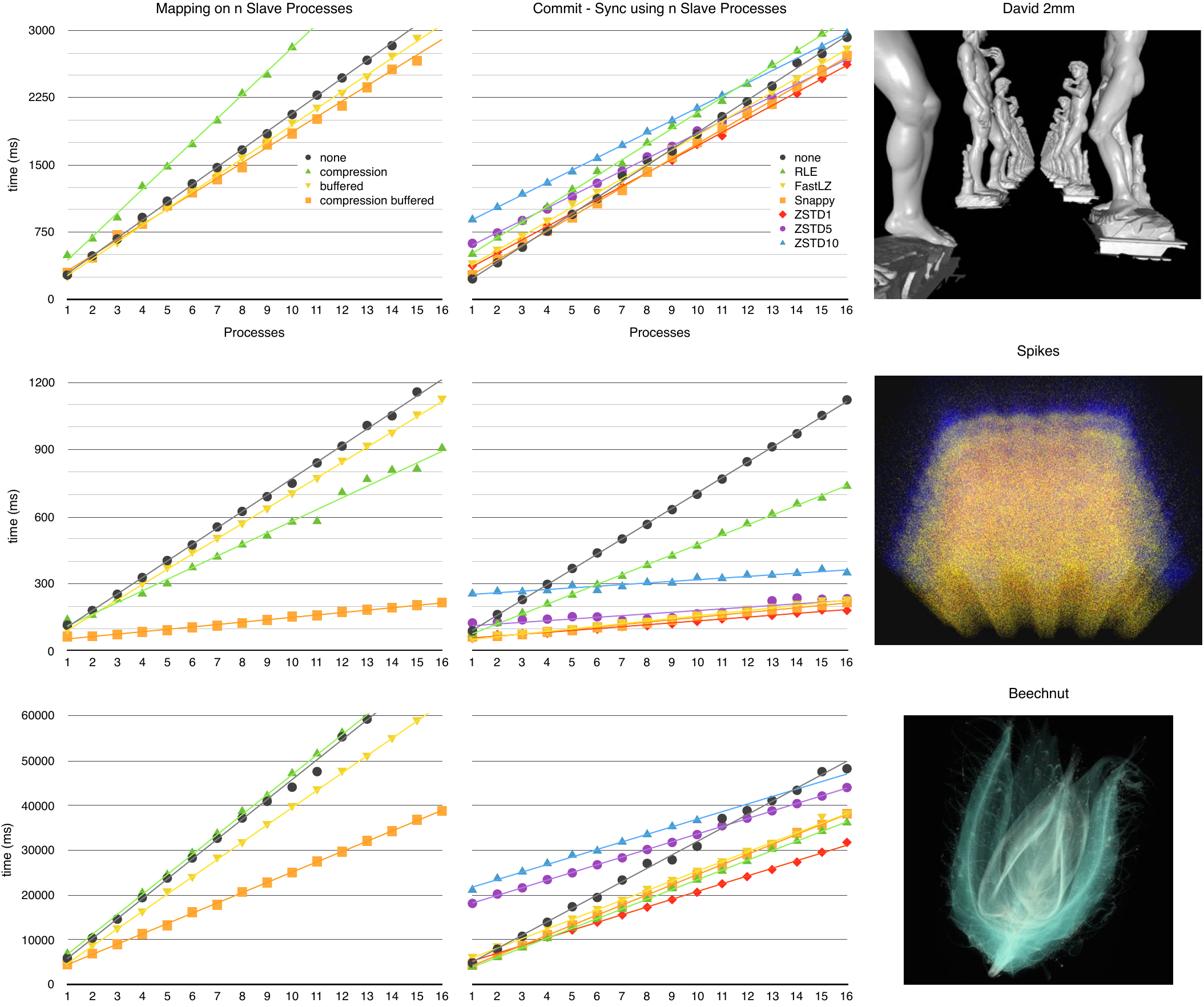}
  \caption{\label{rMapping}Object Mapping and Synchronisation}
\end{benchmark}

We used three different data sets, and ran the benchmark on up to eight
physical nodes, specifically, after eight process nodes start to run two processes
per node, which share CPU, memory and network interface bandwidth. Object
mapping is measured using the following settings: {\em none} distributes the
raw, uncompressed, and unbuffered data, {\em compression} uses the Snappy
compressor to compress and distribute unbuffered data, {\em buffered} reuses
uncompressed, serialised data for mappings from multiple nodes, and
{\em compression buffered} reuses the compressed buffer for multiple nodes.

Unbuffered operations need to reserialise, and potentially recompress, the
master object data for each slave node. Each slave instance needs to
deserialise and decompress the data, which happens naturally in parallel on the
slave nodes. During data synchronisation, the master commits the object data to
all mapped slave instances simultaneously. This is a “push” operation, whereas
the mapping is a slave-triggered “pull” operation. During commit, the buffers
only have to be serialised and compressed once, and can then be sent directly
to all mapped slave nodes. Slave nodes queue this data and consume it during
synchronisation. In contrast, object mapping needs to wait for each slave node
to request the mapping, and then may need to reserialise and compress the
object data. We tested the time to commit and sync the data using the compression
engines discussed above.

The David statue at a 2 mm resolution is organised in a k-d tree for rendering.
Each k-d tree node is a separate distributed object, having two child node
objects. A total of 1023 objects are distributed and synchronised. Due to
limited compressibility of the data, the results are relatively similar.
Compressing the data repeatedly for each client leads to decreased performance,
since the compression overhead cannot be amortised by a decreased
transmission time. Buffering data slightly improves performance by reducing the
CPU and copy overhead. Combining compression and buffering leads to the best
performance, although only by about 10\%. During synchronisation data is
pushed from the master process to all mapped slaves using a unicast connection
to each slave. While the results are relatively close to each other, we can
still observe how the tradeoff between compression ratio and speed influences
overall performance. Better, slower compression algorithms lead to improved
overall performance when amortised over many send operations.

The volume data sets are distributed in a single object, serialising the raw
volume buffer. The Spike volume data set has a significant compression ratio,
which is reflected by the results. Compression for this data is beneficial for
transmitting data over a 10~Gb/s link, even for a single slave process.
Buffering has little benefit, since the serialisation of volume data is
trivial. Buffered compression shows a significant difference, since the
compression cost can be amortised over many nodes, reaching raw data
transmission rates of 3.7~GB/s with the default Snappy compressor, and at best
4.4 GB/s with ZStandard at level 1. The distribution of the beechnut data set
also behaves as expected: Due to the larger object size, uncompressed
transmission is slightly faster compared to the Spike data set at 700~MB/s
since static overheads are comparatively smaller. Compressed transmission does
not improve the mapping performance, likely due to increased memory pressure
caused by the data size. The comparison of the various compression engines is
consistent with the benchmarks in \bench{rCompressorDetail}; RLE, Snappy and
the LZ variants are very close to each other, and ZSTD1 can provide better
performance after four nodes due to the better compression ratio.

\begin{wrapfloat}{benchmark}{O}{.618\textwidth}
  \includegraphics[width=.618\textwidth]{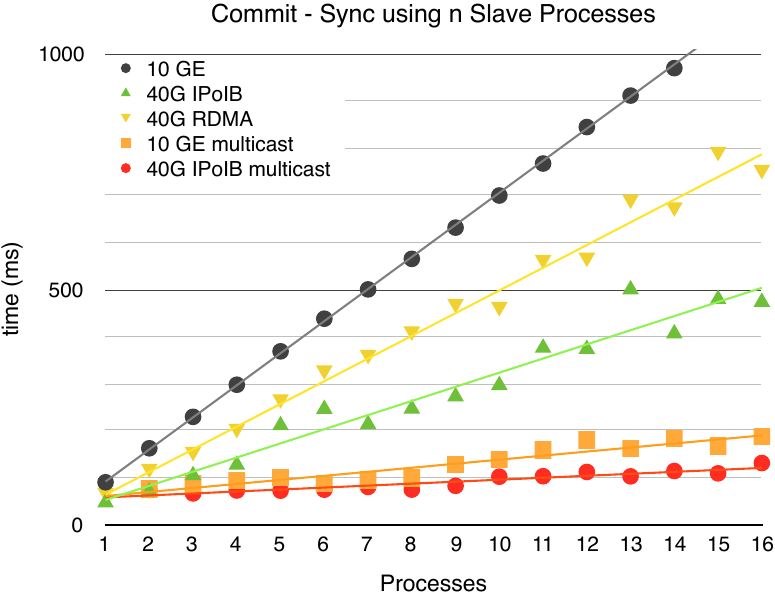}
  {\caption{\label{rNetwork}Synchronisation Performance over different Network Protocols}}
\end{wrapfloat}

\bench{rNetwork} compares data distribution speed using different network
protocols. This benchmark measures the data synchronisation time of the Spike volume
data set. Buffering is enabled, and compression is disabled to
focus on the raw network performance. For the benchmark, eight physical nodes
are used, that is, after eight processes two client processes will run on some
nodes, sharing CPU and network resources.

TCP over the faster InfiniBand link outperforms the cheaper Ten Gigabit
Ethernet link by more than a factor of two. Unexpectedly, the native RDMA
connection performs worse, even though it outperforms IPoIB in a simple
peer-to-peer connection benchmark. This needs further investigation, but we
suspect the abstraction of a byte stream connection chosen by Collage is not
well suited for remote DMA semantics; one needs to design the network
API around zero-copy semantics with managed memory for modern high-speed
transports. Both InfiniBand connections show significant measurement jitter.

RSP multicast performs as expected. Collage starts using multicast to commit
new object versions when two or more clients are mapped, since the transmission
to a single client is faster using unicast. RSP consistently outperforms
unicast on the same physical interface and shows good scaling behaviour (2.5
times slower on 16 vs. 2 clients on Ethernet, 1.8 times slower on InfiniBand).
The scaling is significantly better when only one process per node is used. The
increased transmission time with multiple clients is caused by a higher
probability of packet loss, which increases significantly when using more than
one process per node and network interface. InfiniBand outperforms Ethernet
slightly, but is largely limited by the RSP implementation throughput of
preparing and queueing the datagrams to and from the protocol thread, which we
observed in profiling.

\chapter{Applications}\label{sApplications}

A key performance indicator for a good design of any framework is the
acceptance by developers. A good measure is the adoption by third-party
applications. While the evaluation and architecture of applications build with
Equalizer is outside of the scope of this thesis, we provide a few examples
here to illustrate the variety of use cases supported in our framework.

\section{Livre}

\begin{wrapfloat}{figure}{O}{.618\textwidth}
  \includegraphics[width=.618\textwidth]{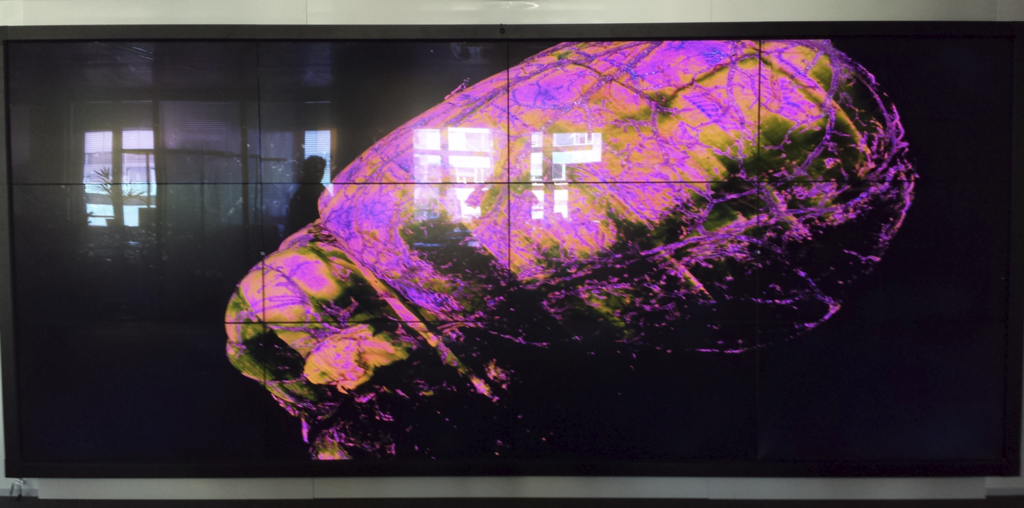}
  {\caption{\label{fLivre}Livre running on a 4x3 Tiled Display Wall}}
\end{wrapfloat}

Livre (Large-scale Interactive Volume Rendering Engine) is a GPU ray-casting
parallel 4D volume renderer, implementing state-of-the-art view-dependent
level-of-detail rendering (LOD) and out-of-core data
management~\cite{EHKRW:06}.

Hierarchical and out-of-core LOD data management is supported by an implicit
volume octree, accessed asynchronously by the renderer from a data source on a
shared file system. Different data sources provide octree-conform access to
RAW or compressed files, as well as to on-the-fly generated volume data (e.g.
such as from event simulations or surface meshes).

High-level state information, e.g., camera position and rendering settings, are
shared in Livre through {\em Collage} objects between the application and
rendering threads. Sort-first decomposition is efficiently supported through
octree traversal and culling, both for scalability, as well as for driving
large-scale tiled display walls.

\section{RTT Deltagen}

RTT Deltagen (now Dassault 3D Excite) is a commercial application for
interactive, high quality rendering of CAD data. The RTT Scale module,
delivering multi-GPU and distributed execution, is based on {\em Equalizer}
and {\em Collage}, and has driven many of the features implemented in
Equalizer.

\begin{wrapfloat}{figure}{O}{.618\textwidth}
  \includegraphics[width=.618\textwidth]{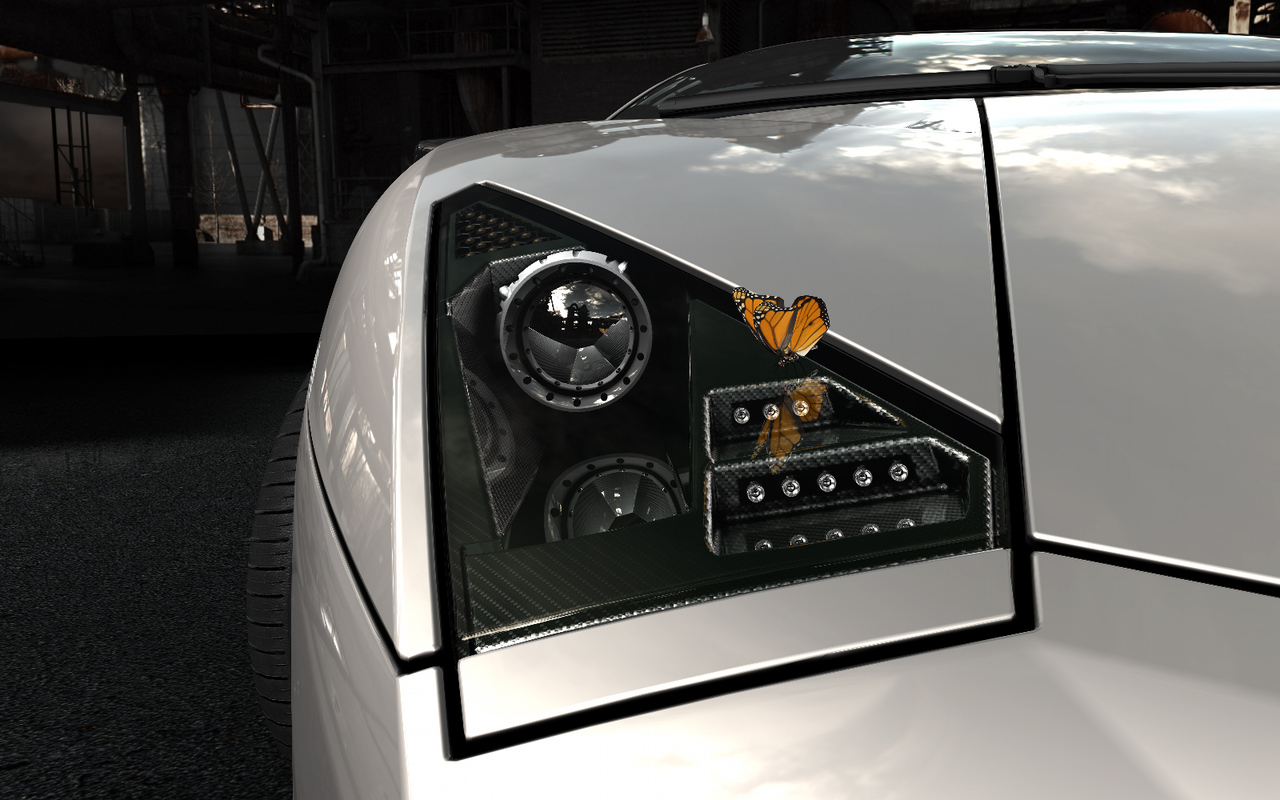}
  {\caption{\label{fDeltaGen}RTT Deltagen mixing OpenGL Rendering and Raytracing (for the head light)}}
\end{wrapfloat}

RTT Scale uses a master-slave execution mode, were a single Deltagen instance
can go into ``Scale mode'' at any time by launching an {\em Equalizer}
configuration. Consequently, the internal representation needed for rendering
is based on a {\em Collage}-based data distribution. The rendering clients
are separate, smaller applications which will map their scenes during startup.
At runtime any change performed in the main application is committed as a
delta at the beginning of the next frame. Multicast is used to keep data
distribution times during session launch reasonable for larger cluster sizes
(tens to hundreds of nodes).

RTT Scale supports a wide variety of use cases. In virtual reality, the
application is used for virtual prototyping and design reviews in front of
high-resolution display walls and CAVEs. It is also used for virtual prototyping of
human-machine interactions in CAVEs and HMDs. For scalability, sort-first and
tile compounds are used to achieve fast, high-quality rendering, primarily for
interactive raytracing, both based on CPUs and GPUs. For CPU-based raytracing,
often Linux-based rendering clients are used with a Windows-based application
node.

\section{RTNeuron}\label{sRTNeuron}

\begin{wrapfloat}{figure}{O}{.618\textwidth}
  \includegraphics[width=.618\textwidth]{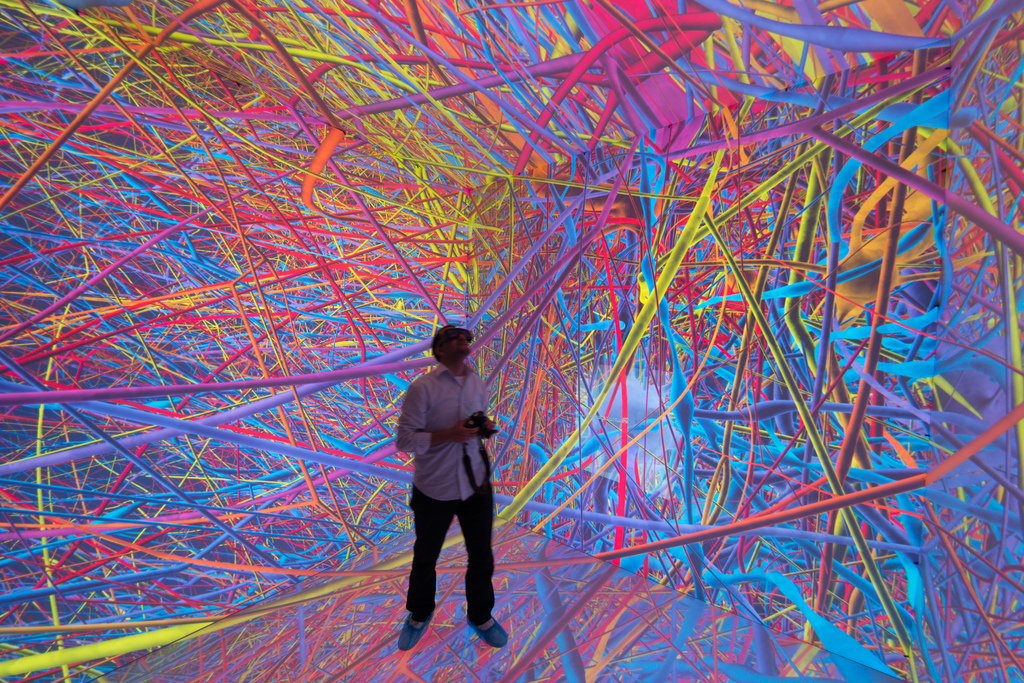}
  {\caption{\label{fRTNeuron}RTNeuron running in a six-sided CAVE}}
\end{wrapfloat}

RTNeuron \cite{HBBES:13} is a scalable real-time rendering tool for the
visualisation of neuronal simulations based on cable models. It uses
OpenSceneGraph for data management and Equalizer for parallel rendering.
The focus is not only on fast rendering times, but also on fast loading times with no
offline preprocessing. It provides level of detail (LOD) rendering, high quality
anti-aliasing based on jittered frusta, accumulation during still views, and
interactive modification of the visual representation of neurons on a per-neuron
basis (full neuron vs. soma only, branch pruning depending on the branch level,
\dots). RTNeuron implements both sort-first and sort-last rendering with order
independent transparency.

\section{RASTeR}

\begin{wrapfloat}{figure}{O}{.618\textwidth}
  \includegraphics[width=.618\textwidth]{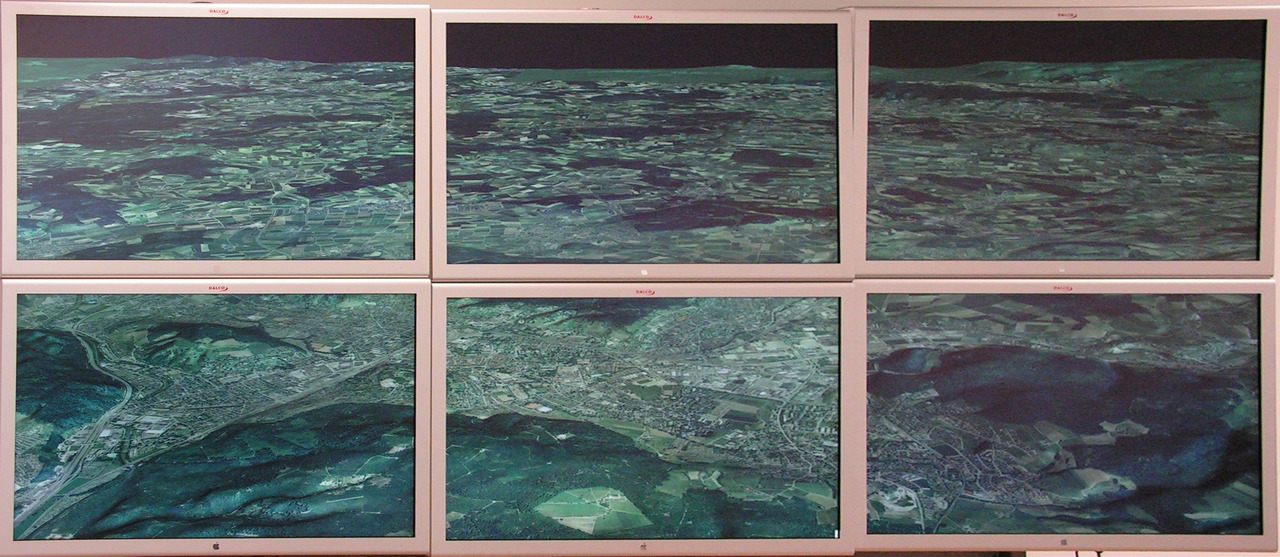}
  {\caption{\label{fRaster}RASTeR running on a 3x2 Tiled Display Wall}}
\end{wrapfloat}

RASTeR~\cite{BGP:09} uses an out-of-core and view-dependent real-time
multi-resolution terrain rendering algorithm. For load balanced parallel rendering~\cite{GMBP:10} it
exploits fast hierarchical view-frustum culling of the level-of-detail (LOD)
quadtree for sort-first decomposition, and uniform distribution of the visible
LOD triangle patches for sort-last decomposition. The latter is enabled by a
fast traversal of the patch-based restricted quadtree triangulation hierarchy,
which results in a list of selected LOD nodes, constituting a view-dependent cut
or \emph{front of activated nodes} through the LOD hierarchy. Assigning and
distributing equally sized segments of this active LOD front to the concurrent
rendering threads results in a near-optimal sort-last decomposition for each
frame.

\section{Bino}

\begin{wrapfloat}{figure}{O}{.618\textwidth}
  \includegraphics[width=.618\textwidth]{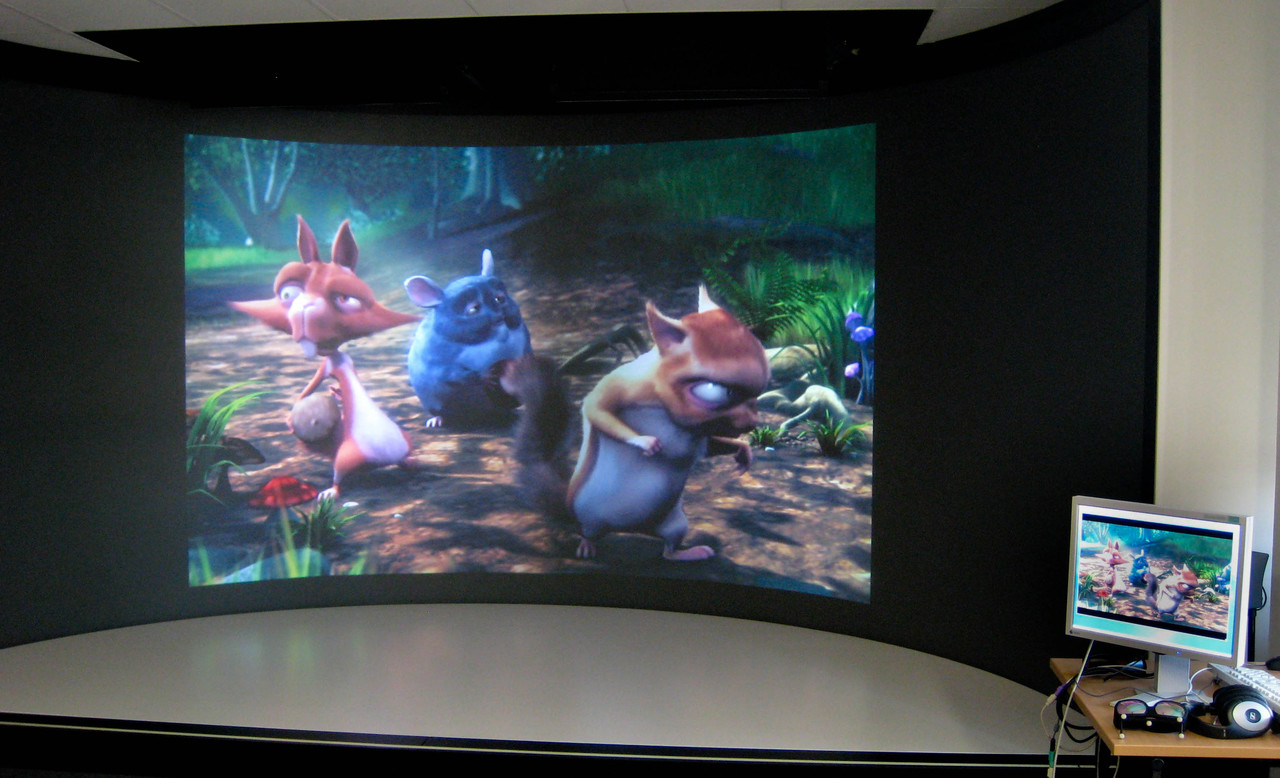}
  {\caption{\label{fBino}Bino on a Semi-Cylindrical Multi-Projector Wall}}
\end{wrapfloat}

Bino is a stereoscopic 3D video player capable of running on very large display
systems. Originally written for the immersive semi-cylindrical projection
system at the University of Siegen, its flexibility enabled its use in many
installations. Bino decodes video on each rendering thread and only
synchronises the time step globally, providing a scalable solution to video
playback. Bino uses the 2D information from the segment viewports to lay out
the video tiles for each projector.

\section{Omegalib}

\begin{wrapfloat}{figure}{O}{.618\textwidth}
  \includegraphics[width=.618\textwidth]{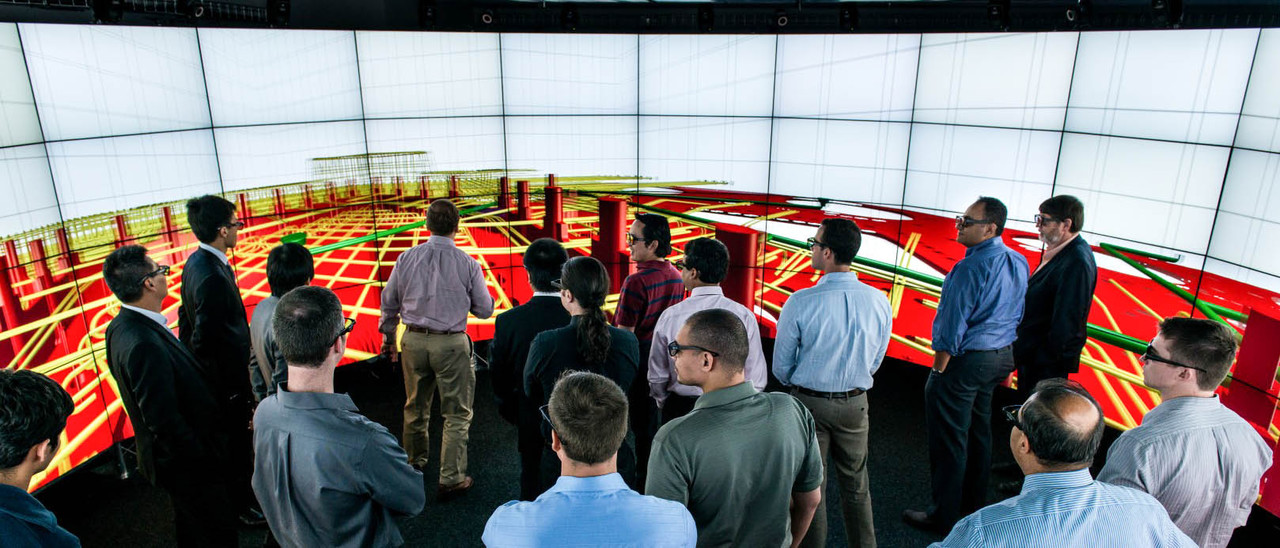}
  {\caption{\label{fOmegalib}An Omegalib Application running in the Cave2}}
\end{wrapfloat}

Omegalib \cite{Omegalib} is a software framework built on top of Equalizer that
facilitates application development for hybrid reality environments, like the
Cave~2. Hybrid reality environments aim to create a seamless 2D/3D environment
supporting both information-rich analysis (traditionally done on tiled
display wall), as well as virtual reality simulation exploration (traditionally
done in VR systems) at a resolution matching human visual acuity. Omegalib
supports dynamic reconfigurability of the display environment, so that areas of
the display can be interactively allocated to 2D or 3D workspaces as needed. It
is possible to have multiple immersive applications running on a
cluster-controlled display system, have different input sources dynamically
routed to applications, and have rendering results optionally redirected to a
distributed compositing manager. Omegalib supports pluggable front-ends to
simplify the integration of third-party libraries like OpenGL, OpenSceneGraph,
and the Visualisation Toolkit (VTK).

\chapter{Conclusion}\label{sConclusion}

\section{Summary}

Formalising, designing and implementing a generic parallel rendering framework,
that can serve both complex applications and research, has been no easy task.
Based on the analysis of Cavelib, practical experience in implementing and
deploying OpenGL Multipipe SDK, we have been in the unique position to make
significant contributions in this area. Equalizer, our parallel rendering
framework, allowed us to take parallel rendering research to a new level. It
enabled us to easily implement new decomposition algorithms, many
improvements for result composition, novel load balancing schemes, and numerous
whole system optimisations, all of which are much harder to research without such a
framework and associated applications. This is not only supported by the
contributions of this thesis, but by other publications and doctoral theses
completed using Equalizer. This research has not only been performed in the
original research group; Equalizer has also been picked up by other
laboratories, e.g., the Electronic visualisation Laboratory at the University
of Illinois at Chicago for Cave2 research.

Beyond the core system design, we have incorporated many new parallel rendering
algorithms into our framework. Most notably, cross-segment load balancing
provides a novel approach to better assign multiple rendering resources to
multi-display systems. It maximises rendering locality for the display GPUs and
is not limited to planar displays, compared to other approaches.

Having a fully-featured rendering framework and real-world applications enabled
us to implement many algorithmic improvements and optimisations, and evaluate
them in a holistic and realistic setup. The results of this work advance
parallel rendering with new decomposition modes, compositing algorithms, better
load balancing and an asynchronous rendering pipeline. Last, but not least, a
network library for distributed, interactive visualisation applications greatly
facilitates the task to distribute and synchronise application state in a
parallel rendering system.

Beyond the scope of this thesis, Equalizer has influenced the field and has
been used in various commercial and research applications. These applications
span a wide field of domains, from virtual prototyping, interactive raytracing,
large-scale volume rendering, terrain rendering, neuroscience applications, to
next-generation visualisation systems such as collaborative tiled display walls
and hybrid 2D/3D setups such as the Cave2.

\section{Future Work}

We consider the core parallel rendering framework largely feature complete,
with the exception of keeping up with new technologies, e.g., providing glue
code for the Vulkan API or exploiting new Multi-GPU extensions. There remains a
large amount of work to make parallel rendering more accessible. This may be
addressed by simplified APIs layered on top of Equalizer, and through
integrations with popular rendering toolkits. Future work should also address
operators and users of visualisation systems through simplified configuration,
monitoring and administration tools.

There is still a significant amount of research in automatically selecting the
best decomposition and recomposition algorithm, as well as the resources used
for a given application. This task becomes even more challenging when
considering changes in the rendering load and algorithm during the runtime of
an application. Furthermore, implementing load balancing for the compositing
task is an area largely unexplored, in particular in combination with state of
the art optimisations.

We foresee an increasing importance for interactive raytracing, which has its
own set of challenges for parallel rendering. In particular for large data
rendering, there are a number of open questions, like out-of-core parallel
raytracing and data-parallel decomposition with global illumination.

Load balancing for better utilisation of available resources, and increased
scalability to higher node counts remains an open area of research. While this
thesis provides many new results in this area, a comprehensive benchmark and
study of different algorithms and applications would be very valuable,
which may lead to the discovery of new load-balancing algorithms.

One of the remaining challenges is to make interactive supercomputing
accessible. Significant research has been performed on how to link simulations
with visualisation, and how to use this monitoring to interactively steer the
simulation. These advances now need to be translated into easily usable
software components, integrated well with existing resource management systems.



\definecolor{numColor}{rgb}{0.22,0.37,0.56}


\makeatletter
\def\@makechapterhead#1{%

    \line(1,0){0}
    \newline
    {
	 \begin{tabular}{@{}lr@{}@{}} 
	 \linethickness{ 4px }\color{numColor}\line(1,0){260} 
	 & \multirow{2}{*}{\fontsize{100}{62}\usefont{OT1}{ptm}{m}{n}\selectfont \color{numColor} \thechapter}\\ 
	 & \\ 
	 \scshape \LARGE \usefont{T1}{fvs}{sc}{n}\selectfont \letterspace to 2.5\naturalwidth{APPENDIX} \hspace{19mm}
	 & \\ 
	 \end{tabular}

         \vskip 100\p@
         \raggedleft
         \interlinepenalty\@M
         \scshape \fontsize{24}{30} \usefont{T1}{fvs}{n}{n}\selectfont \scshape \MakeUppercase{#1}\par\nobreak
         \vskip 80\p@
    }
}

\appendix

\bibliographystyle{apalike}
\bibliography{references/references}

\nociteconf{EAA:17, SPEP:16, EDB:16, HBBES:13, EBAHMP:12, EEP:11, MEP:10, EP:07, BRE:05}
\bibliographystyleconf{apalike}
\bibliographyconf{references/references}                   

\nocitejournal{ESP:18, EMP:09}
\bibliographystylejournal{apalike}
\bibliographyjournal{references/references}                   

\markboth{CURRICULUM VITAE}{}

\chapter*{Curriculum Vitae}
\addcontentsline{toc}{chapter}{Curriculum Vitae}

\setlength{\parindent}{0em}
\newcommand{\tind}{\hspace{-\tabcolsep}}
\newcommand{\exrule}{\vspace{.2em}\hrule\vspace{.2em}}
\def\Cplusplus{{\rm C\raise.2ex\hbox{\small ++}}}
\sloppy

\setlength{\parskip}{1.5em plus .5em minus .5em}

\parbox[t]{2.5cm}{\scshape Particulars}
\parbox[t]{12.5cm}{
  \begin{tabularx}{12.5cm}[t]{lX}
    \tind Date of Birth & 9th August 1975, Wittenberg, Germany\\
    \tind Nationality   & German, Swiss\\
    \tind Languages     & German (native), English (fluent), French (fluent)\\
    \tind Open Source Profile &
\htmladdnormallink{github.com/eile}{https://github.com/eile}\\
  \end{tabularx}
}\\

\parbox[t]{2.5cm}{\scshape Profile}
\parbox[t]{12.5cm}{Senior software engineer and technical team lead, with a
  specialization in interactive large data visualization, \Cplusplus, parallel
  and distributed programming. Successful track record of building and leading
  engineering teams to success.
}\\

\parbox[t]{2.5cm}{\scshape Expertise}
\parbox[t]{12.5cm}{
  Technical leadership for high performance C++ applications, parallel
    programming, distributed systems, Virtual Reality and collaborative
    visualization\vspace{.5em}\\
  Software and library design, test driven development and maintenance
    using \Cplusplus, Typescript, Python, CMake and git\vspace{.5em}\\
  Software development methodology during the whole lifecycle,
    ranging from requirements analysis, specification, design,
    implementation to documentation, education, debugging, optimization and
    support\vspace{.5em}\\
  Broad knowledge of operating systems: Mac OS X, Linux, Windows, Irix
}\\

\parbox[t]{2.5cm}{\scshape Experience}
\parbox[t]{12.5cm}{
  {\em Frontend Software Engineer} \hfill {\bf ESRI R\&D Center}\\
  Z{\"u}rich, Switzerland  \hfill {\bf Nov 2017 -- current}\exrule

  Development of frontend APIs and rendering algorithms for 3D mapping.}\\

\parbox[t]{2.5cm}{\hspace{1pt}}
\parbox[t]{12.5cm}{
  {\em Researcher, Parallel Rendering} \hfill {\bf University of Z"urich}\\
  Z"urich, Switzerland \hfill {\bf 2005 -- 2007, October 2015 -- current}\exrule

  Research new algorithms for large data visualization, in particular the
  parallelization, load-balancing and data distribution of parallel OpenGL
  applications on graphics clusters. Invented and developed Equalizer, a
  framework for scalable, distributed OpenGL applications.}\\

\parbox[t]{2.5cm}{\hspace{1pt}}
\parbox[t]{12.5cm}{
  {\em Visualization Team Manager} \hfill {\bf Blue Brain Project, EPFL}\\
  Lausanne, Switzerland  \hfill {\bf May 2011 -- Sep 2017}\exrule

  Built a team of seven software engineers, one post-doc, one PhD student and one
  media designer to deliver innovative visualization software as well as media
  for communication and scientific publications. Developed the long-term
  interactive supercomputing vision and the corresponding medium-term roadmap
  with the team, motivated and lead the implementation based on modular software
  components. Drove the implementation of software engineering best practices
  for the whole project.}\\

\parbox[t]{2.5cm}{\hspace{1pt}}
\parbox[t]{12.5cm}{
  {\em CEO and Founder} \hfill {\bf Eyescale Software GmbH}\\
  Neuch\^atel, Switzerland  \hfill {\bf January 2007 -- current}\exrule

  Co-founder of Eyescale and lead developer of the Equalizer parallel rendering
  framework and related libraries. Deploying Equalizer in existing ISV
  applications to scale display size, performance and visual quality. Software
  architecture, design and development, hardware and software consulting for
  multi-GPU workstations, visualization clusters and Virtual Reality.  }\\

\parbox[t]{2.5cm}{\hspace{1pt}}
\parbox[t]{12.5cm}{
  {\em Senior Software Engineer, 3D Graphics} \hfill {\bf Tungsten Graphics}\\
  Neuch\^atel, Switzerland  \hfill {\bf January 2007 -- June 2007}\exrule
  {\em Senior Software Engineer} \hfill {\bf Esmertec AG}\\
  Neuch\^atel, Switzerland  \hfill {\bf January 2004 -- September 2005}\exrule

  Job position details available on demand.
}\\

\parbox[t]{2.5cm}{\hspace{1pt}}
\parbox[t]{12.5cm}{
  {\em Senior Software Engineer}  \hfill {\bf Silicon Graphics, Inc.}\\
  Neuch\^atel, Switzerland \hfill {\bf August 2000 -- December 2003}\exrule

  Worked in SGI's advanced graphics division as technical lead for OpenGL
  Multipipe SDK (MPK), a framework to develop high performance, scalable
  visualization software. Worked on DataSync, a distributed shared memory API
  for clusters.  }\\

\parbox[t]{2.5cm}{\hspace{1pt}}
\parbox[t]{12.5cm}{
  {\em Software Engineer} \hfill {\bf Freelancer}\\
  Munich, Germany         \hfill {\bf April 2000 -- July 2000}\exrule
  {\em Software Engineer} \hfill {\bf Intec GmbH}\\
  Wessling, Germany       \hfill {\bf October 1998 -- March 2000}\exrule

  Job position details available on demand.
}\vspace{1em}\\

\parbox[t]{2.5cm}{\scshape Education}
\parbox[t]{12.5cm}{
  University of Zurich\\
  PhD student in Computer Science, Fall 2015 - Spring 2019
  \vspace{.3em}\\
  \`Ecole Polytechnique F\`ed\`erale de Lausanne\\
  Master in Computer Science, October 2015, Grade 5.6/6.0
  \vspace{.3em}\\
  Berufsakademie Heidenheim\\
  Dipl.-Ing. (eq BS) in Computer Science, September 1998
}\vspace{1em}

\end{document}